\documentclass[useAMS,usenatbib]{mnras}
\usepackage{times}
\usepackage{graphicx, latexsym, amssymb, amscd, psfrag, cancel}
\usepackage{epsfig}
\usepackage{color}
\usepackage{amsmath}
\usepackage[upgreek, LGRgreek]{mathastext}
\usepackage{ulem}
\title[Star formation and gas content in CLoGS BGEs]{The Complete Local Volume Groups Sample - IV. Star formation and gas content in group-dominant galaxies}

\author[Konstantinos Kolokythas et al.]{Konstantinos Kolokythas,$^{1}$\thanks{E-mail: k.kolok@nwu.ac.za (KK)} Sravani Vaddi,$^{2}$ Ewan O'Sullivan,$^{3}$ Ilani Loubser,$^{1}$ \newauthor{Arif Babul,$^{4,5}$ Somak Raychaudhury,$^{6,7,8}$ Patricio Lagos$^{9}$ and Thomas H. Jarrett}$^{10}$\\ \\ 
$^{1}$Centre for Space Research, North-West University, Potchefstroom 2520, South Africa\\
$^{2}$ Arecibo Observatory, NAIC, HC3 Box 53995, Arecibo, Puerto Rico, PR 00612, USA\\
$^{3}$Center for Astrophysics $|$ Harvard \& Smithsonian, 60 Garden Street,
Cambridge, MA 02138, USA\\
$^{4}$Department of Physics and Astronomy, University of Victoria, Victoria, BC V8P 1A1, Canada\\
$^{5}$Center for Theoretical Astrophysics and Cosmology, Institute for Computational Science, University of Zurich, Winterthurerstrasse 190, 8057 Zurich, Switzerland\\
$^{6}$Inter-University Centre for Astronomy and Astrophysics, Pune University Campus, Ganeshkhind, Pune, Maharashtra 411007, India\\
$^{7}$School of Physics and Astronomy, University of Birmingham, Birmingham B15~2TT, UK\\
$^{8}$Department of Physics, Presidency University, 86/1 College Street, Kolkata 700073, India\\
$^{9}$Instituto de Astrof\'{i}sica e Ci\^{e}ncias do Espa\c{c}o, Universidade do Porto, CAUP, Rua das Estrelas, 4150-762 Porto, Portugal\\
$^{10}$Department of Astronomy, University of Cape Town, Private Bag X3, Rondebosch 7701, South Africa}

\newcommand{\kms}{~km~s$^{-1}$}

\newcommand\kmsmpc{km~s$^{-1}$~Mpc$^{-1}$}
\newcommand{\sfrunit}{M$_{\odot}~yr^{-1}$}
\newcommand{\msol}{M$_{\odot}$}

\begin{document}

\date{Accepted 2021 December 14. Received 2021 November 16; in original form 2021 August 25}

\pagerange{\pageref{firstpage}--\pageref{lastpage}} \pubyear{2021}
\maketitle

\label{firstpage}
 \begin{abstract}
Using multi-band data we examine the star formation activity of the nearby group--dominant early--type galaxies of the Complete Local-volume Groups Sample (CLoGS), and the relation between star formation, gas content, and local environment. Only a small fraction of the galaxies (13\%; 6/47) are found to be Far-Ultraviolet (FUV) bright, with FUV to near-infrared colours indicative of recent active star formation (NGC~252, NGC~924, NGC~940, NGC~1106, NGC~7252 and ESO~507-25). These systems are lenticulars presenting the highest FUV specific star-formation rates in the sample (sSFR$_{\rm FUV}>$5$\times$10$^{13}$~yr$^{-1}$), significant cold gas reservoirs (M(H$_2$)=0.5-61$\times$10$^8$~M$_\odot$), reside in X-ray faint groups, and none hosts a powerful radio AGN (P$_{1.4GHz}$ $<$10$^{23}$ W~Hz$^{-1}$). The majority of the group-dominant galaxies (87\%; 41/47) are FUV faint, with no significant star formation, classified in most cases as spheroids based on their position on the infrared star-forming main sequence (87\%; 46/53). Examining the relationships between radio power, SFR$_{FUV}$ and stellar mass we find a lack of correlation that suggests a combination of origins for the cool gas in these galaxies, including stellar mass loss, cooling from the intra-group medium (IGrM) or galaxy halo, and acquisition through mergers or tidal interactions. X-ray bright systems, in addition to hosting radio powerful AGN, have a range of SFRs but, with the exception of NGC~315, do not rise to the highest rates seen in the FUV bright systems. We suggest that central group galaxy evolution is linked to gas mass availability, with star formation favoured in the absence of a group-scale X-ray halo, but AGN jet launching is more likely in systems with a cooling IGrM.

 \end{abstract}
 
 \begin{keywords}
   galaxies: groups: general --- galaxies: star-formation ---  galaxies: active --- galaxies: jets --- galaxies: groups: individual ---  radio continuum: galaxies
 \end{keywords}

 
 \section{Introduction} 

The hierarchical nature of structure formation means that the position of a galaxy in the larger dark matter halo, either central or as a satellite, is closely related to its evolutionary path. Central early-type galaxies grow through both mergers and gas inflows over time and eventually become the brightest and most massive galaxies in the halo \citep{WhiteRees78,Broughetal06}. It has long been held that most central galaxies are relatively passive, with low star-formation rates and limited cold gas reservoirs. However, various studies have revealed that both lenticular and elliptical central galaxies are capable of harboring significant cold gas masses, capable of fuelling an active galactic nucleus (AGN) or star formation (e.g., \citealt{Bildfelletal08,Morgantietal06,Combesetal07,Pipinoetal09,Oosterlooetal10,Loubseretal13,Loubseretal16}, see also review by \citealt{Oppenheimeretal21} specifically for simulations of groups). Galaxies in higher density environments (groups or clusters)
present different properties than their equivalents in the field \citep[e.g.,][]{Baloghetal04,Baldryetal06,SkibbaSheth09} with central galaxies in low-mass groups found to possess higher cold gas content than similar galaxies in isolation \citep{Janowieckietal17}. Therefore the properties that galaxies present have a strong dependence on the local environment in which galaxies reside and evolve \citep[e.g.,][]{Pengetal10,Wetzeletal12,Darvishetal17}.

A variety of mechanisms can affect the star-formation rate and cold gas content of galaxies in groups and clusters. These can be considered in terms of two separate pathways: i) either internal evolutionary processes (secular processes) that are only related to the galaxy itself or ii) mechanisms induced by their environment (such as mergers). Disentangling the relevant contribution from each mechanism and comprehending the nature and impact of the various environmental processes on galaxy evolution is of significant importance.

Observations suggest that the most likely mechanism for star formation quenching in galaxies with a mass greater than 10$^{10}$~M$_{\odot}$ is in general associated with their internal evolution \citep{Donnarietal21}. One of the most important “secular” processes is feedback by AGN which is known to play an important role in influencing galaxy evolution by regulating or quenching star formation \citep[e.g.,][]{Babuletal02,DiMatteoetal05,Hopkinsetal08,Hickoxetal09,Silk13,Hopkinsetal14,SomervilleDave15,Beckmannetal17,Oppenheimeretal21} transforming galaxies into red and passive systems \citep[e.g.,][]{KormendyHo13,HeckmanBest14}. The energy released from the central AGN is capable of expelling gas from the galaxy (radiative mode feedback) or depositing energy into the environment and heating up the intergalactic medium (mechanical feedback; \citealt{Nusseretal06,Fabian12,Oppenheimeretal21}). AGN feedback has also been suggested to be responsible for
the observed correlations between the evolution of the host galaxy and the growth of the central SMBH \citep[e.g.,][]{Magorrianetal98,SilkRees98,FerrareseMerritt00} although it has also been argued that these correlations may also be a product of a merger related/induced galaxy growth \citep{Peng07,JahnkeMaccio11}.
However, in order to trigger an AGN its fueling is essential and depends i) on the available gas supply (mainly affected by the large-scale environment in which a galaxy resides), and ii) the efficiency with which the gas is transported to the central SMBH, with the gas transfer efficiency to the centre being closely linked to the triggering AGN mechanisms, i.e., mergers \citep[e.g.,][]{Hopkinsetal08b}, secular processes (e.g., bar instabilities; \citealt{Gouldingetal14}), or direct precipitation from the surrounding gas \citep{Prasadetal15,Prasadetal17,Prasadetal18}.

In addition to secular processes, it is evident that galaxy mergers and interactions can also play an important role in the evolution of a galaxy affecting its evolutionary path and properties (e.g., gas consumption via starbursts; \citealt{vanderWeletal09}, gas transfer efficiency to the AGN etc) as they are powerful mechanisms that can also induce star formation \citep{Kennicutt98}. The local environment and the time that these galaxy mergers take place have an impact on the evolution of galaxies. Since galaxy-galaxy mergers and tidal interactions occur more often in galaxy groups than in clusters \citep{McIntoshetal08} and since galaxy groups are the most common environment where galaxies reside \citep[e.g.,][]{Tully87,Ekeetal04,Knobeletal15,Saulderetal16} it is important to examine and understand how galaxies evolve within such structures.


In this paper we investigate the star formation properties and evolution of the dominant early-type galaxies in the Complete Local-volume Groups Sample (CLoGS), in relation to their gas content and large-scale environment. We try to disentangle the relevant contribution of the various galaxy evolution mechanisms using multiple wavelengths to examine the star-formation rate, specific star-formation rate, central AGN radio power and stellar mass, and to classify the state of BGEs based on diagnostics and providing a qualitative comparison between early-types in galaxy groups, galaxy clusters and the field.

The paper is organized as follows. In Section 2 we present the sample of CLoGS galaxy groups and briefly describe the selection criteria. In Section 3 we describe the radio, X-ray, UV and mid-infrared data along with the methods used to retrieve the magnitude in the UV and for correcting the galactic extinction correction, as well as the method followed for the corrected  mid-infrared photometry. In Section 4 we present and discuss our results along with their implications on galaxy groups. The summary and the conclusions are given in Section 5. Throughout the paper we adopt the $\Lambda$CDM cosmology with $H_o=71$ \kmsmpc, $\Omega_m$ = 0.27, and $\Omega_\Lambda$ = 0.73.

\section{The Sample}

We use the Complete Local-volume Groups Sample (CLoGS), an optically-selected, statistically-complete sample of groups in the nearby Universe targeted to facilitate detailed studies in the radio, X-ray and optical bands. A detailed description of the CLoGS sample selection criteria along with the X-ray properties of the high--richness sub-sample are provided in \citet{OSullivanetal17}, and the radio properties of the group--dominant galaxies are discussed in \citet{Kolokythasetal18,Kolokythasetal19}. The cold gas content of the dominant galaxies (molecular and H\textsc{i}) is described in \citet{OSullivanetal15,OSullivanetal18b}.  We provide here a brief description of the sample.

CLoGS consists of 53 groups, all in the local Universe ($\leq$80 Mpc), drawn from the shallow, all-sky Lyon Galaxy Group catalog (LGG; \citealt{Garciaetal93}), with each group selected to have a minimum of 4 members and at least one luminous early-type galaxy (L$_B$ $>$ 3$\times$10$^{10}$ L$_\odot$). Declination was required to be $>$30$^\circ$ in order to ensure visibility from the GMRT and Very Large Array (VLA). 

The brightest group--member early--type (BGE) galaxy was assumed to be the dominant galaxy of each group, and in X--ray bright systems the BGE typically lies at the center of the hot intra--group medium. A richness $R$ was estimated for each group, defined as the number of member galaxies with log L$_B$ $\geq$ 10.2. Systems with $R>10$ were excluded as they correspond to known galaxy clusters, as were groups with $R$ = 1, since they are not rich enough to give a trustworthy determination of their physical parameters. The final CLoGS sample consists of 53 groups which can be divided into two sub-samples: i) the 26 high-richness groups with $R$ = 4$-$8, and ii) the 27 low-richness groups $R$ = 2$-$3.  In this paper we examine the properties of all 53 BGEs.

\section{The DATA}

To examine the environment and star formation in the 53 BGEs, we combine data from multiple wavelengths, from the radio, through infrared and UV, to the X-ray. We use the radio emission at 1.4~GHz to study the AGN activity of the galaxies, X-rays for the properties of the large-scale environment, UV and mid-infrared to examine their star-formation rates (SFRs), and K$_{s}$ band to determine the galaxy stellar mass. The selection of the 1.4~GHz frequency for the radio data was based on the ability to compare our results with previous relevant studies. In Table~\ref{Sourcetable} we summarize the details of the available information for each dominant galaxy.

\subsection{Radio data} 

The 1.4~GHz data were drawn primarily from the NRAO VLA Sky Survey (NVSS, \citealt{Condonetal98}), the Faint Images of the Radio Sky at Twenty Centimeters Survey (FIRST, \citealt{Beckeretal95})  and the study of \citet{Brownetal11}. For the galaxies NGC~193, NGC~4261 and NGC~5903 we used the studies of \citet{Condonetal02}, \citet{Kuhretal81} and \citep{OSullivanetal18b} respectively. The angular resolution of NVSS is 45$''$ and the detection limit of the catalog at 1$\sigma$ level of significance is 0.45~mJy/beam, while the FIRST survey has a resolution of 5$''$ and an r.m.s. of 0.2~mJy/beam. For 4 galaxies (NGC~924, NGC~940, NGC~978 and NGC~5061), with no previous measurements at 1.4~GHz we extrapolated their radio flux densities from 610~MHz GMRT values drawn from the studies of \citet{Kolokythasetal18} and \citet{Kolokythasetal19}, using the formula:

\begin{equation}
\label{Flux1400}
S_{1.4GHz}=S_{610\small{MHz}}\left(\frac{1.4~GHz}{610~MHz}\right)^{-0.8}.
\end{equation}

Using the measured or derived radio flux densities, we calculated the radio power at 1.4~GHz in each BGE as: 

\begin{equation}
\label{Power1400}
P_{1.4~GHz}=4 \pi D^2 (1+{\rm z})^{(\alpha-1)}S_{1.4~GHz},
\end{equation}

where D is the distance to the source, $\alpha$ is the spectral index, $z$ the redshift and S$_{1.4~GHz}$ is the flux density of the source at 1.4~GHz. In the cases of NGC~924, NGC~978 and NGC~5061 where no spectral index is known for the sources, $\alpha$ was by default set to $0.8$, which is the typical value for extragalactic radio sources \citep{Condon92}. For the BGEs with no radio flux density available at any frequency, we adopt and upper limit at 5$\times$ the typical rms noise of the NVSS catalog, i.e., $<$2.25~mJy/beam (see Table~\ref{Sourcetable} and Table~\ref{infotable}).

\subsection{X-rays} 
 X-ray data are available for the high--richness CLoGS sub-sample based on observations from \textit{Chandra} and \textit{XMM-Newton} observatories. \citet{OSullivanetal17} show that 14/26 systems of the high--richness sub-sample have a detected intra-group medium (IGrM) containing X-ray emitting gas that extends $>65$~kpc, with a luminosity $>10^{41}$~erg~s$^{-1}$. We will refer to these systems as X-ray bright. The typical halo temperatures for these X-ray bright systems are found to be in the range $\sim0.4-1.4$ keV, corresponding to masses in the range $M_{500}\sim0.5-5\times 10^{13} M_{\odot}$, with X-ray luminosities in the $0.5-7$ keV band being between $L_{X,R500}\sim2-200\times10^{41}$ erg s$^{-1}$. 
 

  \subsection{Infrared data} 
  
  \label{WISE}

\subsubsection{K$_s$-band Magnitude}

The near-infrared K$_s$-band magnitudes were taken from the Two Micron All-Sky Survey (2MASS; \citealt{Jarrettetal03}), which has a sensitivity of $\sim$1~mJy at the 10$\sigma$ level of significance for point sources, and a resolution of 2$-$3$''$. To determine the absolute K$_s$-band magnitude (M$_{K_s}$) for each CLoGS dominant galaxy, we use as in \citet{Maetal14,Vealeetal17} and \citet{Loubseretal18} the total extrapolated K$_s$-band magnitude from the extended source catalog (XSC, parameter k\_m\_ext), which is measured in an aperture that includes the isophotal aperture along with the extrapolation of the surface-brightness profile based on a single S$\acute{e}$rsic fit to the inner profile \citep{Jarrettetal03}. We calculate the absolute K$_s$-band magnitudes using the method followed in \citet{Maetal14} (see their equation~1). The calculated values of K\_m\_ext, A${_v}$ (for Galactic extinction correction), and M$_{K_s}$ for our systems are listed in Table~\ref{Sourcetable}. These K$_s$-band magnitudes are then used to estimate the stellar masses of the CLoGS dominant galaxies (\S~\ref{stellarmass}). We do not expect any satellite contributions in our results.

The two main sources of uncertainty on M$_{K_s}$ that can impact the estimation of M$_{stellar}$ are i) the possible underestimation of the \textit{2MASS} K$_s$-band magnitudes due to the shallowness of the survey \citep{Laueretal07,SchombertSmith12} and ii) the selection of the distance estimation and extinction (for more details see \citealt{Maetal14}). Based on the nine galaxies in common with CLoGS that are listed in MASSIVE (from \citealt{Maetal14}) and the five galaxies in common  with ATLAS$^{3D}$ \citep{Cappellarietal11}, we compare with our CLoGS absolute K$_s$-band luminosities as all these studies have used the 2MASS XCS to obtain the K-band magnitudes. We find that on average our values differ  around $\sim$0.14~mag (ranging from 0.01 $-$ 0.35~mag). The only exception is NGC~4697, for which a $\sim$1.0 mag difference is found. This can be attributed to the different distance used between the CLoGS sample (\citealt{Garciaetal93}; 18~Mpc) and the one used by ATLAS$^{3D}$ study (\citealt{Tonryetal01}, that was corrected by subtracting 0.06 mag to the distance modulus of \citealt{Meietal07}; 11~Mpc).

\subsubsection{WISE}

Mid-infrared data were drawn from the \textit{WISE} mission catalog \citep{Wrightetal10} at 3.4 $\mu$m (W1), 4.6 $\mu$m (W2), 12 $\mu$m (W3) and 22 $\mu$m (W4) with an angular resolution of 6.1$''$, 6.4$''$, 6.5$''$ and 12$''$ respectively. The sensitivity of \textit{WISE} at 5$\sigma$ level of significance for a point source is 0.08~mJy at W1, 0.11~mJy at W2, 1~mJy at W3 and 6~mJy at W4 wavelengths.

Following the study of \citet{Vaddietal16} we performed photometry to obtain magnitudes for our galaxies in each of the \textit{WISE} bands, since extended source measurements in the \textit{WISE} catalogue are measured using \textit{2MASS} apertures. This can lead to either \textit{i)} an underestimation of the total flux of about 30-40\% \citep{Cutrietal12}, or \textit{ii)} a total flux overestimation, since no masking of nearby sources was applied and fluxes can be contaminated by stars or close galaxy companions.

The technique used is similar to the one described in \citet{Vaddietal16}. Surface photometry is performed in PyRAF using the ELLIPSE task.  The task fits isophotes, the lines of constant surface brightness, to the input galaxy images.  The isophote at which the intensity is 1$\sigma$ above the mean sky background is taken as the galaxy aperture that encloses all of the galaxy light; background subtracted integrated light within this aperture is taken as the total flux of the galaxy.  The sky background is estimated using the task FITSKY.  During the isophotal fitting, masks are used to exclude contaminating light from foreground stars and companion galaxies. We note that particular caution should be taken with NGC~2292, which is part of a very close interacting galaxy pair with tidal tails and a surrounding ring structure.  The estimated WISE magnitudes for this galaxy use an aperture that encloses the bulge and not the tidal ring.

 \subsection{UV data} 
Magnitude values for Far-UV (FUV) and Near-UV (NUV) data were retrieved from the \textit{Galaxy Evolution Explorer} (\textit{GALEX}; \citealt{Martinetal05}) GR6 data release\footnote{\href{https://galex.stsci.edu/GR6}{https://galex.stsci.edu/GR6}}. The \textit{GALEX} mission conducted an all--sky survey using wide-field UV imaging at FUV  ($\lambda_{eff}  = 1539$~\r{A}, $\Delta\lambda= 1344 - 1786$~\r{A}) and NUV ($\lambda_{eff} = 2316$~\r{A}, $\Delta\lambda = 1771 - 2831$~\r{A}) wavelengths, with limiting magnitudes of 24.7 and 25.5 respectively. The survey angular resolution for FUV is 4.2$''$ and for NUV 5.3$''$. 

\subsection{Galactic extinction correction} 
All magnitudes in K$_{\huge{s}}$, FUV and NUV bands were corrected for galactic extinction using the method of \citet{Wyderetal05}. 
For the calculation of the extinction we used the ratio: $\frac{A_\lambda}{E(B-V)},$ where A$_{\lambda}$ is the extinction correction at wavelength $\lambda$, and has values for K$_{\huge{s}}$, FUV and NUV of 0.347, 8.376 and 8.741 respectively. The values for the colour excess $E(B-V)$ were drawn from the \textit{GALEX} GR6 catalog.

\begin{table*} 
\caption{Observational properties of the BGEs of our sample. The columns list the BGE name, redshift, the morphology type, the Ks band magnitude (K\_m\_ext), the FUV magnitude, the E(B-V), the  flux density at 1.4~GHz (drawn from the literature) and the W1, W2, W3 and W4 photometry corrected magnitudes.  The references for the 1.4~GHz flux densities are listed at the bottom of the table.  \label{Sourcetable}}
\begin{center}
\begin{tabular}{lcccccccccc}
\hline 
 BGE& Redshift&T$_{type}$ , Morph & K$_{\huge{s}}$  & FUV & E($B-V$) & S$_{1.4GHz}$ & W1 & W2 & W3 & \textsc{w4} \\ 
      & ($z$) &     & (mag) &   (mag)   & (mag)   &   (mJy)     & (mag)  & (mag) & (mag)  & (mag) \\
\hline
\multicolumn{9}{l}{\textsc{High--Richness Subsample}}\\
 NGC 193  & 0.014723 & -3.0 , E & 9.21$\pm0.03$ & 19.53 & 0.0222 & 1710$\pm102^a$ & 8.82$\pm$0.03 & 8.81$\pm$0.02 & 8.31$\pm$0.01 & 6.95$\pm$0.02 \\
 NGC 410   &  0.017659 & -4.4 , E & 8.38$\pm0.02$ & 18.44 & 0.0581 & 6.3$\pm0.6^b$  & 8.07$\pm$0.03 & 8.08$\pm$0.03 &  7.67$\pm$0.01 & 6.48$\pm$0.01 \\
 NGC 584   &  0.006011 & -4.6 , E & 7.30$\pm0.03$ & 18.01 & 0.0424 & 0.6$\pm0.5^c$  & 7.27$\pm$0.02 & 7.31$\pm$0.02 &     6.68$\pm$0.01 & 5.73$\pm$0.01 \\
 NGC 677   &  0.017012 & -4.9 , E & 9.30$\pm0.03$ & 19.06 & 0.0887 & 20.6$\pm1.6^b$ & 8.59$\pm$0.02 & 8.85$\pm$0.02 &  8.98$\pm$0.01 & 8.98$\pm$0.04 \\
 NGC 777   &  0.016728 & -4.8 , E & 8.37$\pm0.02$ & 18.18 & 0.0466 & 7.0$\pm0.5^c$  & 8.36$\pm$0.04 & 8.27$\pm$0.02 & 8.01$\pm$0.01 & 7.01$\pm$0.02\\
 NGC 940   &  0.017075 &  -2.0 , S0 & 9.36$\pm0.02$ & 17.80 & 0.0897 &      -     &   9.11$\pm$0.04   & 9.12$\pm$0.04 &  7.14$\pm$0.01 & 5.64$\pm$0.02 \\
 NGC 924   &  0.014880 &  -2.0 , S0 & 9.50$\pm0.03$ & 17.76 & 0.1500 &      -     &   9.26$\pm$0.04   & 9.27$\pm$0.04 & 8.26$\pm$0.01 & 6.81$\pm$0.01  \\
 NGC 978   &  0.015794 &  -3.0 , E/S0 & 9.11$\pm0.02$ & 19.24 & 0.0911 &      -     &   8.77$\pm$0.04   & 8.79$\pm$0.03 & 8.52$\pm$0.01 & 7.20$\pm$0.02  \\
 NGC 1060  &  0.017312 &  -3.0 , E/S0 & 8.20$\pm0.02$ & 18.62 & 0.1916 & 9.2$\pm0.5^c$  & 7.75$\pm$0.02 & 7.65$\pm$0.02 & 8.16$\pm$0.01 & 7.21$\pm$0.02  \\
 NGC 1167  &  0.016495 & -2.4 , S0 & 8.64$\pm0.02$ &   -   & 0.0914 & 1700$\pm100^c$ & 8.35$\pm$0.03 & 8.34$\pm$0.03 & 7.04$\pm$0.01 & 5.89$\pm$0.01  \\
 NGC 1453  &  0.012962 & -4.7 , E & 8.12$\pm0.02$ & 18.77 & 0.1006 & 28.0$\pm1^c$   & 7.82$\pm$0.02 & 7.87$\pm$0.02 & 7.21$\pm$0.01 & 6.81$\pm$0.03  \\
 NGC 1587 & 0.01230& -4.8 , E & 8.51$\pm0.03$ & 18.77 & 0.0721 &  131$\pm5^c$   & 8.34$\pm$0.03 & 8.35$\pm$0.03 & 7.91$\pm$0.01 & 7.09$\pm$0.02  \\
 NGC 2563  &  0.014944 & -2.1 , S0 & 9.01$\pm0.02$ & 18.85 & 0.0432 & 0.3$\pm0.5^c$  & 8.68$\pm$0.03 & 8.54$\pm$0.02 & 8.59$\pm$0.01 & 8.36$\pm$0.03  \\
 NGC 3078  &  0.008606 & -4.7 , E & 7.88$\pm0.02$  & 18.40 & 0.0718 & 310$\pm10^c$   & 7.65$\pm$0.03 & 7.63$\pm$0.02 & 7.27$\pm$0.01 & 6.23$\pm$0.02  \\
 NGC 4008  &  0.012075 & -4.3 , E & 8.83$\pm0.02$  & 19.57 & 0.0230 & 10.9$\pm0.5^c$ & 8.61$\pm$0.03 & 8.62$\pm$0.02 & 8.43$\pm$0.01 & 8.31$\pm$0.04  \\
 NGC 4169  &  0.012622 & -2 , S0  & 9.08$\pm0.02$  &   -  & 0.0211 & 1.07$\pm0.15^d$ &  8.82$\pm$0.03 & 8.95$\pm$0.04 & 7.72$\pm$0.01 & 6.41$\pm$0.02  \\   
 NGC 4261 & 0.007465& -4.8 , E & 7.26$\pm0.03$ & 16.79 &  0.0204  & 19510$\pm410^e$ & 6.93$\pm$0.01 & 7.09$\pm$0.01 & 6.31$\pm$0.01 & 4.98$\pm$0.01  \\
 ESO 507-25 & 0.010788 & -3.0 , E/S0 & 8.31$\pm0.02$ & 16.58 & 0.0894 & 24.0$\pm2^c$  & 8.07$\pm$0.03 & 8.09$\pm$0.03 & 6.80$\pm$0.01 & 5.89$\pm$0.02  \\
 NGC 5044 & 0.009280 & -4.8 , E& 7.71$\pm0.02$  & 17.42  & 0.0701 & 35$\pm1^c$  & 7.31$\pm$0.02 & 7.38$\pm$0.02 & 7.02$\pm$0.01 & 6.53$\pm$0.02  \\
 NGC 5084  &  0.005741 & -2 , S0   & 7.06$\pm0.03$ & 16.85 & 0.1170  & 46.6$\pm1.8^b$ & 7.05$\pm$0.02 & 6.96$\pm$0.02 & 6.21$\pm$0.01 & 4.82$\pm$0.01  \\
 NGC 5153  &  0.014413 & -4.8 , E  & 9.15$\pm0.03$  & 20.29 & 0.0571  &     -    &    9.27$\pm$0.06 & 9.42$\pm$0.07 & 8.99$\pm$0.02 & 8.06$\pm$0.03  \\
 NGC 5353  &  0.007755 & -2.1 , S0 & 7.63$\pm0.01$ & 17.82 & 0.0128 & 41.0$\pm1.3^b$ & 7.41$\pm$0.02 & 7.38$\pm$0.03 & 6.98$\pm$0.01 & 6.00$\pm$0.02  \\ 
 NGC 5846  &  0.005717 & -4.7 , E  & 6.94$\pm0.02$ & 16.86 & 0.0561 & 21.0$\pm1^c$  & 6.65$\pm$0.02 & 6.73$\pm$0.01 & 6.39$\pm$0.01 & 5.83$\pm$0.01  \\ 
 NGC 5982  &  0.010064 & -4.8, E  & 8.15$\pm0.02$ & 18.13 & 0.0175 &  0.5$\pm0.5^c$  & 7.83$\pm$0.03 & 7.85$\pm$0.02 & 7.63$\pm$0.01 & 6.88$\pm$0.03  \\ 
 NGC 6658  &  0.014243 & -1.4 , S0 & 9.51$\pm0.03$ &  -  &  0.1230 &      -      &   9.40$\pm$0.06 & 9.42$\pm$0.06 & 9.10$\pm$0.02 & 9.81$\pm$0.08  \\ 
 NGC 7619$^f$ & 0.012549 & -4.7 , E   & 8.03$\pm0.02$  & 18.44 & 0.0823 & 20$\pm1^c$ &  7.70$\pm$0.02 & 7.75$\pm$0.02 & 7.42$\pm$0.01 & 7.32$\pm$0.02  \\ 
\hline
 \multicolumn{9}{l}{\textsc{Low--Richness Subsample}}\\
\hline
 NGC 128 & 0.012549 & -2.0 , S0 & 8.52$\pm0.02$ & 18.86 & 0.0287 &  1.5$\pm0.5^c$   & 8.38$\pm$0.03 & 8.47$\pm$0.03 & 8.11$\pm$0.01 & 7.47$\pm$0.03  \\ 
 NGC 252 & 0.014723 & -1.3 , S0 & 9.05$\pm0.03$ & 17.25 & 0.5039 &  2.5$^a$   & 8.85$\pm$0.04 & 8.85$\pm$0.04 & 6.70$\pm$0.01 & 7.94$\pm$0.07  \\ 
 NGC 315 & 0.017659 & -4.0 , E  & 7.96$\pm0.02$ & 17.48 & 0.0600 & 1800$\pm100^c$    & 7.89$\pm$0.03 & 7.66$\pm$0.02 & 6.73$\pm$0.01 & 4.53$\pm$0.01  \\ 
 NGC 524 & 0.006011 & -1.2 , S0 & 7.16$\pm0.01$ & 18.28 & 0.0826 &  3.1$\pm0.4^c$   & 6.90$\pm$0.02 & 7.00$\pm$0.02 & 6.11$\pm$0.01 & 5.50$\pm$0.01  \\ 
 NGC 1106 & 0.017012 & -1.6 , S0 & 9.21$\pm0.02$ & 17.14& 0.0914 & 132$\pm$4$^a$    & 9.00$\pm$0.05 & 8.92$\pm$0.05 & 6.04$\pm$0.01 & 3.30$\pm$0.01  \\ 
 NGC 1395 &  0.016728 & -4.8 , E & 6.89$\pm0.03$ & 16.74& 0.0231 & 1.1$\pm0.5^c$    & 6.79$\pm$0.02 & 6.88$\pm$0.02 & 6.36$\pm$0.01 & 5.67$\pm$0.01  \\ 
 NGC 1407 &  0.017075 & -4.5 , E & 6.70$\pm0.03$ & 16.15& 0.0689 &  38$\pm2^f$   & 6.37$\pm$0.01 & 6.46$\pm$0.01 & 5.98$\pm$0.01 & 5.75$\pm$0.02  \\ 
 NGC 1550 &  0.014880 & -4.1 , E & 8.77$\pm0.03$ & 18.01& 0.1318 & 17$\pm2^b$   & 8.53$\pm$0.04 & 8.58$\pm$0.03 & 8.46$\pm$0.01 & 8.00$\pm$0.03  \\ 
 NGC 1779 &  0.015794 & -0.3 , S0/a& 8.77$\pm0.02$ & - &    -  & 5.4$\pm0.6^b$    & 8.53$\pm$0.03 & 8.63$\pm$0.04 & 6.87$\pm$0.01 & 5.64$\pm$0.02  \\ 
 NGC 2292 &  0.017312 & -2.0 , S0&  -  &  -   &     -  &   -  &      8.87$\pm$0.04    &     8.86$\pm$0.01     &     8.41$\pm$0.01     &    7.40$\pm$0.05      \\ 
 NGC 2768 &  0.016495 & -4.4 , E & 7.00$\pm0.03$ & 17.45& 0.0448& 14$\pm1^c$   & 6.89$\pm$0.02 & 6.94$\pm$0.02 & 6.68$\pm$0.01 & 5.74$\pm$0.02  \\ 
 NGC 2911 &  0.012962 & -2.0 , S0& 8.71$\pm0.03$ & 18.68& 0.0300& 56$\pm2^c$   & 8.32$\pm$0.03 & 8.45$\pm$0.02 & 7.43$\pm$0.01 & 6.55$\pm$0.02  \\ 
 NGC 3325 & 0.012300  & -4.9 , E & 9.89$\pm0.04$& 20.77& 0.0552&  -  & 9.34$\pm$0.03 & 9.70$\pm$0.05 & 9.54$\pm$0.01 & 10.53$\pm$0.07  \\ 
 NGC 3613 &  0.014944 & -4.7 , E & 8.00$\pm0.02$& 18.83& 0.0125 &  0.3$\pm0.3^c$  & 7.70$\pm$0.02 & 7.80$\pm$0.03 & 7.51$\pm$0.01 & 7.05$\pm$0.03  \\ 
 NGC 3665 &  0.008606 & -2.1 , S0& 7.68$\pm0.01$& 18.25& 0.0190 &  113.2$\pm3.8^b$ & 7.25$\pm$0.02 & 7.23$\pm$0.01 & 5.97$\pm$0.01 & 4.52$\pm$0.02  \\ 
 NGC 3923 &  0.012075 & -4.7 , E & 6.50$\pm0.03$& 16.62& 0.0807 & $1\pm0.5^c$    & 6.24$\pm$0.01 & 6.28$\pm$0.01 & 5.65$\pm$0.01 & 4.94$\pm$0.01  \\ 
 NGC 4697 &  0.012622 & -4.4 , E & 6.37$\pm0.03$ & 16.72& 0.0295&  $0.6\pm0.5^c$  & 6.29$\pm$0.01 & 6.34$\pm$0.02 & 6.02$\pm$0.01 & 5.58$\pm$0.01  \\ 
 NGC 4956 &  0.017659 & -2.1 , S0& 9.60$\pm0.01$&  -  &   -    &  -  & 9.16$\pm$0.05 & 8.13$\pm$0.02 & 8.02$\pm$0.01 & 7.17$\pm$0.03  \\ 
 NGC 5061 &  0.006011 & -4.3 , E & 7.29$\pm0.01$ & 17.85& 0.0686&  -  & 7.03$\pm$0.02 & 7.10$\pm$0.02 & 6.43$\pm$0.01 & 4.69$\pm$0.01  \\ 
 NGC 5127 &  0.017012 & -4.8 , E & 9.36$\pm0.03$ & 19.70& 0.0137& 1980$^a$   & 8.90$\pm$0.03 & 9.21$\pm$0.04 & 8.89$\pm$0.01 & 7.74$\pm$0.03  \\ 
 NGC 5322 &  0.016728 & -4.8 , E & 7.16$\pm0.03$ & 18.39& 0.0138& 79.3$\pm2.8^b$   & 7.05$\pm$0.02 & 6.96$\pm$0.02 & 6.47$\pm$0.01 & 5.63$\pm$0.02  \\ 
 NGC 5444 &  0.017075 & -4.1 , E & 8.84$\pm0.02$ & 19.50& 0.0088&  - \large{$^{*}$}  & 8.52$\pm$0.03 & 8.57$\pm$0.03 & 8.04$\pm$0.01 & 7.88$\pm$0.04  \\ 
 NGC 5490 &  0.014880 & -4.9 , E & 8.92$\pm0.02$& 18.94& 0.0263&  1300$\pm100^c$  & 8.64$\pm$0.03 & 8.74$\pm$0.03 & 8.55$\pm$0.01 & 7.77$\pm$0.03  \\ 
 NGC 5629 &  0.015794 & -2.0 , S0& 9.22$\pm0.02$& 19.13& 0.0230& -   & 8.92$\pm$0.04 & 8.95$\pm$0.03 & 8.75$\pm$0.01 & 7.83$\pm$0.03  \\ 
 NGC 5903 &  0.017312 & -4.7 , E & 8.10$\pm0.02$& 19.41& 0.1527&  321.5$\pm16.0^g$  & 7.79$\pm$0.02 & 7.92$\pm$0.03 & 7.64$\pm$0.01 & 7.84$\pm$0.02  \\ 
 NGC 7252 &  0.016495 & -2.0 , S0& 9.31$\pm0.03$& 16.86& 0.0302&  25.3$\pm1.2^b$  & 8.99$\pm$0.04 & 8.85$\pm$0.03 & 5.65$\pm$0.01 & 3.39$\pm$0.01  \\ 
 NGC 7377 &  0.012962 & -1.0 , S0/a& 8.18$\pm0.03$ & 18.06& 0.0299&  3.4$\pm0.5^b$   & 7.93$\pm$0.02 & 7.92$\pm$0.02 & 6.22$\pm$0.01 & 5.58$\pm$0.01  \\ 
\hline
\end{tabular}
\end{center}
$^a$ \citet{Condonetal02}, $^b$  \citet{Condonetal98}, $^c$ \citet{Brownetal11}, $^d$  \citet{Beckeretal95}, $^e$ \citet{Kuhretal81}, $^f$ \citet{Giacintuccietal12},  $^g$ \citet{OSullivanetal18b} \mbox{}  \large{$^{*}$}\footnotesize{In \citet{Kolokythasetal19} detailed GMRT radio study no radio emission was detected in NGC~5444 reporting that the 1.4~GHz values provided at NED from \citet{Condonetal98} (NVSS) and \citet{Brownetal11} for this system come from a nearby 4C radio source.}
\end{table*}


\begin{table} 
\caption{Calculated properties of the BGEs. The columns list the BGE name, the star-formation rate (SFR$_{FUV}$) from FUV, the stellar mass M$_{stellar}$, the power at 1.4~GHz, P$_{1.4~GHz}$ and the absolute magnitude M$_{K_{\huge{s}}}$ for each galaxy. All upper limits on the radio power are based on the 5 $\times$ r.m.s. of the NVSS catalog.  \label{infotable}}
\begin{center}
\begin{tabular}{lcccc}
\hline 
 BGE & SFR$_{FUV}$ $\pm5\%$ & M$_{\tiny{stellar}}$ & P$_{1.4~GHz}$ & M$_{K_{\huge{s}}}$ \\ 
     &  (M$_\odot$~yr$^{-1}$) & (10$^{11}$ M$_\odot$) &  (W~Hz$^{-1}$) &(mag)     \\
\hline
\multicolumn{5}{l}{\textsc{High--Richness Sub-sample}}\\
 NGC 193   &  0.047 & 3.34  &  1.12$\times$10$^{24}$& -25.15 \\
 NGC 410   &  0.117 & 8.55  &  4.47$\times$10$^{21}$& -26.07\\
 NGC 584   &  0.018 & 2.13 & 4.49$\times$10$^{19}$&  -24.70\\
 NGC 677   &  0.068 & 3.52 & 1.50$\times$10$^{22}$&  -25.20\\
 NGC 777   &  0.134 & 7.68&  4.46$\times$10$^{21}$ & -25.97\\
 NGC 940$^a$   &  0.195 & 2.94 & 3.56$\times$10$^{21}$& -25.02\\
 NGC 924$^a$   &  0.152 & 1.88 & 4.17$\times$10$^{20}$& -24.58\\
 NGC 978$^a$   &  0.045 & 3.23 &3.70$\times$10$^{20}$ &  -25.11\\
 NGC 1060  &  0.097 & 5.91 & 10.44$\times$10$^{21}$& -26.27\\
 NGC 1167  &  - & 3.41    & 5.91$\times$10$^{24}$ & -25.71\\
 NGC 1453  &  0.058 & 7.27 & 1.33$\times$10$^{22}$ & -25.91\\
 NGC 1587  &  0.038 & 3.03& 4.08$\times$10$^{22}$& -25.05\\
 NGC 2563  &  0.057 & 3.10 & 1.52$\times$10$^{20}$& -25.07\\
 NGC 3078  &  0.024 & 2.36 & 4.29$\times$10$^{22}$&  -24.80\\
 NGC 4008  &  0.020 & 2.45 & 3.80$\times$10$^{21}$& -24.84 \\
 NGC 4169  &  - & 1.27 &  2.59$\times$10$^{20}$& -24.19 \\
 NGC 4261  &  0.092 & 3.79 & 2.39$\times$10$^{24}$& -25.27 \\
 ESO 507-25& 0.221 & 2.84 & 5.81$\times$10$^{21}$& -24.99\\
 NGC 5044  & 0.073 & 3.58 & 6.05$\times$10$^{21}$& -25.21\\
 NGC 5084  &  0.045 &2.33&  2.95$\times$10$^{21}$& -24.79\\
 NGC 5153  &  0.013 & 2.27 & $\leq$9.69$\times$10$^{20}$& -24.76\\
 NGC 5353  &  0.043 & 3.19 &  6.01$\times$10$^{21}$& -25.10\\
 NGC 5846  &  0.057 &3.39 &   1.70$\times$10$^{21}$& -25.16\\
 NGC 5982  &  0.051 & 3.10&  1.16$\times$10$^{20}$ & -25.07\\
 NGC 6658  &  - & 1.79 & $\leq$1.07$\times$10$^{21}$& -24.53\\
 NGC 7619& 0.058 & 5.61 & 6.98$\times$10$^{21}$ & -25.66\\
\hline
 \multicolumn{5}{l}{\textsc{Low--Richness Sub-sample}}\\
\hline
 NGC 128  & 0.049 & 4.24  &  6.46$\times$10$^{20}$ & -25.38 \\
 NGC 252 &  0.307 & 3.77  &  1.55$\times$10$^{21}$  & -25.26\\
 NGC 315 &  0.255 & 11.70 & 1.15$\times$10$^{24}$ & -26.38 \\
 NGC 524 &  0.027 & 4.90 & 4.29$\times$10$^{20}$ &  -25.52\\
 NGC 1106 &  0.267 & 2.49& 6.47$\times$10$^{22}$ & -24.85\\
 NGC 1395 &  0.042 &  2.18  & 5.80$\times$10$^{19}$ & -24.73 \\
 NGC 1407 &  0.086 & 3.29&  5.57$\times$10$^{21}$& -25.13 \\
 NGC 1550 &  0.083 & 2.59 & 5.71$\times$10$^{21}$& -24.89 \\
 NGC 1779 &   - & 1.74  & 1.31$\times$10$^{21}$ & -24.50\\
 NGC 2292 &   - & -  & $\leq$2.42$\times$10$^{20}$ & -\\
 NGC 2768 &  0.026 & 2.42 & 8.86$\times$10$^{20}$& -24.83\\
 NGC 2911 &  0.032 & 1.86 & 1.36$\times$10$^{22}$& -24.57\\
 NGC 3325 & 0.015 &  2.02 & $\leq$1.72$\times$10$^{21}$& -24.65\\
 NGC 3613  &  0.014& 1.79 &  3.68$\times$10$^{19}$& -24.53\\
 NGC 3665  &  0.024 & 2.48 & 1.39$\times$10$^{22}$& -24.85  \\
 NGC 3923  &  0.042 & 2.98 &4.79$\times$10$^{19}$& -25.03\\
 NGC 4697  &  0.031 & 2.66 & 2.33$\times$10$^{19}$& -24.92 \\
 NGC 4956  & - & 2.03 &   $\leq$1.36$\times$10$^{21}$& -24.65\\
 NGC 5061$^a$  & 0.027 & 2.80 & 3.85$\times$10$^{19}$& -24.97\\
 NGC 5127 & 0.032 & 2.69&   1.23$\times$10$^{24}$& -24.93\\
 NGC 5322 &  0.017 & 3.38&  7.98$\times$10$^{21}$& -25.16\\
 NGC 5444 &  0.027 & 3.05 & $\leq$9.69$\times$10$^{20}$& -25.06\\
 NGC 5490  &  0.063 & 4.10& 7.84$\times$10$^{23}$& -25.35\\
 NGC 5629 &  0.047 & 2.65 &  $\leq$1.21$\times$10$^{21}$& -24.92\\
 NGC 5903  & 0.011 & 2.21 & 4.65$\times$10$^{21}$& -24.74\\
 NGC 7252  &  0.368 & 2.35& 1.32$\times$10$^{22}$& -24.80\\
 NGC 7377 & 0.060 & 3.33  & 8.61$\times$10$^{20}$& -25.14\\

\hline
\end{tabular}
\end{center}
$^a$ The 1.4~GHz radio power in these systems was extrapolated by their GMRT 610~MHz flux densities from \citet{Kolokythasetal18,Kolokythasetal19}. 

\end{table}

\section{Results} 

 \subsection{Stellar Mass} 

\label{stellarmass}


As K$_s$ band is less sensitive to dust than bluer bands, it provides an excellent tracer of stellar luminosity and thus the stellar mass of a galaxy (M$_{stellar}$; \citealt{Belletal03}). We estimate the stellar mass of the CLoGS dominant galaxies as in \citet{Maetal14} (their equation~2) using the conversion between K-band luminosity and stellar mass for early-type galaxies in the ATLAS$^{3D}$ sample \citep{Cappellari13}. This relation comes from the fit between the total extinction-corrected \textit{2MASS} K$_s$-band magnitudes and the stellar masses derived from dynamics using the Jeans Anisotropic Multi-Gaussian Expansion method \citep{Emsellemetal94}. 


We find the stellar masses of our group dominant galaxies fall within a narrow range between $\sim 10^{11}$ to $\sim10^{12}~M_{\odot}$, as expected given our sample selection criteria \citep[L$_B$ $>$ 3$\times$10$^{10}$ L$_\odot$, for the dominant early--type galaxy,][]{OSullivanetal17}.


\subsection{FUV - $K_{s}$ colour as a star formation indicator}
\label{sec:FUV-Ks}

\begin{figure}
\centering
\includegraphics[width=0.49\textwidth]{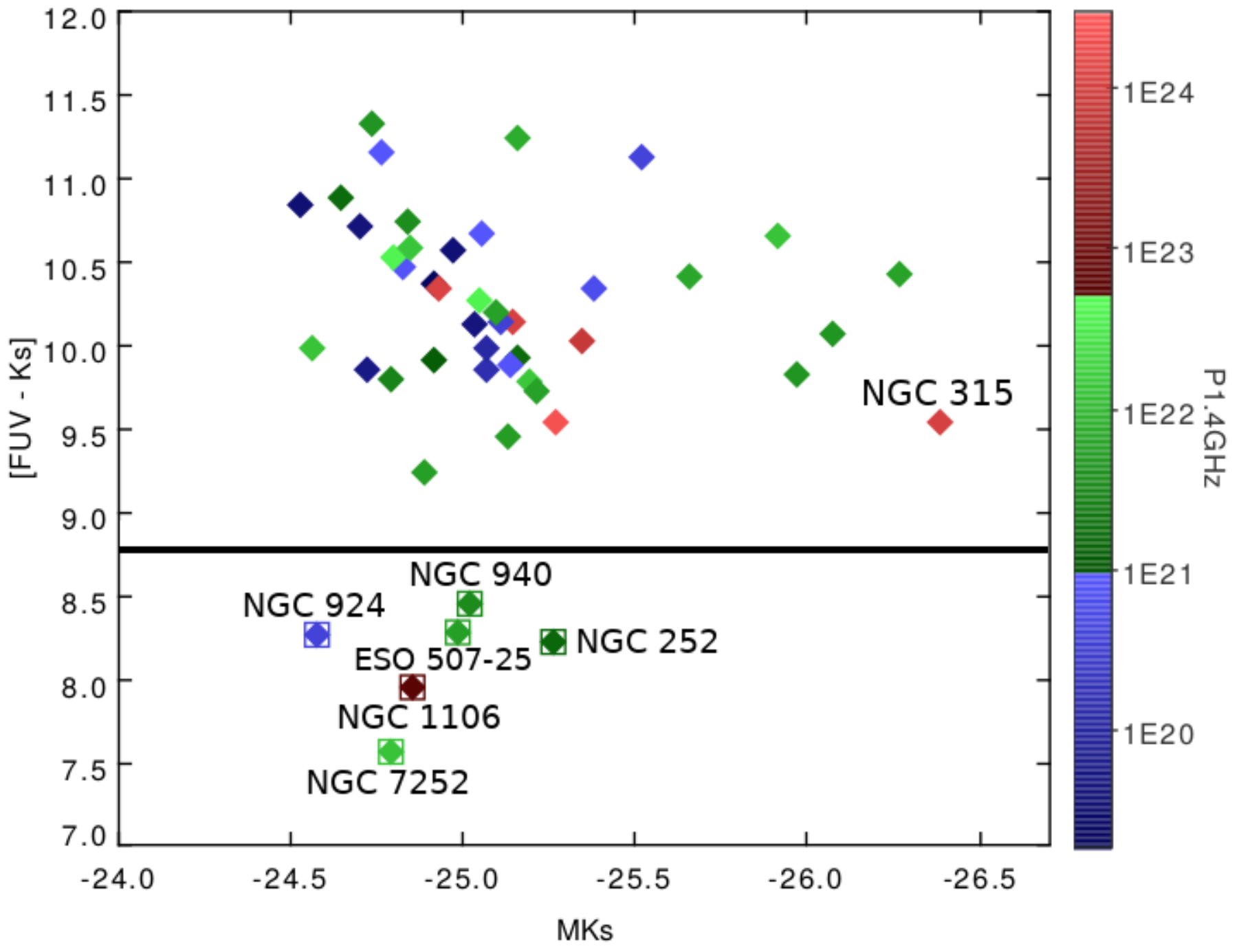}
\caption{Plot of [FUV - K$_s$] vs absolute K$_s$-band magnitude for the CLoGS dominant galaxies. Symbol colour indicates 1.4~GHz radio power. The horizontal black line at 8.8 mag is the boundary between star-forming and non-star-forming galaxies as defined in \citet{GildePaz07}. The 6 galaxies that are bluer than 8.8 mag are considered to be actively star-forming.}
\label{FUVKsvsMKs}
\end{figure}

FUV emission is a good tracer of unobscured star formation and can be combined with NIR emission (which traces the emission of the bulk of the stellar population) to provide an initial window on the star-forming status of the BGEs. A cutoff at [FUV $-$ K$_s$] $<$ 8.8 mag is often taken as a reliable discrimination point between active star-forming and quiescent galaxies \citep{GildePaz07}. In Figure \ref{FUVKsvsMKs} we plot the [FUV $-$ K$_s$] vs M$_{K_s}$ colour-absolute magnitude diagram for the 47/53 group dominant galaxies for which FUV magnitude values are available. The black horizontal line denotes the transition point of [FUV $-$ K$_s$] $=$ 8.8 mag. We find that the majority of the group dominant galaxies (87\%; 41/47) have [FUV $-$ K$_s$] $\sim$9 $-$ 11, in agreement with typical colours of passive early-type galaxies that are found by previous studies \citep{GildePaz07,Vaddietal16}. Only $\sim$13\% (6/47) of the BGEs are found to be bluer than an [FUV $-$ K$_s$]$=$8.8, namely NGC~252, NGC~924, NGC~940, NGC~1106, NGC~7252 and ESO507-25. This suggests that these six galaxies are likely to be actively star-forming, and we will refer to them as such in the rest of the paper. These actively star-forming galaxies have M$_{K_s}>-25.3$, whereas the more passive BGEs in the remainder of the sample have M$_{K_s}$ in the range -24.4 to -26.4. We note here that CLoGS sample selection affects the proportion of actively star-forming systems that we find here.


\subsection{FUV Star-formation rates} 
\label{SFRFUV}

We can also estimate the star-formation rate in each galaxy from its FUV flux. We use the relation of \citet{Salim07},

\begin{equation}
    \centering
    SFR_{\rm FUV} \,(M_{\odot} yr^{-1}) = 1.08 \times 10^{-28}L_{\rm FUV} \, (erg~s^{-1}~Hz^{-1})
\end{equation}
where $L_{FUV}$ is the galactic extinction corrected FUV luminosity. \citet{Salim07} derived this relation for the \textit{GALEX} wavebands and it rests on the constant star formation hypothesis, in which the SFR is considered to remain constant over the lifetime of the UV-emitting population ($<$10$^8$ years). In addition, it assumes a Salpeter Initial Mass Function (IMF; \citealt{Salpeter55} ) with mass limits in the range of 0.1 to 100 M$_{\odot}$. The uncertainties on the SFR$_{FUV}$ were estimated to be $\sim$5\%. 

In rapidly star--forming galaxies FUV luminosity mainly originates from the young stellar population but older evolved populations, in particular extreme horizontal branch stars, can contribute \citep[see e.g., ][]{Kavirajetal07,Loubseretal11,Werleetal20}. This component of the FUV luminosity may be important in our early-type galaxies. FUV emission can also originate from the bright accretion disks of unobscured Seyfert or quasar-type AGN. Contamination from this source is less likely in group--dominant early--type galaxies, which typically host radiatively inefficient AGN.

From Figure~\ref{FUVKsvsMKs} we can estimate the lowest meaningful SFR$_{FUV}$ value that can be measured taking into account the contribution from the evolved stellar population. As in \citet{Vaddietal16} we select the BGEs that have [FUV $-$ K$_s$]~$>$~8.8~mag, assume them to be completely non-star-forming, and calculate their median SFR$_{FUV}$; this is 3$\times$10$^{-2}$ \sfrunit. \citet{Vaddietal16} also provide estimates of the SFR calculated from the young stellar population alone. For the two galaxies we have in common with the Vaddi sample (NGC~1550 and NGC~1587) we can subtract the estimated SFR$_{FUVyoung}$ from our SFR$_{FUV}$ values to find the apparent SFR associated with the evolved stellar population. These are 2.80$\times$10$^{-2}$ \sfrunit for NGC~1550 and 2.96$\times$10$^{-2}$ \sfrunit for NGC~1587. Hence we consider the lowest meaningful SFR$_{FUV}$ value to be $\sim$3$\times$10$^{-2}$ \sfrunit.


\begin{figure}
\centering
\includegraphics[width=0.47\textwidth]{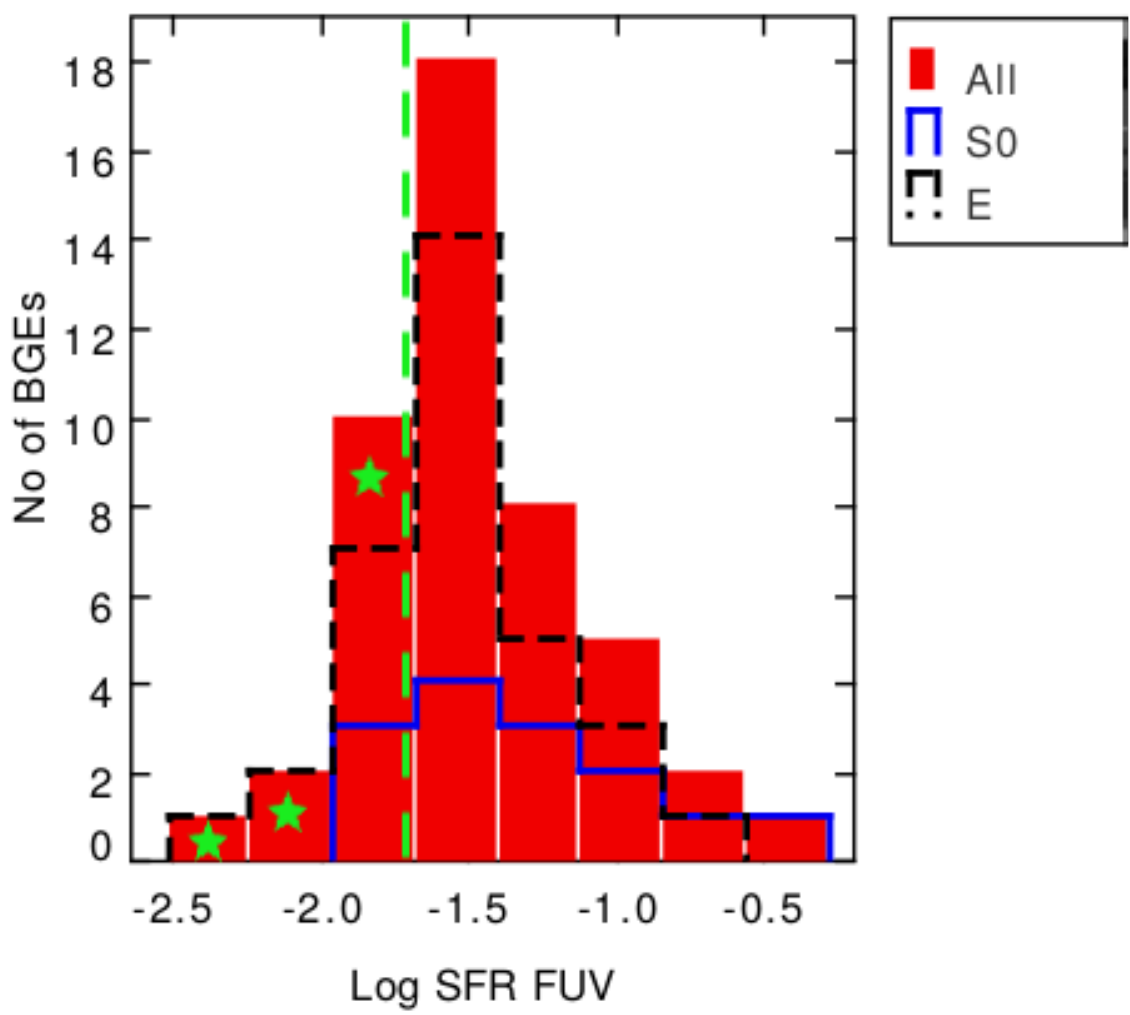}
\caption{Histogram showing the number and type of galaxies with respect to their log SFR from FUV. The dashed green line defines the limit below which the FUV can be explained entirely by passive evolution, with green stars showing the relevant columns that contain these galaxies.}
\label{HistSFR}
\end{figure}

Figure~\ref{HistSFR} shows the SFR$_{FUV}$ distribution for the sample as a whole, as well as histograms for the E and S0 BGEs. The SFR$_{FUV}$ values in our sample, as shown in Table~\ref{infotable}, range between 0.01 $-$ 0.4 \sfrunit with a mean of 0.070$\pm0.005$ \sfrunit. The mean SFR$_{FUV}$ for elliptical galaxies (32/47 BGEs in the \textit{Galex} GR6 catalogue) is 0.060$\pm0.004$ \sfrunit, and for the lenticulars (15/47) slightly higher, 0.080$\pm0.005$ \sfrunit. All BGEs in the CLoGS sample without a detection in the FUV wavelength are S0 type galaxies. We note that Figure~\ref{HistSFR}, and the calculated means, are not corrected for the contribution from the evolved stellar population, since this is relatively small, and will vary between individual galaxies depending on their star formation histories. Anything above the lowest meaningful SFR$_{FUV}$ limit is considered to be excess FUV flux (thus from young stars) in addition to the passively evolving stellar populations.


Comparing SFR$_{FUV}$ with the [FUV $-$ K$_s$] colour for each galaxy, we find that the six actively star-forming galaxies have some of the highest star-formation rates in the sample, $>$0.15~\sfrunit. Only one other galaxy has SFR$_{FUV}$ above that boundary, NGC~315.

\subsection{Dust obscured star formation}

\begin{figure}
\centering
\includegraphics[width=0.48\textwidth]{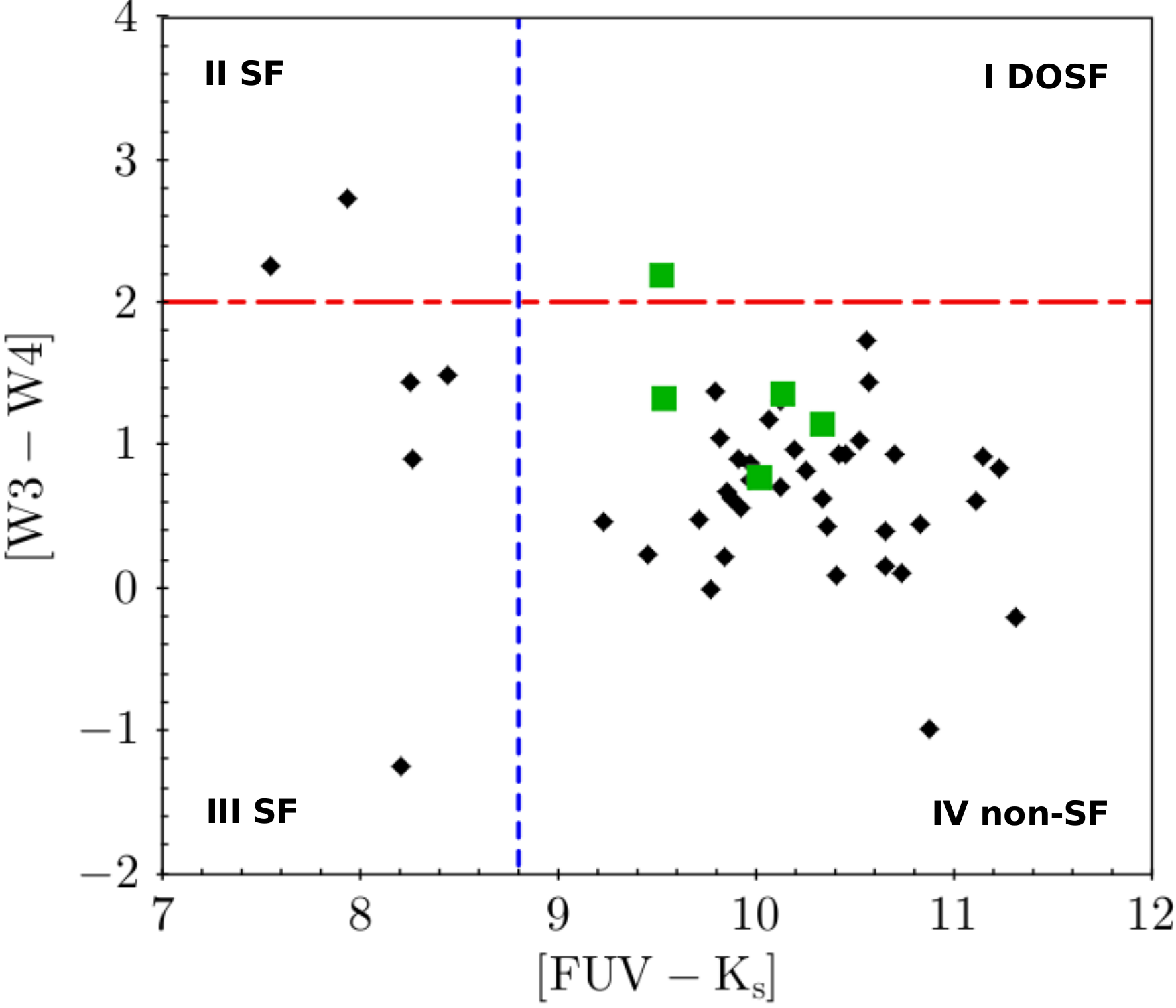}
\caption{WISE [12-22$\mu$m] vs [FUV - K$_s$] colour-colour plot of CLoGS dominant galaxies. The red horizontal dashed line marks the cutoff for  galaxies considered to show a mid-infrared excess, while the vertical dashed blue line marks the cutoff between actively star-forming and non-star-forming galaxies as in Figure~\ref{FUVKsvsMKs}. Systems in quadrant I likely host dust obscured star formation, those in quadrants II and III unobscured, recent star formation, and those in quadrant IV are considered non-star forming, passive systems. Green squares indicate systems with the highest radio power (P$_{1.4~GHz}>10^{23}$ W~Hz$^{-1}$).}
\label{WISEFUVKPradio}
\end{figure}

While FUV emission is a good tracer of unobscured young stellar populations, the mid-infrared can provide information on obscured star formation. We can combine FUV and mid-IR to help disentangle unobscured and obscured SF in the CLoGS BGEs. Figure~\ref{WISEFUVKPradio} shows a colour-colour diagram with \textit{WISE} 12-22~$\mu$m (or [W3 $-$ W4]) plotted against [FUV $-$ K$_s$]. The vertical blue dashed line at [FUV $-$ K$_s$] $=$ 8.8 mag is the same as that used in Figure~\ref{FUVKsvsMKs} to separate star-forming and passive/non-star-forming galaxies. The red horizontal line at [12-22$\mu$m] = 2 mag is chosen to identify galaxies with a mid-IR excess, as in \citet{Vaddietal16}. These two lines separate the diagram into four regions that classify the galaxies based on their position: \textit{I. Dust obscured star formation (DOSF)}. The galaxies in this region show an excess in mid-infrared but are faint in the FUV, indicating star formation that is obscured by dust. \textit{II. and III. Star-forming (SF)}. In these regions the galaxies are FUV bright, indicating recent, unobscured star formation. \textit{IV. non-star-forming (non-SF)}. This region contains passive galaxies with no indication of strong SF in mid-IR or FUV. 

The six galaxies that occupy regions II and III are the actively star-forming systems (NGC~252, NGC~924, NGC~940, NGC~1106, NGC~7252 and ESO507-25) identified in Section~\ref{sec:FUV-Ks}. To these we can add one galaxy that falls in region I, and may host obscured SF, NGC~315. While NGC~315 shows no signs of jet emission in the IRAC bands \citep{Tangetal09b}, the presence of nuclear mid-infrared excess emission at 8~$\mu$m \citep{Guetal07b} may suggest that the observed [12-22$\mu$m] excess in this system includes contamination from an AGN with the system presenting simultaneously both star formation and AGN activity.



\subsection{Mid-infrared colour-colour classification of BGEs}

\begin{figure*}
\centering
\includegraphics[width=0.75\textwidth]{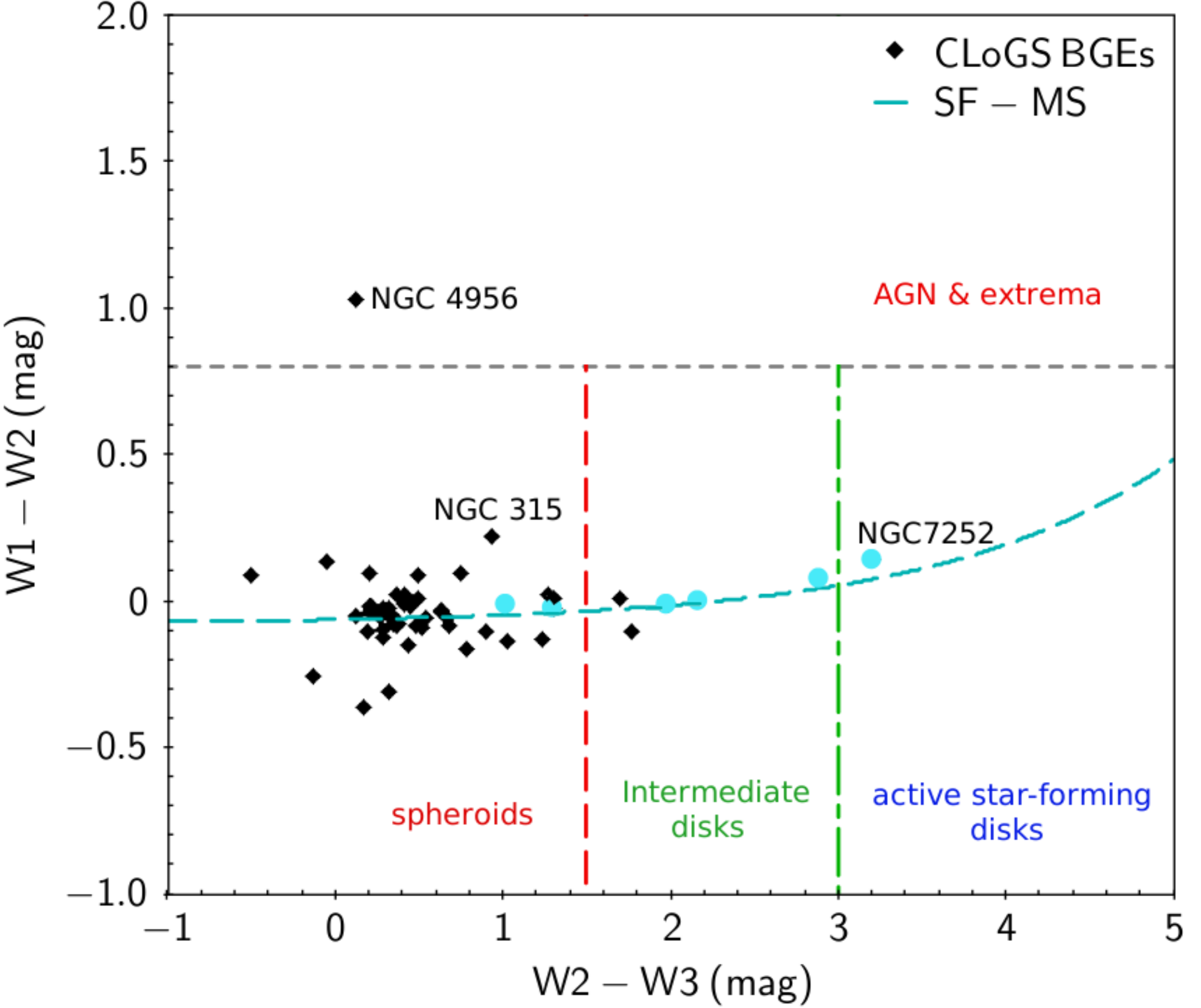}
\caption{WISE [W1 $-$ W2]  vs [W2 $-$ W3] colour-colour diagram used to classify our galaxies into four distinct categories: spheroid-dominated, intermediate disk, star formation dominated, and AGN-dominated populations following \citet{Jarrettetal17,Jarrettetal19}. Higher values of the [W1 $-$ W2] colour indicate the presence of radiatively efficient AGN (Seyferts and quasars), whereas larger [W2 $-$ W3] values trace star formation. The dashed cyan line denotes the `star formation WISE colour-colour main sequence' (SF - MS), and represents the track followed by normal galaxies, from quiescent to star-forming, in the infrared colour-colour space. Only NGC~7252 falls in the star formation dominated category, only NGC~4956 in AGN-dominated. In cyan circles are noted the six FUV bright active star-forming systems identified earlier using [FUV $-$ K$_s$] color.}
\label{Wisecolour}
\end{figure*}

As well as tracing obscured star formation, mid-infrared WISE colours are often used to separate AGN-dominated galaxies from their star-forming and passive counterparts (e.g., \citealt{Jarrettetal11,Sternetal12,Cluveretal14, Cluveretal17,Jarrettetal17,Jarrettetal19}). We follow the approach described by \citet{Jarrettetal17,Jarrettetal19}, plotting the [W1 $-$ W2]  vs [W2 $-$ W3] colour-colour diagram in Figure~\ref{Wisecolour}. Larger values of the [W1 $-$ W2] colour (``redder" values) are known to indicate the presence of radiatively efficient AGN (Seyferts and quasars), while larger [W2 $-$ W3] values trace star formation.

The colour plane is divided into four regions which have been shown to accurately classify systems by previous studies comparing colours with infrared morphology and bulge, disk and nuclear stellar population properties  \citep{Johnsonetal07,Walkeretal10,Jarrettetal11,Jarrettetal13}. We use these regions to categorise our BGEs as either: 1. Spheroid dominated, 2. Intermediate disks, 3. Actively star-forming disks, or 4. AGN dominated populations. The dashed cyan line in Figure~\ref{Wisecolour} marks the `star formation WISE colour-colour sequence', and represents the track in the infrared colour-colour space that normal galaxies, from quiescent to star-forming, would follow. It is defined as \citep{Jarrettetal19}:

\begin{equation}
    (W1 - W2) = [0.015 \times e^{\frac{(W2-W3)}{1.38}}]-0.08
\end{equation}

The sequence is relatively flat in the spheroid and intermediate disk regions, but bends upwards in the W1 $-$ W2 colour with increasing star formation, as the contribution from warm dust in the W2 (4.6~$\mu$m) band becomes more important.

Figure~\ref{Wisecolour} shows that the majority (46/53, 87\%) of our CLoGS BGEs are categorised as spheroids, with minimal star formation. 
The galaxies categorised as spheroids include some of the most radio-luminous AGNs in the CLoGS sample, such as NGC~193 (see \citealt{Bogdanetal14}), NGC~315 (see \citealt{Giacintuccietal11}) and NGC~4261 (see \citealt{Kolokythasetal15}) confirming that while they may launch powerful jets, their nuclei are not radiatively bright. NGC~315 falls some way off the star-forming sequence, but not significantly farther than other, radio-faint systems.

Five galaxies (NGC~252, NGC~940, NGC~1106, NGC~1779, and NGC~7377) fall into the region of intermediate disks, indicating that they have similar mid-infrared colours to early-type spirals with semi-quiescent star-forming disks and large bulges \citep{Jarrettetal19}. We will discuss these systems further in \S\ref{Intermediate Mid-IR}. The actively star-forming systems identified from [FUV - K$_s$] colours are indicated by cyan circles; three of these galaxies are classed as intermediate disks.


The only BGE which is classed as an active star-forming disk based on mid-infrared colours is NGC~7252, which is also identified as actively star-forming based on its FUV colours and SFR$_{FUV}$. This is consistent with previous works, which have shown it to be a post-merger starburst galaxy.

Only one galaxy has mid-IR colours that may indicate dust heating by a radiatively powerful AGN. NGC~4956 has [W1 $-$ W2] $>$ 0.8 (horizontal dashed line; \citealt{Sternetal12}), comparable to high-excitation radio galaxies (HERGs), Seyferts and quasars \citep{Jarrettetal19}. The galaxy is one of the few for which no FUV information is available from \textit{Galex}, and is not detected in the radio \citep{Kolokythasetal19}, but is relatively luminous in the far-infrared \citep{OSullivanetal18b}. This suggests that if it does host a radiatively efficient AGN, it has not launched jets and is strongly obscured.

\section{Discussion} 

Our FUV and mid-IR indicators provide insight into star formation in the CLoGS dominant galaxies. We can now examine how this relates to the environment, gas content, and nuclear activity of the galaxies. A number of processes can affect the fuelling and triggering of star formation and AGN, including: i) galaxy interactions and mergers \citep{Fernandezetal14}, ii) gas cooling from the interstellar medium (ISM) of the galaxy, or a surrounding hot X-ray emitting IGrM \citep[e.g.,][]{McNamaraNulsen07, Fabian12,Oppenheimeretal21} and iii) mass loss from stars (e.g., \citealt{HeckmanBest14}). However, the exact mechanisms of SF and AGN fuelling, particularly in group and cluster-dominant galaxies, are still a subject of active research (e.g., \citealt{Raffertyetal08,ODeaetal08,OSullivanetal18b}). In this section we will compare our star formation indicators with radio power (P$_{1.4~GHz}$), stellar mass, gas content, environmental richness, and other parameters, with the goal of gaining insight into the processes driving SF and AGN activity in group-dominant galaxies.

\subsection{Radio power and star formation}
 \label{sec:Prad_Mstar}

\begin{figure}
\centering
\includegraphics[width=0.48\textwidth]{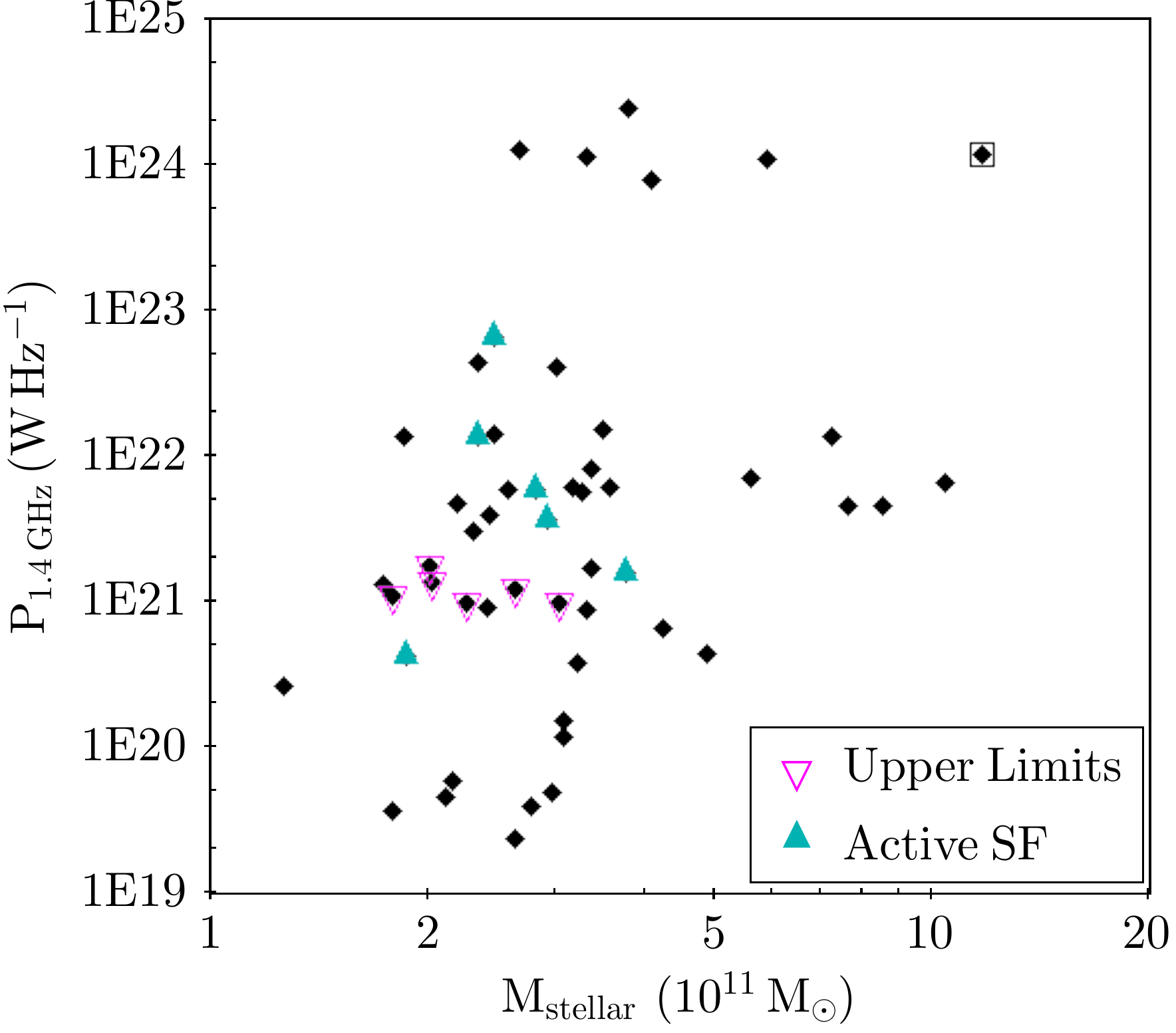}
\caption{Radio power of the CLoGS group dominant galaxies compared to stellar mass. Cyan triangles mark the active star-forming galaxies (from Figure~\ref{FUVKsvsMKs}) and a square marks NGC~315. }
\label{PradMstellar}
\end{figure}

We can complement the use of mid-IR colours to identify radiatively powerful AGN by using 1.4~GHz radio power as a tracer of AGN jet activity. Both AGN and star formation produce radio emission, but star formation emission scales relatively slowly with mass, and AGN emission begins to dominate in galaxies with masses $\ga$10$^{11}$\msol\ \citep[e.g.,][]{Gurkanetal18}. The AGN of massive early-type galaxies such as the CLoGS BGEs produce radio emission (either from the core or from jets) much of the time \citep[e.g.,][]{Sabateretal19}, suggesting that in these systems the AGN is always being fuelled, if only at low levels. \citet{Kolokythasetal18,Kolokythasetal19} provide a full discussion of the radio properties of the CLoGS BGEs; almost all are detected at radio wavelengths with a subset hosting active jets, the largest of which are luminous enough for the galaxies to be classed as radio loud (P$_{1.4~GHz} \geq10^{23}~W~Hz^{-1}$).


The relationship between radio power with M$_{stellar}$  for our galaxies is shown in Figure~\ref{PradMstellar}. P$_{1.4~GHz}$ ranges from 3.5$\times$10$^{19} - 2.5\times 10^{24} W~Hz^{-1}$, with the mean radio power being  P$_{1.4~GHz} \sim 1.5\times10^{23} W~Hz^{-1}$ and the median P$_{1.4~GHz} \sim 3\times10^{22} W~Hz^{-1}$. Six systems meet the criterion to be considered radio loud. A weak correlation between radio power and stellar mass is seen in Figure~\ref{PradMstellar} which is insufficient to draw definitive conclusions (Spearman rank test coefficient $\rho\approx0.36$). In part this reflects the large scatter in P$_{1.4~GHz}$ at fixed mass in the sample. The degree of scatter is roughly consistent with the range seen at these masses in larger samples \citep{Sabateretal19}, and indicates that galaxy stellar mass is not the only driving factor in determining radio power. The narrow stellar mass range of the CLoGS BGEs makes it difficult to probe the known correlation between stellar mass and radio luminosity, but this also means that variation in stellar mass among CLoGS galaxies does not introduce a strong bias on radio power.



\begin{figure}
\centering
\includegraphics[width=0.47\textwidth]{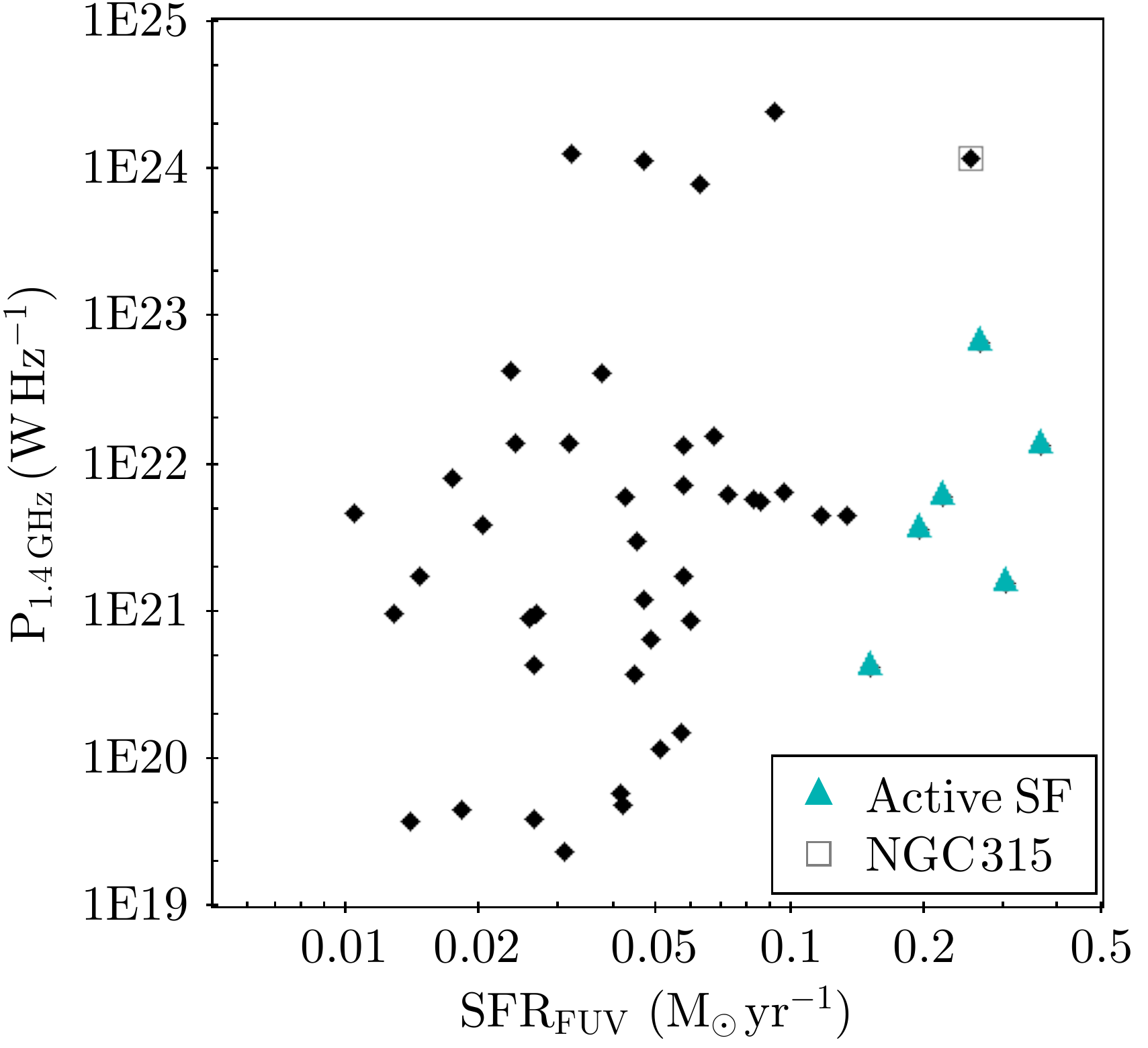}
\caption{Radio power of the CLoGS BGEs in relation to SFR$_{FUV}$. The active star-forming galaxies are shown in cyan and the galaxy in square is NGC~315. }
\label{PradSFRFUV}
\end{figure}

Figure~\ref{PradSFRFUV} shows the relation between radio power and SFR$_{FUV}$. We find a weak correlation with a large scatter (Spearman rank test coefficient $\rho\approx0.31$), with the scatter on the star-formation rates increasing for systems with P$_{1.4~GHz}\geq10^{21} W~Hz^{-1}$. The degree of scatter in radio power is much greater than that seen in star formation dominated galaxies \citep[$\sim$6 orders of magnitude compared to $\sim$2 for SF galaxies in the sample of][]{Gurkanetal18}. The weak correlation and large scatter suggests that, while SF may be a significant source of radio emission in a subset of our galaxies, and it is possible that gas reservoirs may fuel both SF and AGN activity, generally speaking the two processes are not strongly connected in our sample. AGN fuelling and jet launching are complex processes, affected by the availability and dynamics of the gas near the central engine \citep{Tadhunteretal11,Kavirajetal15}, the mass and the spin of the central black hole, and the external environment \citep[e.g.,][]{Baumetal1995, Meieretal1999,Nemmenetal07,Woldetal2007,BensonBabul09}. Even when jets are active, AGN are variable \citep[e.g.,][]{Hickoxetal14} and these factors will produce scatter in radio power even among galaxies with similar properties. We do not see evidence of radio mode AGN feedback suppressing or enhancing star formation in our BGEs. This suggests that where jets are present they generally do not disturb any cool gas component that may be forming stars \citep[though this may be occurring in at least one of our galaxies, NGC~1167;][]{Shulevskietal12,Brienzaetal18}. NGC~4956, the only system identified as a radiatively powerful AGN from its mid-IR colours, is undetected at 1.4~GHz and has no measured FUV flux, so we cannot speculate on the impact of its AGN on any star formation it may host.

 \subsection{Star formation and stellar mass} 
 \label{sec:SFR_Mstar}

\begin{figure}
\centering
\includegraphics[width=0.48\textwidth]{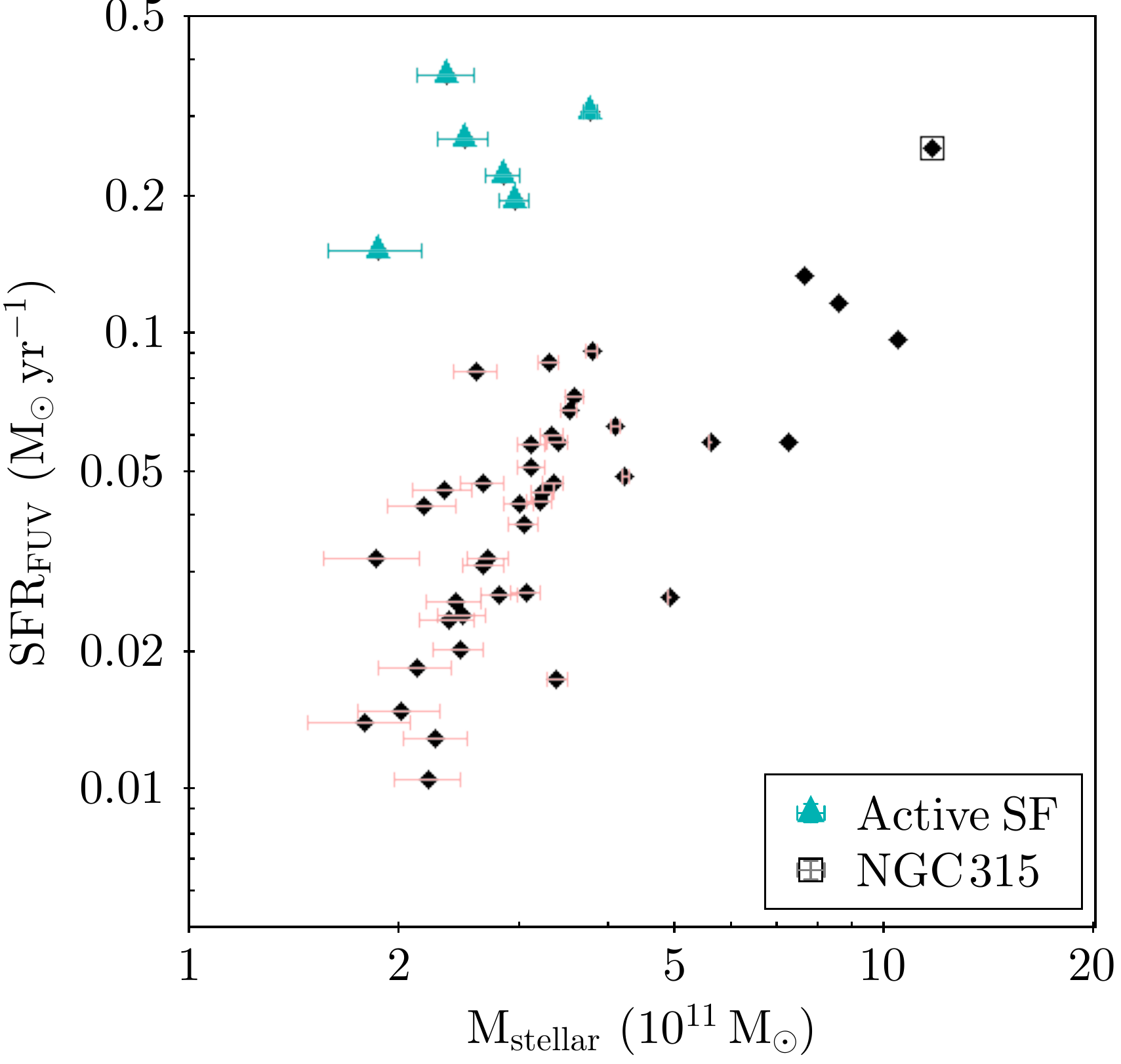}
\caption{SFR$_{FUV}$ of the CLoGS BGEs plotted against M$_{stellar}$. The active star-forming galaxies are shown as cyan triangles, and the galaxy NGC~315 marked with a square. }
\label{SFRFUVMstellar}
\end{figure}

Figure~\ref{SFRFUVMstellar} shows the relation between SFR$_{FUV}$ and stellar mass. A strong correlation is seen for the majority of systems which have passive [FUV $-$ K$_s$] colours (black rhombus points; Spearman rank test coefficient $\rho\approx0.74$). This is similar to the trend found in \citet{Vaddietal16} for a sample of isolated early-type galaxies, and resembles to the known correlation between SFR and stellar mass \citep[see][]{Brinchmannetal04,Elbazetal07,Salmonetal15}. However, the 6 active star-forming systems appear to be detached from the observed correlation, with high SFR for their stellar mass. 

\begin{figure}
\centering
\includegraphics[width=0.48\textwidth]{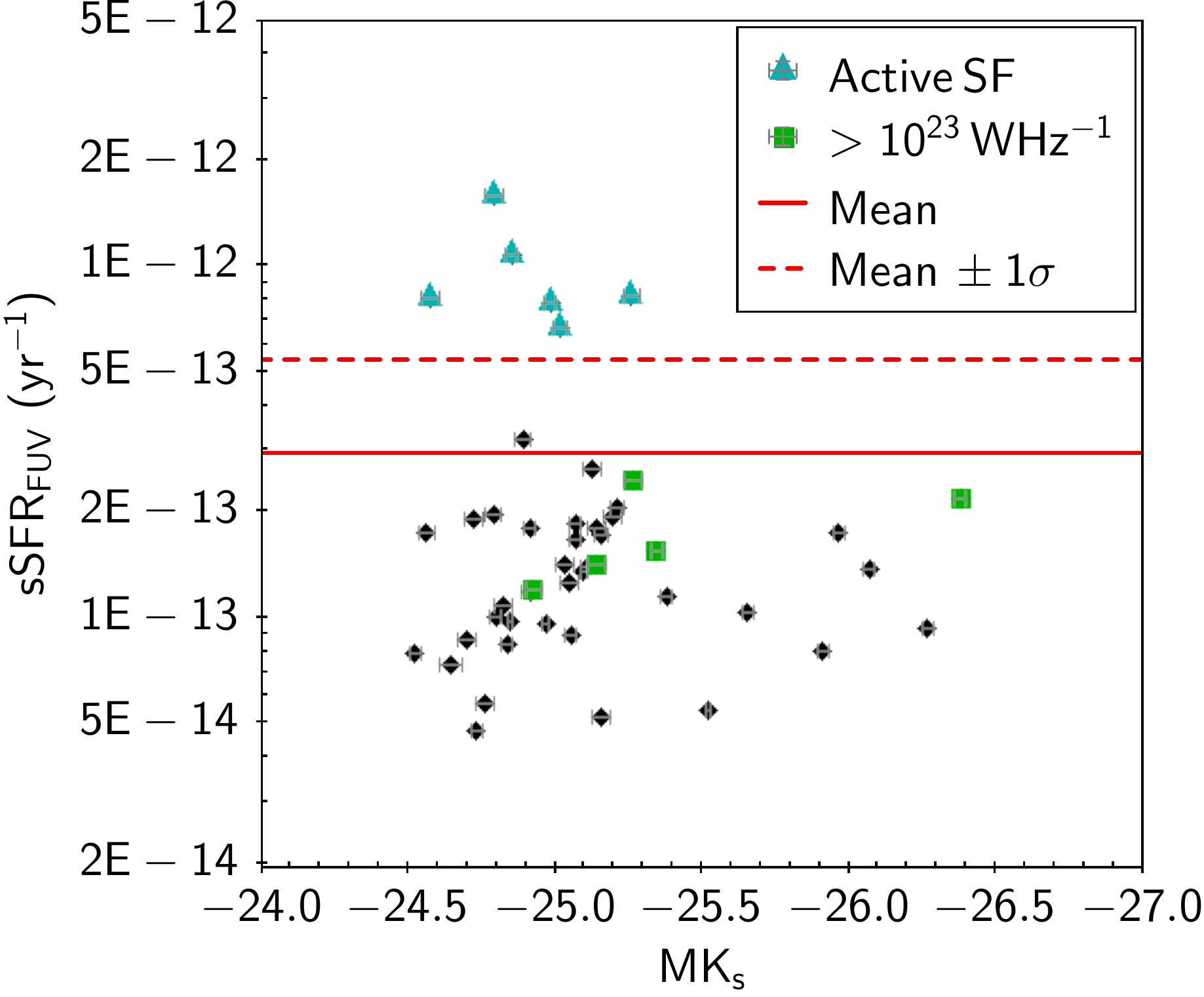}
\caption{Specific star-formation rate (sSFR$_{FUV}$) plotted against absolute K-band magnitude M$_{K_s}$. The active star-forming galaxies are shown in cyan and the systems with P$_{1.4~GHz} \geq10^{23} W~Hz^{-1}$ as green squares.}
\label{sSFRFUVMKs}
\end{figure}

Figure~\ref{sSFRFUVMKs} shows specific star-formation rate (sSFR$_{FUV}$=SFR$_{FUV}$/M$_{stellar}$) for our systems, plotted against stellar mass. Specific star-formation rate is a measure of star formation efficiency and the fractional growth rate of the galaxies. As expected, the majority of the BGEs have comparable sSFR$_{FUV}$, generally within a factor of $\sim$2-3 of 10$^{-13}$~yr$^{-1}$, without any evidence of a trend with M$_{K_s}$, indicating that no other factors beyond stellar mass are  driving the trend seen in Figure~\ref{SFRFUVMstellar}. 


The actively star-forming systems have values about an order of magnitude higher, scattering within a factor $\sim$2 of 10$^{-12}$~yr$^{-1}$. The mean value for the sample as a whole is $\sim$3$\times$10$^{-13}$~yr$^{-1}$, and the actively star-forming systems fall at least 1$\sigma$ above this. The only passive system barely above the mean sSFR$_{FUV}$ of the whole sample is NGC~1550 \citep{Kolokythasetal20} which falls about a factor of $\sim$3 below the actively star-forming galaxies. The mean sSFR$_{FUV}$ value for the passive systems alone is $\sim$1.4$\times$10$^{-13}$~yr$^{-1}$ with a standard deviation of $\sim$0.6$\times$10$^{-13}$~yr$^{-1}$. Compared to this value, the actively star-forming systems are $>$7$\sigma$ outliers.

The five radio-loud BGEs with measured sSFR$_{FUV}$ are also marked on Figure~\ref{sSFRFUVMKs}. We see no evident trend in their sSFR, and they are consistent with the rest of the passive population. Again, there is no evidence of suppression or enhancement of star formation by the jets in these systems.

\subsection{The FUV star formation main sequence}

Having examined the sSFR of our galaxies and shown that the actively star-forming systems fall at higher values, we can consider how they compare to other star-forming galaxies. It is well established that most spiral galaxies follow a star formation main sequence, with similar sSFR. \citet{OSullivanetal15,OSullivanetal18b} examined the location of the CLoGS BGEs with respect to the far-Infrared (FIR) version of this relation \citep{Wuytsetal11} and found that almost all of them fell well below it. Only NGC~4169 (for which we have no \textit{GALEX} FUV detection) and NGC~7252  were found to be consistent (to within a factor of 5) with the SF main sequence.
 
\begin{figure}
\centering
\includegraphics[width=0.48\textwidth]{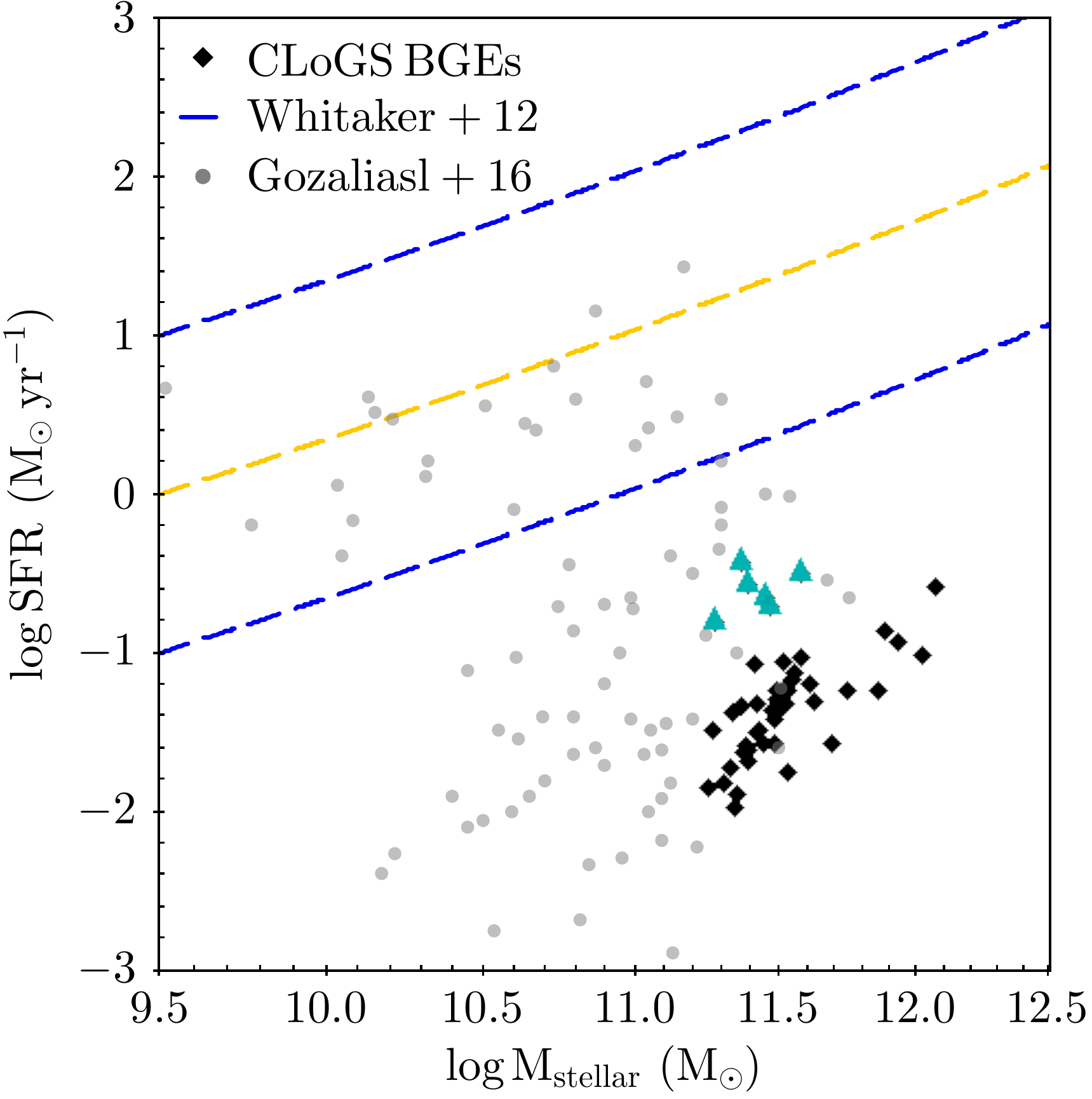}
\caption{SFR$_{FUV}$ versus M$_{stellar}$ low \textit{z} galaxy groups. CLoGS passive BGEs are shown by black diamonds, actively star-forming galaxies by cyan triangles. The grey circles show group-dominant galaxies from sub-samples S-I and S-II of \citet{Gozaliasletal16}. The dashed yellow line represents the star formation main sequence of \citet{Whitakeretal12} with blue dashed lines indicating the factor of 10 dispersion around the relation. }
\label{GozWhitSFRFUVMstellar}
\end{figure}

Figure~\ref{GozWhitSFRFUVMstellar} shows the FUV star-forming main sequence, as defined by \citet{Whitakeretal12} for 0$\leq$z$\leq$0.5 galaxies. CLoGS passive BGEs are marked as black diamonds, and actively star-forming galaxies as cyan triangles. We have also included the low-z sub-sample (S-I and S-II) of \citet{Gozaliasletal16}, consisting of group-dominant galaxies drawn from the COSMOS, AEGIS and XMM-LSS surveys. While some of the Gozaliasl group-dominant galaxies are consistent with the star formation main sequence, none of the CLoGS BGEs are. Even the actively star-forming systems fall below the relation by $\sim$2~dex, which can be attributed to the CLoGS sample selection criteria that excludes massive spirals as central group galaxies.


This suggests that the majority of CLoGS BGEs lie in the red sequence of quenched galaxies, in good agreement with our results from Figures~\ref{FUVKsvsMKs} and ~\ref{Wisecolour}. Nearby Brightest Cluster Galaxies (BCGs; e.g., \citealt{FraserMcKelvieetal14}) fall at a similar location on the plot, suggesting the evolutionary paths of group and cluster-dominant ellipticals lead to effective quenching (see e.g., \citealt{Barroetal13,Bellietal15}). Comparison of our actively star-forming systems with a sample of 52 star-forming S0s extracted from the SDSS-IV MaNGA sample (\citealt{Keetal2021}, their figure 4) shows that these star-forming S0s typically fall around the star-formation main sequence, well above our group dominant galaxies. Only one lies in the same area as our actively star-forming galaxies.

It is notable that some of the CLoGS BGEs fall closer to the FIR star formation main sequence than they do to its FUV equivalent. This could be an indication that these systems host dust-obscured star formation, or that their infrared emission is boosted by AGN dust heating. However, Figure~\ref{WISEFUVKPradio} suggests that neither factor is likely to have a strong impact in the majority of the galaxies.

\subsection{Properties of mid-infrared intermediate disk systems and comparison with active SF systems}
 \label{Intermediate Mid-IR}
 
As noted earlier, of the 6 actively SF BGEs, one (NGC~7252) is classed as an active star-forming disk in Figure~\ref{Wisecolour}, three (NGC~252, NGC~940 and NGC~1106) are intermediate disks, and two are classed as spheroids (NGC~924 and ESO~507-25), though they are located closer to the boundary with intermediate disks than the majority of BGEs in the spheroid region. Two additional BGEs fall into the intermediate disk region, NGC~1779 and NGC~7377. NGC~1779 has no available FUV flux measurements, but NGC~7377 has [FUV - K$_s$] colours consistent with being a passive, non-star-forming galaxy. Such inconsistencies between different star formation indicators are to be expected; FUV bright systems probably host younger, less obscured star formation, while mid-IR colours may be better at tracing older, dustier star-forming regions.
 
The five mid-IR intermediate disk systems share several common features. All five are lenticular galaxies, classed as either S0 or S0/Sa (NGC~1779 and NGC~7377). All five host very weak central radio point sources \citep{Kolokythasetal18,Kolokythasetal19}, and are H$_{2}$-rich with M(H$_{2}$) $>$ 10$^8$ \msol  \citep{OSullivanetal18b}. Three of the five are identified as hosting cold gas disks, though the variable quality of the available data means that this fraction must be considered a lower limit.

Generally speaking, galaxies are expected to evolve along the sequence from active star-forming disks, through intermediate disks, eventually reaching the spheroid category as their star formation is quenched. NGC~7252, as a post-merger galaxy, is dynamically the youngest of our galaxies, consistent with its position  on the mid-IR colour-colour diagram. However, for the intermediate disks, it is unclear whether these are systems slowly moving toward the spheroid category as they age, or systems ``restored to youth" by an uptick in star formation fuelled by their cold gas reservoirs.

\subsection{Cold gas content} 

\citet{OSullivanetal15,OSullivanetal18b} describe the cold gas content of the CLoGS BGEs. A survey of CO emission using the IRAM~30m and APEX telescopes was used to determine molecular gas content, and information on H\textsc{i} content was assembled for a subset of galaxies from the literature. In some cases the morphology of the gas is known from prior studies, but in many systems only a mass is available.

\begin{figure}
\centering
\includegraphics[width=0.48\textwidth]{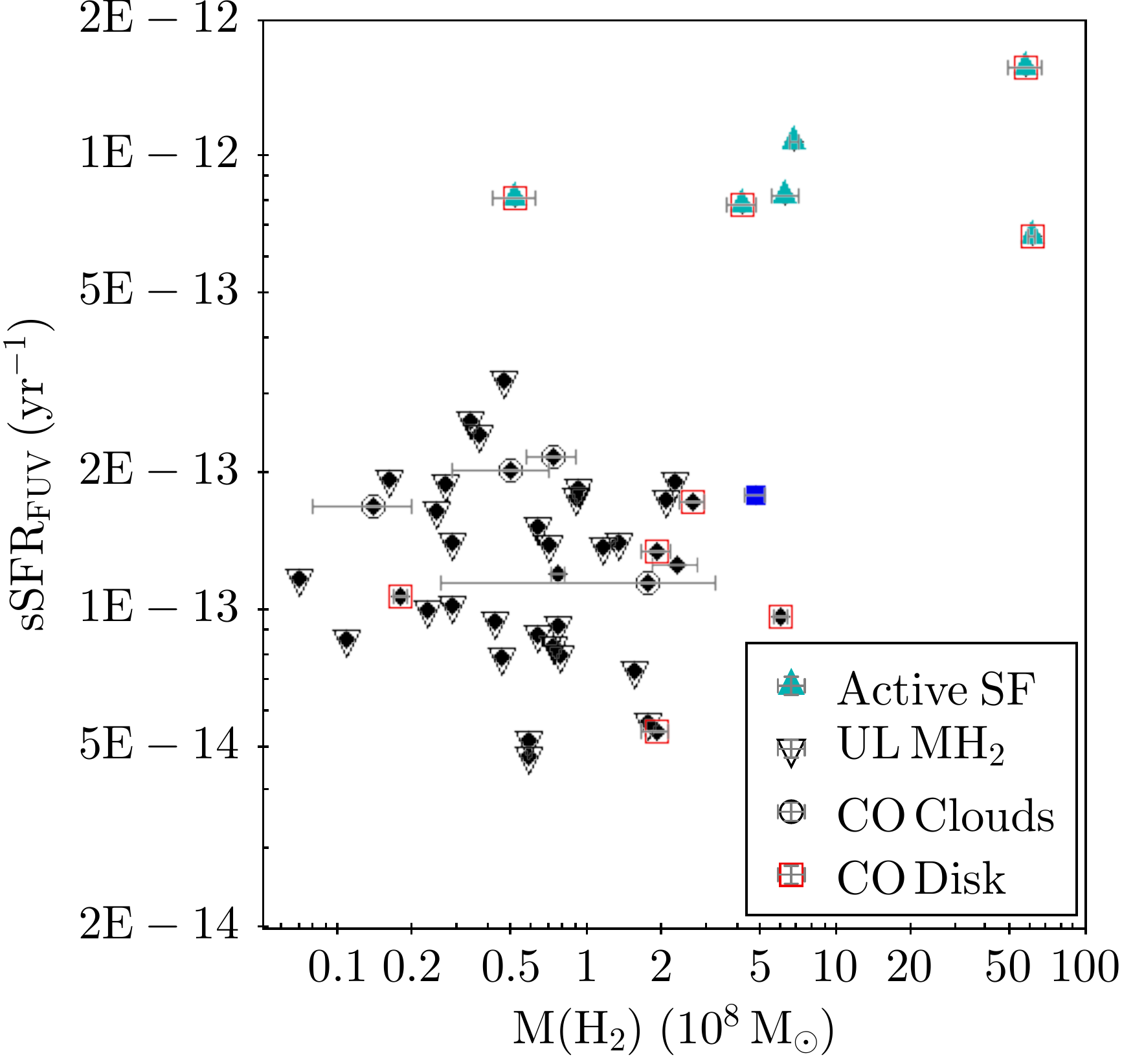}
\caption{Specific star-formation rate (sSFR$_{FUV}$) plotted against molecular gas mass, M(H$_{2}$). The active star-forming galaxies are shown as cyan upward-pointing triangles, the WISE intermediate-disk system NGC~7377 as a blue square. Galaxies for which only upper limits (UL) on M(H$_2$) are available are marked with black, downward-pointing triangles. Galaxies identified as hosting cold gas disks by \citet{OSullivanetal18b} are marked with open red squares and systems whose cold gas is located in clouds with an open circle.}
\label{sSFRFUVMH2}
\end{figure}

Figure~\ref{sSFRFUVMH2} shows the molecular gas content of the BGEs plotted against their sSFR, for the 47 CLoGS BGEs that have measured SFR$_{FUV}$. Note that many of the galaxies have only upper limits on M(H$_2$). Although the two most cold gas rich systems in our sample (M(H$_2$)$>$50$\times$10$^8$ M$_\odot$) are classed as actively star-forming (NGC~940 and NGC~7252), the molecular gas masses of the remaining actively-SF galaxies overlaps with those of the passive galaxies, in the range 0.5$-$7$\times$10$^8$ M$_\odot$.

Examining the cold gas morphology of the systems we find that four of the six actively-SF systems host a molecular gas disk. However, we also find five systems with molecular gas disks among the passive systems. The presence of an extended gas disk might promote star formation, providing a relatively stable environment in which gas clouds have time to undergo gravitational collapse and form stars. However, it is clear that some AGN are fuelled by smaller-scale disks, as in NGC~4261 and NGC~315 \citep{Boizelleetal21}, both passive galaxies hosting powerful radio jets. Systems whose cold gas appears to be located in clouds, or in some cases filaments, are among the passive galaxies, though with a wide spread of molecular gas masses. These include systems like NGC~5044 and NGC~5846, where we have strong evidence that the molecular gas is material that has cooled from the hot IGrM \citep{Temietal17,Schellenbergeretal20}. Both systems host filamentary H$\alpha$ nebulae which closely resemble those in strong cool core galaxy clusters, with the molecular gas located within these nebulae, with low velocity relative to their galaxies and no sign of rotation or streaming motions. The molecular gas may have cooled out of the IGrM at its current location, or condensed closer to the core of the galaxies and then been uplifted to its current position by rising radio bubbles, but it is very unlikely to be material captured directly from other galaxies.

\begin{figure}
\centering
\includegraphics[width=0.48\textwidth]{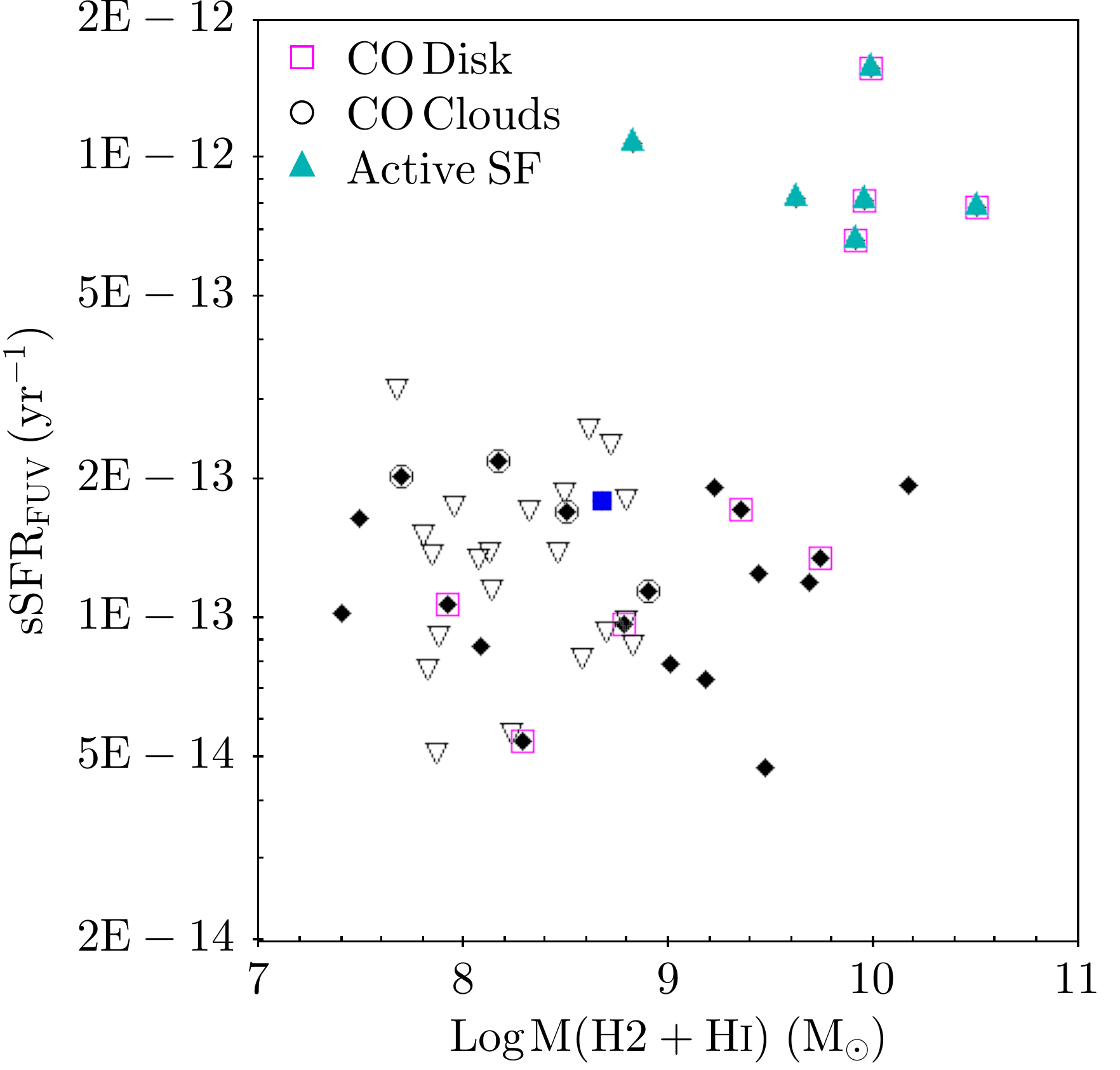}
\caption{Specific star-formation rate (sSFR$_{FUV}$) plotted against total cold gas mass, M(H$_2$+H\textsc{i}). The active star-forming galaxies are shown as cyan triangles, with galaxies identified as hosting cold gas disks by \citet{OSullivanetal18b} marked with open magenta squares and systems with cold gas cloud morphology with an open circle. The systems with upper limits on M(H$_2$+H\textsc{i}) are noted with open downward-pointing triangles.}
\label{sSFRFUVMH2HI}
\end{figure}

H\textsc{i} masses drawn from the literature are available for a subset of the BGEs. As they are derived from surveys and targeted observations with a range of spatial resolutions and depths, the total cold gas mass we estimate from them, M(H$_2$+H\textsc{i}), has a large degree of scatter. We nonetheless examined a plot of M(H$_2$+H\textsc{i}) vs sSFR in Figure~\ref{sSFRFUVMH2HI}, but found that the separation between actively SF and passive galaxies was not further clear. A uniform survey of H\textsc{i} in CLoGS BGEs is probably required if we are to examine total cold gas mass in more detail.

\subsection{Influence of the group environment}

\begin{figure}
\centering
\includegraphics[width=0.45\textwidth]{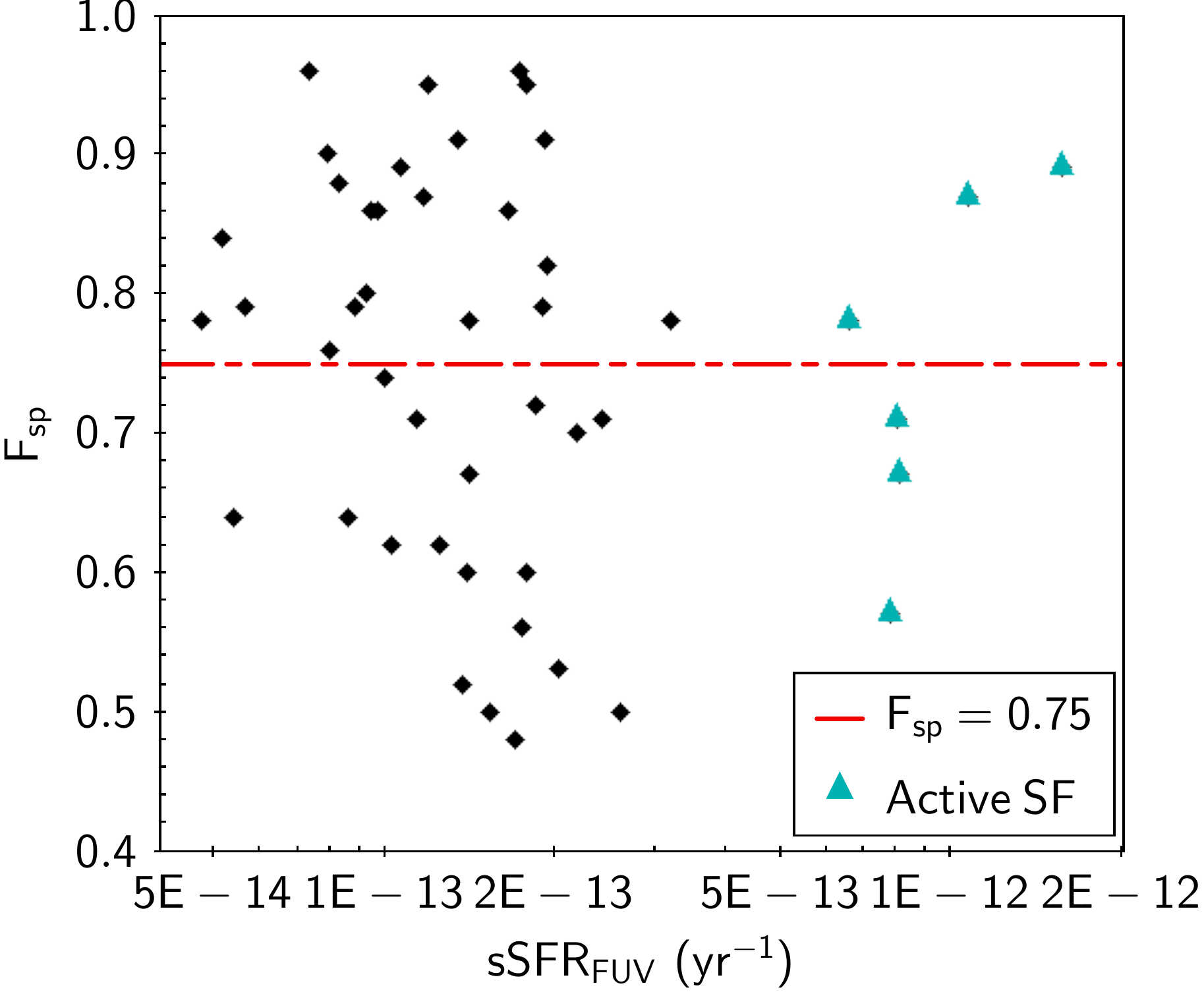}
\caption{Spiral fraction F$_{sp}$ of CLoGS groups in relation to the specific star-formation rate, sSFR$_{FUV}$, that their dominant galaxies present.}
\label{sSFRFUVFsp}
\end{figure}

\label{sec:environment}

The role of the environment is known to be important for the evolution of galaxies. For group-dominant galaxies, their merger history has been driven by their location in a relatively dense environment, and it is clear that their presence at the centre of a larger system makes them particularly likely to host radio AGN, and to contain cold gas, compared to non-central early-type galaxies \citep{Kolokythasetal18,Kolokythasetal19,OSullivanetal18b}. We can now examine the relation between the environment of the CLoGS group-dominant galaxies and their star formation. We assess the environment of the group dominant galaxies using 3 factors: i) the spiral fraction of each group F$_{sp}$, ii) the richness of the groups, $R$, and iii) the presence of extended X-ray emission tracing a hot IGrM.

The dynamical age of a galaxy group can be determined by the spiral galaxy content of its member galaxies, with dynamically older systems expected to have undergone more galaxy interactions and mergers, and therefore to contain more early--type galaxies. Following \citet{Bitsakisetal10},  \citet{Kolokythasetal18} (see their Table~8) and \citet{Kolokythasetal19} (see their Table~10) determined the fraction of spiral galaxies in each CLoGS group (F$_{sp}$; the ratio of the number of late-type plus unknown galaxies over the total number of galaxies). Groups with F$_{sp}>$ 0.75 were classed as spiral-rich and deemed potentially dynamically young, whereas groups with F$_{sp}<$ 0.75 were considered spiral-poor and dynamically old. Figure~\ref{sSFRFUVFsp} shows that there is no obvious correlation between sSFR$_{FUV}$ of the BGE and F$_{sp}$ of the group. The actively star-forming BGEs are evenly distributed between spiral rich and spiral poor systems, implying that growth of the BGE through star formation is independent of the dynamical evolution of the galaxy population of their group. 

The richness parameter $R$ measures the number of luminous, massive galaxies in the group, a proxy for the stellar mass of the group, and perhaps for its total mass. Comparing specific star-formation rate with $R$, we find that the mean sSFR$_{FUV}$ is similar between low--richness ($\sim4.2\times10^{-13} yr^{-1}$) and high-richness ($\sim3.5\times10^{-13} yr^{-1}$) CLoGS groups. The six actively SF galaxies are again split evenly between the low and high--richness systems. The two additional galaxies identified as intermediate disks on the mid-IR colour--colour diagram are in low--richness groups, as is the only BGE that falls in the dust-obscured SF section of Figure~\ref{WISEFUVKPradio}, NGC~315.

\citet{OSullivanetal17} identify the groups in the high--richness sample which host a hot IGrM, defined as extending $>$65~kpc and X-ray luminosity $>$10$^{41}$~erg~s$^{-1}$. The same classification has been performed for the low--richness groups (O'Sullivan et al., in prep.). We find that all six of the actively-SF BGEs occupy groups which lack a hot IGrM, as do both intermediate-disk systems. NGC~315, by contrast, is the BGE of an X-ray luminous group with an extended hot IGrM. This suggests that the presence of a hot IGrM hinders star formation in the dominant galaxy. The absence of a hot IGrM is not enough to promote SF, however; there are X-ray faint groups whose BGEs are passive systems.

It therefore appears that the growth through star formation of group dominant galaxies is probably not strongly affected by the wider galaxy population of their host group, but that it may be suppressed in the presence of a hot IGrM.

\subsection{Comparison with galaxy clusters}
\label{sec:clusters}

At low redshift, AGN feedback in galaxy clusters is thought to operate through the cavities inflated by, and shocks driven by, the AGN of the dominant elliptical \citep[e.g.,][]{McNamaraNulsen12}. The same conditions in the intra-cluster medium which are associated with jet-mode activity (low central entropy, short cooling times, etc.) also appear to be correlated with star formation in the BCG \citep[e.g.,][]{Bildfelletal08,Hoganetal17,Pulidoetal18,Loubseretal21}. This makes it clear that material cooling from the ICM fuels both SF and AGN in cool-core galaxy clusters.

It should be emphasized that galaxy clusters differ from groups in a number of important ways. All galaxy clusters possess a hot ICM, which is in fact their dominant baryonic component; there are no clusters with properties equivalent to X-ray faint groups. It is also significantly easier for group-dominant galaxies to acquire cold gas from their neighbours than it is for BCGs. The deeper gravitational potentials of galaxy clusters mean that, compared to groups, their member galaxies move at higher velocities and are exposed to stripping by the ICM over longer periods. It is therefore difficult for a cold-gas-rich galaxy to reach the cluster core without losing a large fraction of its gas, and those that do, move at velocities which make it difficult for the BCG to acquire material from them.

\begin{figure}
	\includegraphics[width=\columnwidth]{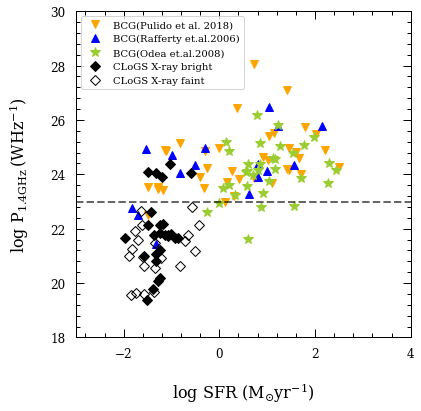}
    \caption{1.4 GHz radio power vs SFR of the BGEs and BCGs from \citet{Raffertyetal06}, \citet{ODeaetal08}, and \citet{Pulidoetal18}.The horizontal dashed line drawn at P$_{1.4GHz}$ = 10$^{23}$ WHz$^{-1}$ separates radio-quiet and  radio-loud systems.}
    \label{fig:clogs_bcg_power_sfr}
\end{figure}

As discussed in \citet{OSullivanetal17}, AGN feedback in our X-ray bright groups appears to function similarly to feedback in cool-core clusters. Jet activity is mainly observed in systems with luminous, extended (group-scale) X-ray halos with cool cores (whether defined by short central cooling times or centrally declining temperature profiles) and in most cases the enthalpy of the lobes appears to be sufficient to balance cooling losses \citep{Kolokythasetal19}. However, the results presented in previous sections suggest some important differences in star formation. It is therefore of interest to compare star formation and AGN activity in our galaxy groups with galaxy clusters. We therefore compare the radio power and star-formation rates of CLoGS BGEs with those of BCGs drawn from \citet{Pulidoetal18}, \citet{ODeaetal08} and \citet{Raffertyetal06} samples. We note that the Rafferty and Pulido samples include a few groups and individual ellipticals and that all three samples include primarily cool-core systems. This is to be expected, as AGN jet activity is most common in the BCGs of cool core clusters. By the widely used definition of central cooling time $\leq$7.7~Gyr, all X-ray bright CLoGS groups are also cool core systems \citep[see][for a full discussion]{OSullivanetal17}. Figure ~\ref{fig:clogs_bcg_power_sfr} shows the comparison, with the BGEs marked by filled or open diamond points depending on whether they occupy X-ray bright or faint groups respectively, as described in \S~\ref{sec:environment}.

We find that the majority of CLoGS dominant galaxies have lower SFR and P$_{1.4GHz}$ than the BCGs in the \citet{Pulidoetal18}, \citet{ODeaetal08} and \citet{Raffertyetal06} cluster samples. Although BCGs present some scatter in radio power, a statistical test of correlation using the non-parametric Spearman
rank test for the CLoGS and BCGs samples indicates a correlation (correlation coefficient 0.7) which is plausible since both SFR and P$_{1.4GHz}$ correlate with stellar mass. The total range of SFR observed in Figure~\ref{fig:clogs_bcg_power_sfr} is almost five orders of magnitude and the total range in radio power is roughly eight orders of magnitude.

As is evident, on average, BCGs present higher radio power (average P$_{1.4GHz}$ = 1.9$\times$10$^{24}$ WHz$^{-1}$) and SFR (average SFR = 0.46 \sfrunit) compared to BGEs. It is notable that the SFRs measured in the group-dominant galaxies are all lower than those measured in the O'Dea BCGs which is plausible since these BCGs are located in cool-core clusters which typically show higher star formation activity as their cooling rates are higher. By contrast, there is significant overlap in the range of radio powers and SFRs between CLoGS groups and clusters from the Rafferty and Pulido samples (which include some less massive systems), despite the fact that the clusters are generally more distant. 




%

With only one exception, radio-loud CLoGS BGEs  (with P$_{1.4GHz}\geq10^{23}$~W~Hz$^{-1}$) reside in X-ray bright groups \citep{OSullivanetal17} and host FR~I radio sources, as do the great majority of cluster dominant galaxies. The exception \citep[the restarted FR~I NGC~1167, which resides in an X-ray faint group,][]{OSullivanetal17} is not shown in Figure~\ref{fig:clogs_bcg_power_sfr} since it is not detected in the FUV and thus was not included in this analysis. The radio-loud systems sit well above the general trend, though as noted above, there are BCGs which seem to be outliers to a similar degree.

For radio-quiet systems, there  is no clear connection observed between P$_{1.4GHz}$ and the presence of an X-ray bright hot IGrM. However, \citet{OSullivanetal17} showed that for the high--richness groups, jet activity in the BGE was almost entirely restricted to systems with either a full-scale IGrM or a smaller X-ray halo associated with the BGE itself. Again, the only exception was NGC~1167.

The six actively SF BGEs sit at the high-SFR end of the distribution for groups. As noted previously, NGC~315 has a similar SFR but significantly lower sSFR. Interestingly, the six actively SF groups fall on the border between groups and clusters in Figure~\ref{fig:clogs_bcg_power_sfr} despite being X-ray faint systems. This indicates that in the absence of a radio loud AGN host, higher SFR can be observed in groups,  although we are constrained from drawing firm conclusions by the low number statistics. The remaining systems are scattered in SFR, with BGEs of X--ray bright groups mostly, but not exclusively, at the higher-SFR end of the range.

The observed correlation between groups and clusters in Figure~\ref{fig:clogs_bcg_power_sfr} may be consistent with a common fuel source and set of processes governing AGN feedback and SF across the full mass range of all these systems. However, interpretation is complex as both SFR and P$_{1.4GHz}$ are likely also correlated with other factors, such as BCG mass, X-ray luminosity, cooling rate, etc., that can diversify the processes involved. For SF, the higher SFR observed in clusters can be attributed to the higher cooling rate from the ICM, whereas in the case of galaxy groups, cooling gas seems to be more likely to feed the central AGN, rather than higher SFR. Mergers may provide an alternative route to supply cold gas to group central galaxies and fuel star formation, but this mechanism is unlikely to be important in clusters. For AGN feedback, the inclusion of X--ray faint groups at the lower-SFR end of the range makes the idea of a common fuel source also less plausible as such systems seem unlikely to be fuelling their central AGN through cooling from a hot IGrM. However the presence of cold gas from a different origin in some of these systems may also trigger AGN activity.

\subsection{Gas content and properties of star-forming systems} 
\label{GAScontent}

To understand how SF is fuelled in the CLoGS BGEs, it is worth examining some of the individual galaxies in more detail. As mentioned above, all six FUV-bright, actively-SF galaxies are classed as lenticulars, though NGC~7252 is a post-merger system with tidal tails, whose morphology may therefore not yet be settled. \citet{OSullivanetal18b} summarizes the cold gas content of the BGEs, finding that 21 of the 53 are detected in their CO survey, and at least 27 of 53 contain H\textsc{i}. Information on the morphology of the gas is only available for a subset of systems, either from reported imaging observations, or based on signatures of rotation. Combining the systems listed by \citet{OSullivanetal18b} with recent ALMA detections of sub-kiloparsec scale CO disks in NGC~315 and NGC4261 \citep{Boizelleetal21}, we know of 12 galaxies with detected or probable rotating disks, three with gas clouds or filaments, and one, NGC~315, which has both a disk and clouds.



NGC~940 and NGC~7252 are the two CO-richest BGEs in CLoGS, with M(H$_2$)$\simeq$6$\times$10$^9$~M$_\odot$. Both are also H\textsc{i}-rich and highly FIR-luminous for early-type galaxies. The cold gas in NGC~7252 is located in the tidal arms and a disk, and the H\textsc{i} and H$_2$ line profiles in NGC~940 are suggestive of a disk. Both galaxies have M(H$_2$)/M(H\textsc{i})$>$1, indicating a large fraction of dense gas in their disks \citep{OSullivanetal18b}. The two galaxies host radio point sources \citep{Kolokythasetal18} with NGC~7252, as mentioned earlier, known to be a post-merger starburst system with evidence of older star formation (e.g., \citealt{Hibbardetal94,Dopitaetal02,ChienBarnes10}) as well as the current phase.

NGC~924 and NGC~252 are also both detected in H\textsc{i} and CO, with NGC~924 having one of the highest cold gas masses, M(H$_{2}$+H\textsc{i}), in our sample. Both host only radio point sources, with NGC~924 being particularly radio-faint (P$_{1.4~GHz} \sim 4\times10^{20}~W~Hz^{-1}$). No information is available on the distribution of cold gas in NGC~252, but NGC~924 has an H\textsc{i} line profile suggestive of a disk.

ESO~507-25 is detected in both CO and H\textsc{i} and is in fact the H\textsc{I}-richest BGE in the sample, with a double-peaked line profile indicative of a rotating disk. As well as a radio point source, extended diffuse radio emission is detected in the galaxy \citep{Kolokythasetal18}, with a flat spectral index suggesting a relatively young electron population. The diffuse emission may be tracing star formation in a disk, in which case its radius would be 5-6~kpc (see Lagos et al. 2021 in prep. for a detailed MUSE analysis on this system). ESO~507-25 is also a relatively FIR-luminous galaxy.

NGC~1106 is also strong in FIR, but only detected in CO. It hosts a relatively strong radio point source (P$_{1.4~GHz} \sim 7\times10^{22}~W~Hz^{-1}$),
and is a Compton-thick Seyfert 2, with a hard X-ray spectrum and strong reflected 6.4~keV Fe-K line, demonstrating the the AGN is radiatively powerful \citep{Tanimotoetal16}. Its FIR luminosity may therefore represent a combination of SF and reprocessed AGN emission, but as the source is very heavily obscured, its FUV emission is still probably dominated by SF.

All six of the actively-SF systems are cold-gas rich, including the H\textsc{i}-richest BGE in the sample, the two H$_2$-richest, and five of the six galaxies with M(H$_2$)$>$5$\times$10$^8$~M$_\odot$. Four of the six show indications of hosting a cold gas disk, and the diffuse radio emission in ESO~507-25 may arise from a star-forming disk. They also include three of the five BGEs with the highest FIR luminosities, L$_{\rm FIR}>$10$^{10}$~L$_\odot$.

We can also consider the two FUV-passive or non-detected intermediate disk galaxies, NGC~1779 and NGC~7377. Both are detected in H$_2$ with masses $\sim$4$\times$10$^8$~M$_\odot$, and NGC~1779 is detected in H\textsc{i}, with a top-hat line profile indicative of a disk. Both host radio point sources. These systems seem similar to the actively-SF systems.

Lastly, there is the only system characterized as dust obscured star-forming from Figure~\ref{WISEFUVKPradio}, NGC~315. Unlike the other star-forming systems, it is a well-known giant plumed FR~I radio galaxy \citep[also known as B2~0055+30,][]{Giacintuccietal11}. It contains clouds of H\textsc{i} detected in  both emission and absorption, as well as a small-scale CO disk \citep{Boizelleetal21}. It has been suggested that the H\textsc{i} may be of external origin, acquired via a gas rich merger \citep{Morgantietal09} and not just a product of cooling, but as noted above, NGC~315 is the only one of the systems identified as strongly star-forming that resides in a group with a full-scale hot IGrM. This, and its powerful radio jets, separate it from the other BGEs discussed above.

In summary, it seems clear that the actively-SF systems are more likely to be particularly gas-rich compared to the rest of the CLoGS BGEs, and, although only partial gas morphology data is available, they may be more likely to contain gas disks. As noted previously, only NGC~315 manages to combine indications of strong star formation with a hot IGrM and radio jets; the others all occupy X-ray faint groups and have only radio point sources, or diffuse emission that may arise from star formation.




\subsection{Environment, interactions and gas supply} 

While our group-dominant galaxies show a range of radio powers and star-formation rates, our results show that the sample includes a handful of exceptional systems in each category: radio-loud FR~I galaxies with P$_{1.4GHz}>$10$^{23}$~W~Hz$^{-1}$, and the actively star-forming systems with sSFR$_{FUV}>$5$\times$10$^{-13}$~yr$^{-1}$. If our understanding of AGN processes is correct, both sets of galaxies require cold, dense molecular gas as fuel, and we see evidence of that gas in some systems. The actively-SF systems have sSFRs a factor of $\sim$5 greater than the rest of the sample. The radio powers of the radio-loud systems are at least a factor of 10 above the radio-quiet systems, and a factor of $\sim$100 greater than the typical radio luminosity for the sample. The question of how BGEs acquire their gas and why it produces stronger or weaker nuclear activity or star formation in particular systems is of obvious interest.

There are three possible sources of gas. It may be produced internally, through stellar mass loss, or acquired externally through gas-rich (``wet") mergers, or by cooling from a hot IGrM. Stellar mass loss rates depend on the age of the stellar population, but since large early-type galaxies such as our BGEs are typically dominated by old populations, the mass of gas available will scale primarily with mass. It is thus difficult to explain systems whose sSFR is exceptionally high as being fuelled by stellar mass loss.

There is extensive evidence that cooling from the ICM can drive both star formation and AGN activity in cluster-dominant galaxies. The fact that most (5/6) of our radio-loud BGEs occupy X-ray bright groups supports cooling as the most likely fuel source for their activity. However, it is notable that, with the incomplete data available on cold gas morphology and dynamics for our sample, the systems which most resemble the BCGs of rapidly cooling clusters host AGN with radio jets, but not the most powerful examples \citep[e.g., NGC~5044 and NGC~5846,][]{Schellenbergeretal20, Temietal17}. As described in \S~\ref{GAScontent} NGC~315 hosts a small molecular gas disk and H\textsc{i} clouds that may indicate a past gas-rich merger. NGC~4261 also hosts a sub-kiloparsec rotating disk \citep{Boizelleetal21}. Neither of these radio-loud systems shows evidence of the filamentary nebulae characteristic of ICM cooling in clusters. For the high sSFR systems, IGrM cooling appears to be ruled out, since they occupy X-ray faint groups.

Several CLoGS BGEs show evidence of ongoing or past galaxy-galaxy interactions, some clear examples being the post-merger galaxy NGC~7252, the interacting pairs NGC~1587/88, NGC~2292/92 and NGC5353/54, the tidal H\textsc{i} filament extending across NGC~5903 \citep{OSullivanetal18}. In the case of NGC~7252, its enhanced sSFR is almost certainly related to the merger, which was clearly cold gas rich. Gas-rich mergers can potentially affect the AGN as well as the SFR \citep{Ellisonetal19}, but for both AGN and SFR it is clear that the details of the merger or interaction are important in determining whether activity is enhanced \citep{Dargetal10,Weigeletal18,Pearsonetal19}. Our other actively SF systems do not show clear signs of recent mergers or interactions, though NGC~315 may. The presence of large gas disks in some of our actively SF systems may also be an indicator that gas has been acquired from external galaxies, since material falling in along a particular axis may naturally form a disk. By constrast, disks are not generally found in BCGs with clear ICM cooling signatures.

One other factor to consider is the possibility that AGN feedback may suppress star formation, complicating our ability to separate the two processes. There is evidence that small-scale jets can have some impact in this regard, both from observations \citep{Alataloetal11,Alataloetal15,Nylandetal18} and simulations \citep{Mukherjeeetal16}. A number of the low-SFR BGEs in our sample have small-scale jets, and our actively SF systems lack jets. However, the sSFR of the jet systems does not appear unusually low compared to the rest of the sample, and of course the powerful jets in NGC~315 do not seem to have suppressed the dust-obscured SF in that system. Suppression through radiation driven AGN winds is another possibility, but here the lack of radiatively powerful AGN in the sample argues against a significant impact. NGC~1106 appears to be one of the rare examples, but the presence of a Seyfert 2 nucleus alongside a high sSFR suggests that any wind has been unsuccessful in clearing gas from the galaxy.

 \section{Summary and Conclusions}
 \label{conclusions}
 
In this paper we have examined and characterized the star formation and galaxy growth properties of the CLoGS group dominant galaxies using FUV and mid-infrared diagnostics. These indicators were complemented by 1.4~GHz NVSS radio data \citep{Condonetal98}, molecular CO \citep{OSullivanetal15,OSullivanetal18} and X-rays studies \citep{OSullivanetal17} in order to comprehend the role of the available gas content of the BGEs in relation to their star formation activity and environment in which they reside shedding also light onto the relevant processes involved. 

We find that the SFR$_{FUV}$ values of the CLoGS group dominant galaxies range between 0.01 $-$ 0.4 \sfrunit with a mean of 0.070$\pm0.005$ \sfrunit. Their stellar masses fall within a narrow range between $\sim 10^{11}$ to $\sim 12\times10^{11} M_{\odot}$, as expected from our sample selection criteria. Using the [FUV - K$_s$] $<$ 8.8 mag criterion to distinguish  between actively star-forming and quiescent galaxies, we find that only $\sim$13\% (6/47) of the dominant galaxies show signs of active star formation, namely NGC~252, NGC~924, NGC~940, NGC~1106, NGC~7252 and ESO507-25. All six FUV bright systems are lenticular, cold gas rich, host only weak radio sources (P$_{1.4GHz}$ $<$10$^{23}$ W~Hz$^{-1}$) and occupy X-ray faint groups. On the other hand, the majority of the group dominant galaxies (87\%; 41/47) are found to be passive (FUV faint) with no significant star-forming activity in agreement with previous studies \citep{GildePaz07,Vaddietal16}.

Examining the mid-infrared activity of the CLoGS dominant galaxies based on the mid-infrared diagnostics by \citet{Jarrettetal17,Jarrettetal19} we also find that the majority of the CLoGS group dominant galaxies (87\%; 46/53) lie in the spheroid region ([W2 $-$ W3] $<$ 1.5 mag) as expected for early-type galaxies with little star formation. Only NGC~4956 is classed as a mid-IR bright AGN, presenting a very `warm' [W1 $-$ W2] colour. The post-merger starburst galaxy NGC~7252 falls in the region of actively star-forming disks, with the reddest ([W2 $-$ W3] $>$3 mag) colour in the sample. Five systems are classified as mid-infrared intermediate disks, namely NGC~252, NGC~940, NGC~1106, NGC~1779, and NGC~7377. Three of the five are also classed as actively star-forming based on [FUV - K$_s$] colour, and all are lenticular, cold gas rich, host only weak radio sources, and occupy X-ray faint groups.


The mean specific star-formation rate of the sample is found to be $\sim3\times10^{-13}~yr^{-1}$ with the majority of the galaxies exhibiting low sSFRs near $10^{-13}~yr^{-1}$ and only active star-forming galaxies presenting an order of magnitude higher sSFRs around $10^{-12} yr^{-1}$. While information on cold gas morphology is limited, four of the six actively star-forming systems show signs of hosting rotating gas disks, as does one of the two mid-IR intermediate disk systems not classed as actively SF. This may suggest that star formation is driven not merely by the presence of cold gas, but its dynamical state, with disks perhaps providing a relatively stable environment in which the gas has time to collapse and form stars. The fact that all of these systems occupy X-ray faint systems indicates that the cold gas is unlikely to be the product of cooling from a hot intra-group medium, and is more likely to have been acquired through gas-rich mergers or tidal interactions.

By contrast, the most radio powerful galaxies occupy X-ray bright groups, demonstrating the linkage between IGrM cooling and jet-mode feedback. All of these galaxies fall around the relation between stellar mass and SFR, including NGC~315, the most powerful radio galaxy and most massive BGE in the sample. Its SFR$_{FUV}$ is comparable to the FUV-classified actively-SF systems, though its sSFR is significantly lower. It hosts a sub-kiloparsec scale molecular gas disk (as does another radio-powerful BGE, NGC~4261) and H\textsc{i} clouds that may indicate a recent tidal encounter, but is overall significantly poorer in cold gas than the actively SF galaxies. In general, we find no evidence that the presence of radio jets affects star formation in the BGEs, even for the most powerful and long-lived radio galaxies. As expected for early-type galaxies in the local Universe, our sample contains very few radiatively powerful AGN, but it is notable that one example, the Compton-thick Seyfert 2 NGC~1106, is one of the actively SF galaxies.

We conclude that CLoGS sample group dominant galaxies constitute a mosaic of systems, whose evolution is affected by a combination of secular processes and mergers/interactions, regulated by the environment in which they reside. The presence of a large cold gas reservoir, probably in the form of a rotating disk, seems to promote active star formation, while the presence of an X-ray bright IGrM increases the chances of AGN jet activity, but clearly further work is required to disentangle the other factors which affect these processes.

\section*{Acknowledgments}
K. Kolokythas is supported by the Centre of Space Research at North-West University. This work is based on the research supported in part by the National Research Foundation of South Africa (NRF Grant Numbers: 120850). E. O'Sullivan acknowledges support for this work from the National Aeronautics and Space Administration through \textit{XMM-Newton} award 80NSSC19K1056. A.~Babul acknowledges support from the National Science and Engineering Research Council (NSERC) of Canada. P.~Lagos is supported by national funds through FCT and CAUP (contract DL57/2016/CP1364/CT0010). Some of this research was supported by the EU/FP7 Marie Curie award of the IRSES grant CAFEGROUPS (247653). We acknowledge the usage of the HyperLeda database (http://leda.univ-lyon1.fr). This research has made use of the NASA/IPAC Extragalactic Database (NED) which is operated by the Jet Propulsion Laboratory, California Institute of Technology, under contract with the National Aeronautics and Space Administration. This research makes use of data products from the Widefield Infrared Survey Explorer, which is a joint project of the University of California, Los Angeles, and the Jet Propulsion Laboratory/California Institute of Technology, funded by the National Aeronautics and Space Administration. GALEX (\textit{Galaxy Evolution Explorer}) is a NASA Small Explorer, launched in 2003 April. We gratefully acknowledge NASA’s support for construction, operation, and science analysis for the
GALEX mission, developed in cooperation with the Centre National d’Etudes Spatiales of France and the Korean Ministry of Science and Technology. This research makes also use of data products from the Two Micron All Sky Survey, which is a joint project of the University of Massachusetts and the Infrared Processing and Analysis Center/California Institute of Technology, funded by the National Aeronautics and Space Administration and the National Science Foundation.  Opinions, findings and conclusions or recommendations expressed in this publication is that of the author(s), and that the NRF accepts no liability whatsoever in this regard.

\section*{Data Availability}
The data underlying this article are available in the article and can be found in Table~\ref{Sourcetable}.



\bibstyle{mnras}
\bibliography{paper.bib}

\begin{thebibliography}{}
\makeatletter
\relax
\def\mn@urlcharsother{\let\do\@makeother \do\$\do\&\do\#\do\^\do\_\do\%\do\~}
\def\mn@doi{\begingroup\mn@urlcharsother \@ifnextchar [ {\mn@doi@}
  {\mn@doi@[]}}
\def\mn@doi@[#1]#2{\def\@tempa{#1}\ifx\@tempa\@empty \href
  {http://dx.doi.org/#2} {doi:#2}\else \href {http://dx.doi.org/#2} {#1}\fi
  \endgroup}
\def\mn@eprint#1#2{\mn@eprint@#1:#2::\@nil}
\def\mn@eprint@arXiv#1{\href {http://arxiv.org/abs/#1} {{\tt arXiv:#1}}}
\def\mn@eprint@dblp#1{\href {http://dblp.uni-trier.de/rec/bibtex/#1.xml}
  {dblp:#1}}
\def\mn@eprint@#1:#2:#3:#4\@nil{\def\@tempa {#1}\def\@tempb {#2}\def\@tempc
  {#3}\ifx \@tempc \@empty \let \@tempc \@tempb \let \@tempb \@tempa \fi \ifx
  \@tempb \@empty \def\@tempb {arXiv}\fi \@ifundefined
  {mn@eprint@\@tempb}{\@tempb:\@tempc}{\expandafter \expandafter \csname
  mn@eprint@\@tempb\endcsname \expandafter{\@tempc}}}

\bibitem[\protect\citeauthoryear{{Alatalo} et~al.,}{{Alatalo}
  et~al.}{2011}]{Alataloetal11}
{Alatalo} K.,  et~al., 2011, \mn@doi [\apj] {10.1088/0004-637X/735/2/88}, \href
  {https://ui.adsabs.harvard.edu/abs/2011ApJ...735...88A} {735, 88}

\bibitem[\protect\citeauthoryear{{Alatalo} et~al.,}{{Alatalo}
  et~al.}{2015}]{Alataloetal15}
{Alatalo} K.,  et~al., 2015, \mn@doi [ApJ] {10.1088/0004-637X/812/2/117}, \href
  {http://adsabs.harvard.edu/abs/2015ApJ...812..117A} {812, 117}

\bibitem[\protect\citeauthoryear{{Babul}, {Balogh}, {Lewis}  \&
  {Poole}}{{Babul} et~al.}{2002}]{Babuletal02}
{Babul} A.,  {Balogh} M.~L.,  {Lewis} G.~F.,   {Poole} G.~B.,  2002, \mn@doi
  [\mnras] {10.1046/j.1365-8711.2002.05044.x}, \href
  {https://ui.adsabs.harvard.edu/abs/2002MNRAS.330..329B} {330, 329}

\bibitem[\protect\citeauthoryear{{Baldry}, {Balogh}, {Bower}, {Glazebrook},
  {Nichol}, {Bamford}  \& {Budavari}}{{Baldry} et~al.}{2006}]{Baldryetal06}
{Baldry} I.~K.,  {Balogh} M.~L.,  {Bower} R.~G.,  {Glazebrook} K.,  {Nichol}
  R.~C.,  {Bamford} S.~P.,   {Budavari} T.,  2006, \mn@doi [\mnras]
  {10.1111/j.1365-2966.2006.11081.x}, \href
  {https://ui.adsabs.harvard.edu/abs/2006MNRAS.373..469B} {373, 469}

\bibitem[\protect\citeauthoryear{{Balogh}, {Baldry}, {Nichol}, {Miller},
  {Bower}  \& {Glazebrook}}{{Balogh} et~al.}{2004}]{Baloghetal04}
{Balogh} M.~L.,  {Baldry} I.~K.,  {Nichol} R.,  {Miller} C.,  {Bower} R.,
  {Glazebrook} K.,  2004, \mn@doi [\apjl] {10.1086/426079}, \href
  {https://ui.adsabs.harvard.edu/abs/2004ApJ...615L.101B} {615, L101}

\bibitem[\protect\citeauthoryear{{Barro} et~al.,}{{Barro}
  et~al.}{2013}]{Barroetal13}
{Barro} G.,  et~al., 2013, \mn@doi [\apj] {10.1088/0004-637X/765/2/104}, \href
  {https://ui.adsabs.harvard.edu/abs/2013ApJ...765..104B} {765, 104}

\bibitem[\protect\citeauthoryear{{Baum}, {Zirbel}  \& {O'Dea}}{{Baum}
  et~al.}{1995}]{Baumetal1995}
{Baum} S.~A.,  {Zirbel} E.~L.,   {O'Dea} C.~P.,  1995, \mn@doi [\apj]
  {10.1086/176202}, \href
  {https://ui.adsabs.harvard.edu/abs/1995ApJ...451...88B} {451, 88}

\bibitem[\protect\citeauthoryear{{Becker}, {White}  \& {Helfand}}{{Becker}
  et~al.}{1995}]{Beckeretal95}
{Becker} R.~H.,  {White} R.~L.,   {Helfand} D.~J.,  1995, \mn@doi [ApJ]
  {10.1086/176166}, \href {http://adsabs.harvard.edu/abs/1995ApJ...450..559B}
  {450, 559}

\bibitem[\protect\citeauthoryear{{Beckmann} et~al.,}{{Beckmann}
  et~al.}{2017}]{Beckmannetal17}
{Beckmann} R.~S.,  et~al., 2017, \mn@doi [\mnras] {10.1093/mnras/stx1831},
  \href {https://ui.adsabs.harvard.edu/abs/2017MNRAS.472..949B} {472, 949}

\bibitem[\protect\citeauthoryear{{Bell}}{{Bell}}{2003}]{Belletal03}
{Bell} E.~F.,  2003, \mn@doi [ApJ] {10.1086/367829}, \href
  {http://adsabs.harvard.edu/abs/2003ApJ...586..794B} {586, 794}

\bibitem[\protect\citeauthoryear{{Belli}, {Newman}  \& {Ellis}}{{Belli}
  et~al.}{2015}]{Bellietal15}
{Belli} S.,  {Newman} A.~B.,   {Ellis} R.~S.,  2015, \mn@doi [\apj]
  {10.1088/0004-637X/799/2/206}, \href
  {https://ui.adsabs.harvard.edu/abs/2015ApJ...799..206B} {799, 206}

\bibitem[\protect\citeauthoryear{{Benson} \& {Babul}}{{Benson} \&
  {Babul}}{2009}]{BensonBabul09}
{Benson} A.~J.,  {Babul} A.,  2009, \mn@doi [MNRAS]
  {10.1111/j.1365-2966.2009.15087.x}, \href
  {http://adsabs.harvard.edu/abs/2009MNRAS.397.1302B} {397, 1302}

\bibitem[\protect\citeauthoryear{{Bildfell}, {Hoekstra}, {Babul}  \&
  {Mahdavi}}{{Bildfell} et~al.}{2008}]{Bildfelletal08}
{Bildfell} C.,  {Hoekstra} H.,  {Babul} A.,   {Mahdavi} A.,  2008, \mn@doi
  [MNRAS] {10.1111/j.1365-2966.2008.13699.x}, \href
  {http://adsabs.harvard.edu/abs/2008MNRAS.389.1637B} {389, 1637}

\bibitem[\protect\citeauthoryear{{Bitsakis}, {Charmandaris}, {Le Floc'h},
  {D{\'\i}az-Santos}, {Slater}, {Xilouris}  \& {Haynes}}{{Bitsakis}
  et~al.}{2010}]{Bitsakisetal10}
{Bitsakis} T.,  {Charmandaris} V.,  {Le Floc'h} E.,  {D{\'\i}az-Santos} T.,
  {Slater} S.~K.,  {Xilouris} E.,   {Haynes} M.~P.,  2010, \mn@doi [\aap]
  {10.1051/0004-6361/201014102}, \href
  {https://ui.adsabs.harvard.edu/abs/2010A&A...517A..75B} {517, A75}

\bibitem[\protect\citeauthoryear{{Bogd{\'a}n} et~al.,}{{Bogd{\'a}n}
  et~al.}{2014}]{Bogdanetal14}
{Bogd{\'a}n} {\'A}.,  et~al., 2014, ApJ, 782, L19

\bibitem[\protect\citeauthoryear{{Boizelle} et~al.,}{{Boizelle}
  et~al.}{2021}]{Boizelleetal21}
{Boizelle} B.~D.,  et~al., 2021, \mn@doi [\apj] {10.3847/1538-4357/abd24d},
  \href {https://ui.adsabs.harvard.edu/abs/2021ApJ...908...19B} {908, 19}

\bibitem[\protect\citeauthoryear{{Brienza} et~al.,}{{Brienza}
  et~al.}{2018}]{Brienzaetal18}
{Brienza} M.,  et~al., 2018, \mn@doi [\aap] {10.1051/0004-6361/201832846},
  \href {https://ui.adsabs.harvard.edu/abs/2018A&A...618A..45B} {618, A45}

\bibitem[\protect\citeauthoryear{{Brinchmann}, {Charlot}, {White}, {Tremonti},
  {Kauffmann}, {Heckman}  \& {Brinkmann}}{{Brinchmann}
  et~al.}{2004}]{Brinchmannetal04}
{Brinchmann} J.,  {Charlot} S.,  {White} S.~D.~M.,  {Tremonti} C.,  {Kauffmann}
  G.,  {Heckman} T.,   {Brinkmann} J.,  2004, \mn@doi [\mnras]
  {10.1111/j.1365-2966.2004.07881.x}, \href
  {https://ui.adsabs.harvard.edu/abs/2004MNRAS.351.1151B} {351, 1151}

\bibitem[\protect\citeauthoryear{{Brough}, {Forbes}, {Kilborn}  \&
  {Couch}}{{Brough} et~al.}{2006}]{Broughetal06}
{Brough} S.,  {Forbes} D.~A.,  {Kilborn} V.~A.,   {Couch} W.,  2006, \mn@doi
  [\mnras] {10.1111/j.1365-2966.2006.10542.x}, \href
  {https://ui.adsabs.harvard.edu/abs/2006MNRAS.370.1223B} {370, 1223}

\bibitem[\protect\citeauthoryear{{Brown}, {Jannuzi}, {Floyd}  \&
  {Mould}}{{Brown} et~al.}{2011}]{Brownetal11}
{Brown} M.~J.~I.,  {Jannuzi} B.~T.,  {Floyd} D.~J.~E.,   {Mould} J.~R.,  2011,
  \mn@doi [ApJ] {10.1088/2041-8205/731/2/L41}, \href
  {http://adsabs.harvard.edu/abs/2011ApJ...731L..41B} {731, L41}

\bibitem[\protect\citeauthoryear{{Cappellari}}{{Cappellari}}{2013}]{Cappellari13}
{Cappellari} M.,  2013, \mn@doi [\apjl] {10.1088/2041-8205/778/1/L2}, \href
  {https://ui.adsabs.harvard.edu/abs/2013ApJ...778L...2C} {778, L2}

\bibitem[\protect\citeauthoryear{{Cappellari} et~al.,}{{Cappellari}
  et~al.}{2011}]{Cappellarietal11}
{Cappellari} M.,  et~al., 2011, \mn@doi [MNRAS]
  {10.1111/j.1365-2966.2011.18600.x}, \href
  {http://adsabs.harvard.edu/abs/2011MNRAS.416.1680C} {416, 1680}

\bibitem[\protect\citeauthoryear{{Chien} \& {Barnes}}{{Chien} \&
  {Barnes}}{2010}]{ChienBarnes10}
{Chien} L.~H.,  {Barnes} J.~E.,  2010, \mn@doi [\mnras]
  {10.1111/j.1365-2966.2010.16903.x}, \href
  {https://ui.adsabs.harvard.edu/abs/2010MNRAS.407...43C} {407, 43}

\bibitem[\protect\citeauthoryear{{Cluver} et~al.,}{{Cluver}
  et~al.}{2014}]{Cluveretal14}
{Cluver} M.~E.,  et~al., 2014, \mn@doi [\apj] {10.1088/0004-637X/782/2/90},
  \href {https://ui.adsabs.harvard.edu/abs/2014ApJ...782...90C} {782, 90}

\bibitem[\protect\citeauthoryear{{Cluver}, {Jarrett}, {Dale}, {Smith}, {August}
   \& {Brown}}{{Cluver} et~al.}{2017}]{Cluveretal17}
{Cluver} M.~E.,  {Jarrett} T.~H.,  {Dale} D.~A.,  {Smith} J. D.~T.,  {August}
  T.,   {Brown} M.~J.~I.,  2017, \mn@doi [\apj] {10.3847/1538-4357/aa92c7},
  \href {https://ui.adsabs.harvard.edu/abs/2017ApJ...850...68C} {850, 68}

\bibitem[\protect\citeauthoryear{{Combes}, {Young}  \& {Bureau}}{{Combes}
  et~al.}{2007}]{Combesetal07}
{Combes} F.,  {Young} L.~M.,   {Bureau} M.,  2007, \mn@doi [MNRAS]
  {10.1111/j.1365-2966.2007.11759.x}, \href
  {http://adsabs.harvard.edu/abs/2007MNRAS.377.1795C} {377, 1795}

\bibitem[\protect\citeauthoryear{{Condon}}{{Condon}}{1992}]{Condon92}
{Condon} J.~J.,  1992, \mn@doi [ARA\&A] {10.1146/annurev.aa.30.090192.003043},
  \href {http://adsabs.harvard.edu/abs/1992ARA\%26A..30..575C} {30, 575}

\bibitem[\protect\citeauthoryear{{Condon}, {Cotton}, {Greisen}, {Yin},
  {Perley}, {Taylor}  \& {Broderick}}{{Condon} et~al.}{1998}]{Condonetal98}
{Condon} J.~J.,  {Cotton} W.~D.,  {Greisen} E.~W.,  {Yin} Q.~F.,  {Perley}
  R.~A.,  {Taylor} G.~B.,   {Broderick} J.~J.,  1998, \mn@doi [AJ]
  {10.1086/300337}, \href
  {http://adsabs.harvard.edu/cgi-bin/nph-bib_query?bibcode=1998AJ....115.1693C&db_key=AST}
  {115, 1693}

\bibitem[\protect\citeauthoryear{{Condon}, {Cotton}  \& {Broderick}}{{Condon}
  et~al.}{2002}]{Condonetal02}
{Condon} J.~J.,  {Cotton} W.~D.,   {Broderick} J.~J.,  2002, \mn@doi [AJ]
  {10.1086/341650}, \href {http://adsabs.harvard.edu/abs/2002AJ....124..675C}
  {124, 675}

\bibitem[\protect\citeauthoryear{{Cutri} \& {et al.}}{{Cutri} \& {et
  al.}}{2012}]{Cutrietal12}
{Cutri} R.~M.,  {et al.} 2012, VizieR Online Data Catalog, \href
  {https://ui.adsabs.harvard.edu/abs/2012yCat.2311....0C} {p. II/311}

\bibitem[\protect\citeauthoryear{{Darg} et~al.,}{{Darg}
  et~al.}{2010}]{Dargetal10}
{Darg} D.~W.,  et~al., 2010, \mn@doi [\mnras]
  {10.1111/j.1365-2966.2009.15686.x}, \href
  {https://ui.adsabs.harvard.edu/abs/2010MNRAS.401.1043D} {401, 1043}

\bibitem[\protect\citeauthoryear{{Darvish}, {Mobasher}, {Martin}, {Sobral},
  {Scoville}, {Stroe}, {Hemmati}  \& {Kartaltepe}}{{Darvish}
  et~al.}{2017}]{Darvishetal17}
{Darvish} B.,  {Mobasher} B.,  {Martin} D.~C.,  {Sobral} D.,  {Scoville} N.,
  {Stroe} A.,  {Hemmati} S.,   {Kartaltepe} J.,  2017, \mn@doi [\apj]
  {10.3847/1538-4357/837/1/16}, \href
  {https://ui.adsabs.harvard.edu/abs/2017ApJ...837...16D} {837, 16}

\bibitem[\protect\citeauthoryear{{Di Matteo}, {Springel}  \& {Hernquist}}{{Di
  Matteo} et~al.}{2005}]{DiMatteoetal05}
{Di Matteo} T.,  {Springel} V.,   {Hernquist} L.,  2005, \mn@doi [\nat]
  {10.1038/nature03335}, \href
  {https://ui.adsabs.harvard.edu/abs/2005Natur.433..604D} {433, 604}

\bibitem[\protect\citeauthoryear{{Donnari} et~al.,}{{Donnari}
  et~al.}{2021}]{Donnarietal21}
{Donnari} M.,  et~al., 2021, \mn@doi [\mnras] {10.1093/mnras/staa3006}, \href
  {https://ui.adsabs.harvard.edu/abs/2021MNRAS.500.4004D} {500, 4004}

\bibitem[\protect\citeauthoryear{{Dopita}, {Pereira}, {Kewley}  \&
  {Capaccioli}}{{Dopita} et~al.}{2002}]{Dopitaetal02}
{Dopita} M.~A.,  {Pereira} M.,  {Kewley} L.~J.,   {Capaccioli} M.,  2002,
  \mn@doi [ApJS] {10.1086/342624}, \href
  {http://adsabs.harvard.edu/abs/2002ApJS..143...47D} {143, 47}

\bibitem[\protect\citeauthoryear{{Eke} et~al.,}{{Eke} et~al.}{2004}]{Ekeetal04}
{Eke} V.~R.,  et~al., 2004, \mn@doi [MNRAS] {10.1111/j.1365-2966.2004.08354.x},
  \href
  {http://adsabs.harvard.edu/cgi-bin/nph-bib_query?bibcode=2004MNRAS.355..769E&db_key=AST}
  {355, 769}

\bibitem[\protect\citeauthoryear{{Elbaz} et~al.,}{{Elbaz}
  et~al.}{2007}]{Elbazetal07}
{Elbaz} D.,  et~al., 2007, \mn@doi [\aap] {10.1051/0004-6361:20077525}, \href
  {https://ui.adsabs.harvard.edu/abs/2007A&A...468...33E} {468, 33}

\bibitem[\protect\citeauthoryear{{Ellison}, {Viswanathan}, {Patton},
  {Bottrell}, {McConnachie}, {Gwyn}  \& {Cuillandre}}{{Ellison}
  et~al.}{2019}]{Ellisonetal19}
{Ellison} S.~L.,  {Viswanathan} A.,  {Patton} D.~R.,  {Bottrell} C.,
  {McConnachie} A.~W.,  {Gwyn} S.,   {Cuillandre} J.-C.,  2019, \mn@doi
  [\mnras] {10.1093/mnras/stz1431}, \href
  {https://ui.adsabs.harvard.edu/abs/2019MNRAS.487.2491E} {487, 2491}

\bibitem[\protect\citeauthoryear{{Emsellem}, {Monnet}  \& {Bacon}}{{Emsellem}
  et~al.}{1994}]{Emsellemetal94}
{Emsellem} E.,  {Monnet} G.,   {Bacon} R.,  1994, \aap, \href
  {https://ui.adsabs.harvard.edu/abs/1994A&A...285..723E} {285, 723}

\bibitem[\protect\citeauthoryear{{Fabian}}{{Fabian}}{2012}]{Fabian12}
{Fabian} A.~C.,  2012, ARA\&A, 50, 455

\bibitem[\protect\citeauthoryear{{Fern{\'a}ndez-Trincado}, {Forero-Romero},
  {Foex}, {Verdugo}  \& {Motta}}{{Fern{\'a}ndez-Trincado}
  et~al.}{2014}]{Fernandezetal14}
{Fern{\'a}ndez-Trincado} J.~G.,  {Forero-Romero} J.~E.,  {Foex} G.,  {Verdugo}
  T.,   {Motta} V.,  2014, \mn@doi [\apjl] {10.1088/2041-8205/787/2/L34}, \href
  {https://ui.adsabs.harvard.edu/abs/2014ApJ...787L..34F} {787, L34}

\bibitem[\protect\citeauthoryear{{Ferrarese} \& {Merritt}}{{Ferrarese} \&
  {Merritt}}{2000}]{FerrareseMerritt00}
{Ferrarese} L.,  {Merritt} D.,  2000, \mn@doi [\apjl] {10.1086/312838}, \href
  {https://ui.adsabs.harvard.edu/abs/2000ApJ...539L...9F} {539, L9}

\bibitem[\protect\citeauthoryear{{Fraser-McKelvie}, {Brown}  \&
  {Pimbblet}}{{Fraser-McKelvie} et~al.}{2014}]{FraserMcKelvieetal14}
{Fraser-McKelvie} A.,  {Brown} M.~J.~I.,   {Pimbblet} K.~A.,  2014, \mn@doi
  [\mnras] {10.1093/mnrasl/slu117}, \href
  {https://ui.adsabs.harvard.edu/abs/2014MNRAS.444L..63F} {444, L63}

\bibitem[\protect\citeauthoryear{{Garcia}, {Paturel}, {Bottinelli}  \&
  {Gouguenheim}}{{Garcia} et~al.}{1993}]{Garciaetal93}
{Garcia} A.~M.,  {Paturel} G.,  {Bottinelli} L.,   {Gouguenheim} L.,  1993,
  A\&AS, \href {http://adsabs.harvard.edu/abs/1993A\%26AS...98....7G} {98, 7}

\bibitem[\protect\citeauthoryear{{Giacintucci} et~al.,}{{Giacintucci}
  et~al.}{2011}]{Giacintuccietal11}
{Giacintucci} S.,  et~al., 2011, \mn@doi [ApJ] {10.1088/0004-637X/732/2/95},
  \href {http://adsabs.harvard.edu/abs/2011ApJ...732...95G} {732, 95}

\bibitem[\protect\citeauthoryear{{Giacintucci} et~al.,}{{Giacintucci}
  et~al.}{2012}]{Giacintuccietal12}
{Giacintucci} S.,  et~al., 2012, \mn@doi [ApJ] {10.1088/0004-637X/755/2/172},
  \href {http://adsabs.harvard.edu/abs/2012ApJ...755..172G} {755, 172}

\bibitem[\protect\citeauthoryear{{Gil de Paz} et~al.,}{{Gil de Paz}
  et~al.}{2007}]{GildePaz07}
{Gil de Paz} A.,  et~al., 2007, \mn@doi [\apjs] {10.1086/516636}, \href
  {https://ui.adsabs.harvard.edu/abs/2007ApJS..173..185G} {173, 185}

\bibitem[\protect\citeauthoryear{{Goulding} et~al.,}{{Goulding}
  et~al.}{2014}]{Gouldingetal14}
{Goulding} A.~D.,  et~al., 2014, \mn@doi [\apj] {10.1088/0004-637X/783/1/40},
  \href {https://ui.adsabs.harvard.edu/abs/2014ApJ...783...40G} {783, 40}

\bibitem[\protect\citeauthoryear{{Gozaliasl}, {Finoguenov}, {Khosroshahi},
  {Mirkazemi}, {Erfanianfar}  \& {Tanaka}}{{Gozaliasl}
  et~al.}{2016}]{Gozaliasletal16}
{Gozaliasl} G.,  {Finoguenov} A.,  {Khosroshahi} H.~G.,  {Mirkazemi} M.,
  {Erfanianfar} G.,   {Tanaka} M.,  2016, \mn@doi [\mnras]
  {10.1093/mnras/stw448}, \href
  {https://ui.adsabs.harvard.edu/abs/2016MNRAS.458.2762G} {458, 2762}

\bibitem[\protect\citeauthoryear{{Gu}, {Huang}, {Wilson}  \& {Fazio}}{{Gu}
  et~al.}{2007}]{Guetal07b}
{Gu} Q.~S.,  {Huang} J.~S.,  {Wilson} G.,   {Fazio} G.~G.,  2007, \mn@doi
  [\apjl] {10.1086/525018}, \href
  {https://ui.adsabs.harvard.edu/abs/2007ApJ...671L.105G} {671, L105}

\bibitem[\protect\citeauthoryear{{G{\"u}rkan} et~al.,}{{G{\"u}rkan}
  et~al.}{2018}]{Gurkanetal18}
{G{\"u}rkan} G.,  et~al., 2018, \mn@doi [\mnras] {10.1093/mnras/sty016}, \href
  {https://ui.adsabs.harvard.edu/abs/2018MNRAS.475.3010G} {475, 3010}

\bibitem[\protect\citeauthoryear{{Heckman} \& {Best}}{{Heckman} \&
  {Best}}{2014}]{HeckmanBest14}
{Heckman} T.~M.,  {Best} P.~N.,  2014, \mn@doi [\araa]
  {10.1146/annurev-astro-081913-035722}, \href
  {https://ui.adsabs.harvard.edu/abs/2014ARA&A..52..589H} {52, 589}

\bibitem[\protect\citeauthoryear{Hibbard, Guhathakurta, van Gorkom  \&
  Schweizer}{Hibbard et~al.}{1994}]{Hibbardetal94}
Hibbard J.~E.,  Guhathakurta P.,  van Gorkom J.~H.,   Schweizer F.,  1994, AJ,
  107, 67

\bibitem[\protect\citeauthoryear{{Hickox} et~al.,}{{Hickox}
  et~al.}{2009}]{Hickoxetal09}
{Hickox} R.~C.,  et~al., 2009, \mn@doi [\apj] {10.1088/0004-637X/696/1/891},
  \href {https://ui.adsabs.harvard.edu/abs/2009ApJ...696..891H} {696, 891}

\bibitem[\protect\citeauthoryear{{Hickox}, {Mullaney}, {Alexander}, {Chen},
  {Civano}, {Goulding}  \& {Hainline}}{{Hickox} et~al.}{2014}]{Hickoxetal14}
{Hickox} R.~C.,  {Mullaney} J.~R.,  {Alexander} D.~M.,  {Chen} C.-T.~J.,
  {Civano} F.~M.,  {Goulding} A.~D.,   {Hainline} K.~N.,  2014, \mn@doi [\apj]
  {10.1088/0004-637X/782/1/9}, \href
  {https://ui.adsabs.harvard.edu/abs/2014ApJ...782....9H} {782, 9}

\bibitem[\protect\citeauthoryear{{Hogan} et~al.,}{{Hogan}
  et~al.}{2017}]{Hoganetal17}
{Hogan} M.~T.,  et~al., 2017, \mn@doi [ApJ] {10.3847/1538-4357/aa9af3}, \href
  {http://adsabs.harvard.edu/abs/2017ApJ...851...66H} {851, 66}

\bibitem[\protect\citeauthoryear{{Hopkins}, {McClure-Griffiths}  \&
  {Gaensler}}{{Hopkins} et~al.}{2008a}]{Hopkinsetal08}
{Hopkins} A.~M.,  {McClure-Griffiths} N.~M.,   {Gaensler} B.~M.,  2008a,
  \mn@doi [\apjl] {10.1086/590494}, \href
  {https://ui.adsabs.harvard.edu/abs/2008ApJ...682L..13H} {682, L13}

\bibitem[\protect\citeauthoryear{{Hopkins}, {Hernquist}, {Cox}, {Younger}  \&
  {Besla}}{{Hopkins} et~al.}{2008b}]{Hopkinsetal08b}
{Hopkins} P.~F.,  {Hernquist} L.,  {Cox} T.~J.,  {Younger} J.~D.,   {Besla} G.,
   2008b, \mn@doi [\apj] {10.1086/592087}, \href
  {https://ui.adsabs.harvard.edu/abs/2008ApJ...688..757H} {688, 757}

\bibitem[\protect\citeauthoryear{{Hopkins}, {Kere{\v{s}}}, {O{\~n}orbe},
  {Faucher-Gigu{\`e}re}, {Quataert}, {Murray}  \& {Bullock}}{{Hopkins}
  et~al.}{2014}]{Hopkinsetal14}
{Hopkins} P.~F.,  {Kere{\v{s}}} D.,  {O{\~n}orbe} J.,  {Faucher-Gigu{\`e}re}
  C.-A.,  {Quataert} E.,  {Murray} N.,   {Bullock} J.~S.,  2014, \mn@doi
  [\mnras] {10.1093/mnras/stu1738}, \href
  {https://ui.adsabs.harvard.edu/abs/2014MNRAS.445..581H} {445, 581}

\bibitem[\protect\citeauthoryear{{Jahnke} \& {Macci{\`o}}}{{Jahnke} \&
  {Macci{\`o}}}{2011}]{JahnkeMaccio11}
{Jahnke} K.,  {Macci{\`o}} A.~V.,  2011, \mn@doi [\apj]
  {10.1088/0004-637X/734/2/92}, \href
  {https://ui.adsabs.harvard.edu/abs/2011ApJ...734...92J} {734, 92}

\bibitem[\protect\citeauthoryear{{Janowiecki}, {Catinella}, {Cortese},
  {Saintonge}, {Brown}  \& {Wang}}{{Janowiecki}
  et~al.}{2017}]{Janowieckietal17}
{Janowiecki} S.,  {Catinella} B.,  {Cortese} L.,  {Saintonge} A.,  {Brown} T.,
   {Wang} J.,  2017, \mn@doi [\mnras] {10.1093/mnras/stx046}, \href
  {https://ui.adsabs.harvard.edu/abs/2017MNRAS.466.4795J} {466, 4795}

\bibitem[\protect\citeauthoryear{{Jarrett}, {Chester}, {Cutri}, {Schneider}  \&
  {Huchra}}{{Jarrett} et~al.}{2003}]{Jarrettetal03}
{Jarrett} T.~H.,  {Chester} T.,  {Cutri} R.,  {Schneider} S.~E.,   {Huchra}
  J.~P.,  2003, \mn@doi [\aj] {10.1086/345794}, \href
  {https://ui.adsabs.harvard.edu/abs/2003AJ....125..525J} {125, 525}

\bibitem[\protect\citeauthoryear{{Jarrett} et~al.,}{{Jarrett}
  et~al.}{2011}]{Jarrettetal11}
{Jarrett} T.~H.,  et~al., 2011, \mn@doi [\apj] {10.1088/0004-637X/735/2/112},
  \href {https://ui.adsabs.harvard.edu/abs/2011ApJ...735..112J} {735, 112}

\bibitem[\protect\citeauthoryear{{Jarrett} et~al.,}{{Jarrett}
  et~al.}{2013}]{Jarrettetal13}
{Jarrett} T.~H.,  et~al., 2013, \mn@doi [\aj] {10.1088/0004-6256/145/1/6},
  \href {https://ui.adsabs.harvard.edu/abs/2013AJ....145....6J} {145, 6}

\bibitem[\protect\citeauthoryear{{Jarrett} et~al.,}{{Jarrett}
  et~al.}{2017}]{Jarrettetal17}
{Jarrett} T.~H.,  et~al., 2017, \mn@doi [\apj] {10.3847/1538-4357/836/2/182},
  \href {https://ui.adsabs.harvard.edu/abs/2017ApJ...836..182J} {836, 182}

\bibitem[\protect\citeauthoryear{{Jarrett}, {Cluver}, {Brown}, {Dale}, {Tsai}
  \& {Masci}}{{Jarrett} et~al.}{2019}]{Jarrettetal19}
{Jarrett} T.~H.,  {Cluver} M.~E.,  {Brown} M.~J.~I.,  {Dale} D.~A.,  {Tsai}
  C.~W.,   {Masci} F.,  2019, \mn@doi [\apjs] {10.3847/1538-4365/ab521a}, \href
  {https://ui.adsabs.harvard.edu/abs/2019ApJS..245...25J} {245, 25}

\bibitem[\protect\citeauthoryear{{Johnson}, {Hibbard}, {Gallagher}, {Charlton},
  {Hornschemeier}, {Jarrett}  \& {Reines}}{{Johnson}
  et~al.}{2007}]{Johnsonetal07}
{Johnson} K.~E.,  {Hibbard} J.~E.,  {Gallagher} S.~C.,  {Charlton} J.~C.,
  {Hornschemeier} A.~E.,  {Jarrett} T.~H.,   {Reines} A.~E.,  2007, \mn@doi
  [AJ] {10.1086/520921}, \href
  {http://adsabs.harvard.edu/abs/2007AJ....134.1522J} {134, 1522}

\bibitem[\protect\citeauthoryear{{Kaviraj} et~al.,}{{Kaviraj}
  et~al.}{2007}]{Kavirajetal07}
{Kaviraj} S.,  et~al., 2007, \mn@doi [\apjs] {10.1086/516633}, \href
  {https://ui.adsabs.harvard.edu/abs/2007ApJS..173..619K} {173, 619}

\bibitem[\protect\citeauthoryear{{Kaviraj}, {Shabala}, {Deller}  \&
  {Middelberg}}{{Kaviraj} et~al.}{2015}]{Kavirajetal15}
{Kaviraj} S.,  {Shabala} S.~S.,  {Deller} A.~T.,   {Middelberg} E.,  2015,
  \mn@doi [\mnras] {10.1093/mnras/stv1329}, \href
  {https://ui.adsabs.harvard.edu/abs/2015MNRAS.452..774K} {452, 774}

\bibitem[\protect\citeauthoryear{{Kennicutt}}{{Kennicutt}}{1998}]{Kennicutt98}
{Kennicutt} Jr. R.~C.,  1998, \mn@doi [ApJ] {10.1086/305588}, \href
  {http://adsabs.harvard.edu/abs/1998ApJ...498..541K} {498, 541}

\bibitem[\protect\citeauthoryear{{Knobel}, {Lilly}, {Woo}  \&
  {Kova{\v{c}}}}{{Knobel} et~al.}{2015}]{Knobeletal15}
{Knobel} C.,  {Lilly} S.~J.,  {Woo} J.,   {Kova{\v{c}}} K.,  2015, \mn@doi
  [\apj] {10.1088/0004-637X/800/1/24}, \href
  {https://ui.adsabs.harvard.edu/abs/2015ApJ...800...24K} {800, 24}

\bibitem[\protect\citeauthoryear{{Kolokythas}, {O'Sullivan}, {Giacintucci},
  {Raychaudhury}, {Ishwara-Chandra}, {Worrall}  \& {Birkinshaw}}{{Kolokythas}
  et~al.}{2015}]{Kolokythasetal15}
{Kolokythas} K.,  {O'Sullivan} E.,  {Giacintucci} S.,  {Raychaudhury} S.,
  {Ishwara-Chandra} C.~H.,  {Worrall} D.~M.,   {Birkinshaw} M.,  2015, \mn@doi
  [MNRAS] {10.1093/mnras/stv665}, \href
  {http://adsabs.harvard.edu/abs/2015MNRAS.450.1732K} {450, 1732}

\bibitem[\protect\citeauthoryear{{Kolokythas}, {O'Sullivan}, {Raychaudhury},
  {Giacintucci}, {Gitti}  \& {Babul}}{{Kolokythas}
  et~al.}{2018}]{Kolokythasetal18}
{Kolokythas} K.,  {O'Sullivan} E.,  {Raychaudhury} S.,  {Giacintucci} S.,
  {Gitti} M.,   {Babul} A.,  2018, \mn@doi [MNRAS] {10.1093/mnras/sty2030},
  \href {http://adsabs.harvard.edu/abs/2018MNRAS.481.1550K} {481, 1550}

\bibitem[\protect\citeauthoryear{{Kolokythas}, {O'Sullivan}, {Intema},
  {Raychaudhury}, {Babul}, {Giacintucci}  \& {Gitti}}{{Kolokythas}
  et~al.}{2019}]{Kolokythasetal19}
{Kolokythas} K.,  {O'Sullivan} E.,  {Intema} H.,  {Raychaudhury} S.,  {Babul}
  A.,  {Giacintucci} S.,   {Gitti} M.,  2019, \mn@doi [\mnras]
  {10.1093/mnras/stz2082}, \href
  {https://ui.adsabs.harvard.edu/abs/2019MNRAS.489.2488K} {489, 2488}

\bibitem[\protect\citeauthoryear{{Kolokythas} et~al.,}{{Kolokythas}
  et~al.}{2020}]{Kolokythasetal20}
{Kolokythas} K.,  et~al., 2020, \mn@doi [\mnras] {10.1093/mnras/staa1506},
  \href {https://ui.adsabs.harvard.edu/abs/2020MNRAS.496.1471K} {496, 1471}

\bibitem[\protect\citeauthoryear{{Kormendy} \& {Ho}}{{Kormendy} \&
  {Ho}}{2013}]{KormendyHo13}
{Kormendy} J.,  {Ho} L.~C.,  2013, \mn@doi [\araa]
  {10.1146/annurev-astro-082708-101811}, \href
  {https://ui.adsabs.harvard.edu/abs/2013ARA&A..51..511K} {51, 511}

\bibitem[\protect\citeauthoryear{{K\"{u}hr}, {Witzel}, {Pauliny-Toth}  \&
  {Nauber}}{{K\"{u}hr} et~al.}{1981}]{Kuhretal81}
{K\"{u}hr} H.,  {Witzel} A.,  {Pauliny-Toth} I.~I.~K.,   {Nauber} U.,  1981,
  A\&AS, \href {http://adsabs.harvard.edu/abs/1981A\%26AS...45..367K} {45, 367}

\bibitem[\protect\citeauthoryear{{Lauer} et~al.,}{{Lauer}
  et~al.}{2006}]{Laueretal07}
{Lauer} T.~R.,  et~al., 2006, ArXiv Astrophysics e-prints, \href
  {http://adsabs.harvard.edu/cgi-bin/nph-bib_query?bibcode=2006astro.ph..9762L&db_key=PRE}
  {astro-ph/0609762}

\bibitem[\protect\citeauthoryear{{Loubser} \&
  {S{\'a}nchez-Bl{\'a}zquez}}{{Loubser} \&
  {S{\'a}nchez-Bl{\'a}zquez}}{2011}]{Loubseretal11}
{Loubser} S.~I.,  {S{\'a}nchez-Bl{\'a}zquez} P.,  2011, \mn@doi [\mnras]
  {10.1111/j.1365-2966.2010.17666.x}, \href
  {https://ui.adsabs.harvard.edu/abs/2011MNRAS.410.2679L} {410, 2679}

\bibitem[\protect\citeauthoryear{{Loubser} \& {Soechting}}{{Loubser} \&
  {Soechting}}{2013}]{Loubseretal13}
{Loubser} S.~I.,  {Soechting} I.~K.,  2013, \mn@doi [\mnras]
  {10.1093/mnras/stt394}, \href
  {https://ui.adsabs.harvard.edu/abs/2013MNRAS.431.2933L} {431, 2933}

\bibitem[\protect\citeauthoryear{{Loubser}, {Babul}, {Hoekstra}, {Mahdavi},
  {Donahue}, {Bildfell}  \& {Voit}}{{Loubser} et~al.}{2016}]{Loubseretal16}
{Loubser} S.~I.,  {Babul} A.,  {Hoekstra} H.,  {Mahdavi} A.,  {Donahue} M.,
  {Bildfell} C.,   {Voit} G.~M.,  2016, \mn@doi [MNRAS]
  {10.1093/mnras/stv2784}, \href
  {http://adsabs.harvard.edu/abs/2016MNRAS.456.1565L} {456, 1565}

\bibitem[\protect\citeauthoryear{{Loubser}, {Hoekstra}, {Babul}  \&
  {O'Sullivan}}{{Loubser} et~al.}{2018}]{Loubseretal18}
{Loubser} S.~I.,  {Hoekstra} H.,  {Babul} A.,   {O'Sullivan} E.,  2018, \mn@doi
  [\mnras] {10.1093/mnras/sty498}, \href
  {https://ui.adsabs.harvard.edu/abs/2018MNRAS.477..335L} {477, 335}

\bibitem[\protect\citeauthoryear{{Loubser}, {Hoekstra}, {Babul}, {Bah{\'e}}  \&
  {Donahue}}{{Loubser} et~al.}{2021}]{Loubseretal21}
{Loubser} S.~I.,  {Hoekstra} H.,  {Babul} A.,  {Bah{\'e}} Y.~M.,   {Donahue}
  M.,  2021, \mn@doi [\mnras] {10.1093/mnras/staa3530}, \href
  {https://ui.adsabs.harvard.edu/abs/2021MNRAS.500.4153L} {500, 4153}

\bibitem[\protect\citeauthoryear{{Ma}, {Greene}, {McConnell}, {Janish},
  {Blakeslee}, {Thomas}  \& {Murphy}}{{Ma} et~al.}{2014}]{Maetal14}
{Ma} C.-P.,  {Greene} J.~E.,  {McConnell} N.,  {Janish} R.,  {Blakeslee} J.~P.,
   {Thomas} J.,   {Murphy} J.~D.,  2014, \mn@doi [\apj]
  {10.1088/0004-637X/795/2/158}, \href
  {https://ui.adsabs.harvard.edu/abs/2014ApJ...795..158M} {795, 158}

\bibitem[\protect\citeauthoryear{{Magorrian} et~al.,}{{Magorrian}
  et~al.}{1998}]{Magorrianetal98}
{Magorrian} J.,  et~al., 1998, AJ, 115, 2285

\bibitem[\protect\citeauthoryear{{Martin} et~al.,}{{Martin}
  et~al.}{2005}]{Martinetal05}
{Martin} D.~C.,  et~al., 2005, \mn@doi [\apjl] {10.1086/426387}, \href
  {https://ui.adsabs.harvard.edu/abs/2005ApJ...619L...1M} {619, L1}

\bibitem[\protect\citeauthoryear{{McIntosh}, {Guo}, {Hertzberg}, {Katz}, {Mo},
  {van den Bosch}  \& {Yang}}{{McIntosh} et~al.}{2008}]{McIntoshetal08}
{McIntosh} D.~H.,  {Guo} Y.,  {Hertzberg} J.,  {Katz} N.,  {Mo} H.~J.,  {van
  den Bosch} F.~C.,   {Yang} X.,  2008, \mn@doi [MNRAS]
  {10.1111/j.1365-2966.2008.13531.x}, \href
  {http://adsabs.harvard.edu/abs/2008MNRAS.388.1537M} {388, 1537}

\bibitem[\protect\citeauthoryear{{McNamara} \& {Nulsen}}{{McNamara} \&
  {Nulsen}}{2007}]{McNamaraNulsen07}
{McNamara} B.~R.,  {Nulsen} P.~E.~J.,  2007, \mn@doi [ARA\&A]
  {10.1146/annurev.astro.45.051806.110625}, \href
  {http://adsabs.harvard.edu/abs/2007ARA\%26A..45..117M} {45, 117}

\bibitem[\protect\citeauthoryear{{McNamara} \& {Nulsen}}{{McNamara} \&
  {Nulsen}}{2012}]{McNamaraNulsen12}
{McNamara} B.~R.,  {Nulsen} P.~E.~J.,  2012, \mn@doi [New Journal of Physics]
  {10.1088/1367-2630/14/5/055023}, \href
  {http://adsabs.harvard.edu/abs/2012NJPh...14e5023M} {14, 055023}

\bibitem[\protect\citeauthoryear{{Mei} et~al.,}{{Mei} et~al.}{2007}]{Meietal07}
{Mei} S.,  et~al., 2007, \mn@doi [\apj] {10.1086/509598}, \href
  {https://ui.adsabs.harvard.edu/abs/2007ApJ...655..144M} {655, 144}

\bibitem[\protect\citeauthoryear{{Meier}}{{Meier}}{1999}]{Meieretal1999}
{Meier} D.~L.,  1999, \mn@doi [\apj] {10.1086/307671}, \href
  {https://ui.adsabs.harvard.edu/abs/1999ApJ...522..753M} {522, 753}

\bibitem[\protect\citeauthoryear{{Morganti} et~al.,}{{Morganti}
  et~al.}{2006}]{Morgantietal06}
{Morganti} R.,  et~al., 2006, \mn@doi [MNRAS]
  {10.1111/j.1365-2966.2006.10681.x}, \href
  {http://adsabs.harvard.edu/abs/2006MNRAS.371..157M} {371, 157}

\bibitem[\protect\citeauthoryear{{Morganti}, {Peck}, {Oosterloo}, {van
  Moorsel}, {Capetti}, {Fanti}, {Parma}  \& {de Ruiter}}{{Morganti}
  et~al.}{2009}]{Morgantietal09}
{Morganti} R.,  {Peck} A.~B.,  {Oosterloo} T.~A.,  {van Moorsel} G.,  {Capetti}
  A.,  {Fanti} R.,  {Parma} P.,   {de Ruiter} H.~R.,  2009, \mn@doi [A\&A]
  {10.1051/0004-6361/200912605}, \href
  {http://adsabs.harvard.edu/abs/2009A\%26A...505..559M} {505, 559}

\bibitem[\protect\citeauthoryear{{Mukherjee}, {Bicknell}, {Sutherland}  \&
  {Wagner}}{{Mukherjee} et~al.}{2016}]{Mukherjeeetal16}
{Mukherjee} D.,  {Bicknell} G.~V.,  {Sutherland} R.,   {Wagner} A.,  2016,
  \mn@doi [\mnras] {10.1093/mnras/stw1368}, \href
  {https://ui.adsabs.harvard.edu/abs/2016MNRAS.461..967M} {461, 967}

\bibitem[\protect\citeauthoryear{{Nemmen}, {Bower}, {Babul}  \&
  {Storchi-Bergmann}}{{Nemmen} et~al.}{2007}]{Nemmenetal07}
{Nemmen} R.~S.,  {Bower} R.~G.,  {Babul} A.,   {Storchi-Bergmann} T.,  2007,
  \mn@doi [MNRAS] {10.1111/j.1365-2966.2007.11726.x}, \href
  {http://adsabs.harvard.edu/abs/2007MNRAS.377.1652N} {377, 1652}

\bibitem[\protect\citeauthoryear{{Nusser}, {Silk}  \& {Babul}}{{Nusser}
  et~al.}{2006}]{Nusseretal06}
{Nusser} A.,  {Silk} J.,   {Babul} A.,  2006, \mn@doi [\mnras]
  {10.1111/j.1365-2966.2006.11061.x}, \href
  {https://ui.adsabs.harvard.edu/abs/2006MNRAS.373..739N} {373, 739}

\bibitem[\protect\citeauthoryear{{Nyland} et~al.,}{{Nyland}
  et~al.}{2018}]{Nylandetal18}
{Nyland} K.,  et~al., 2018, \mn@doi [\apj] {10.3847/1538-4357/aab3d1}, \href
  {https://ui.adsabs.harvard.edu/abs/2018ApJ...859...23N} {859, 23}

\bibitem[\protect\citeauthoryear{{O'Dea} et~al.,}{{O'Dea}
  et~al.}{2008}]{ODeaetal08}
{O'Dea} C.~P.,  et~al., 2008, \mn@doi [ApJ] {10.1086/588212}, \href
  {http://adsabs.harvard.edu/abs/2008ApJ...681.1035O} {681, 1035}

\bibitem[\protect\citeauthoryear{{O'Sullivan}, {Combes}, {Hamer}, {Salom{\'e}},
  {Babul}  \& {Raychaudhury}}{{O'Sullivan} et~al.}{2015}]{OSullivanetal15}
{O'Sullivan} E.,  {Combes} F.,  {Hamer} S.,  {Salom{\'e}} P.,  {Babul} A.,
  {Raychaudhury} S.,  2015, \mn@doi [A\&A] {10.1051/0004-6361/201424835}, \href
  {http://adsabs.harvard.edu/abs/2015A\%26A...573A.111O} {573, A111}

\bibitem[\protect\citeauthoryear{{O'Sullivan} et~al.,}{{O'Sullivan}
  et~al.}{2017}]{OSullivanetal17}
{O'Sullivan} E.,  et~al., 2017, \mn@doi [MNRAS] {10.1093/mnras/stx2078}, 472,
  1482

\bibitem[\protect\citeauthoryear{{O'Sullivan}, {Kolokythas}, {Kantharia},
  {Raychaudhury}, {David}  \& {Vrtilek}}{{O'Sullivan}
  et~al.}{2018a}]{OSullivanetal18}
{O'Sullivan} E.,  {Kolokythas} K.,  {Kantharia} N.~G.,  {Raychaudhury} S.,
  {David} L.~P.,   {Vrtilek} J.~M.,  2018a, \mn@doi [MNRAS]
  {10.1093/mnras/stx2702}, \href
  {http://adsabs.harvard.edu/abs/2018MNRAS.473.5248O} {473, 5248}

\bibitem[\protect\citeauthoryear{{O'Sullivan} et~al.,}{{O'Sullivan}
  et~al.}{2018b}]{OSullivanetal18b}
{O'Sullivan} E.,  et~al., 2018b, \mn@doi [\aap] {10.1051/0004-6361/201833580},
  \href {https://ui.adsabs.harvard.edu/abs/2018A&A...618A.126O} {618, A126}

\bibitem[\protect\citeauthoryear{{Oosterloo} et~al.,}{{Oosterloo}
  et~al.}{2010}]{Oosterlooetal10}
{Oosterloo} T.,  et~al., 2010, \mn@doi [MNRAS]
  {10.1111/j.1365-2966.2010.17351.x}, \href
  {http://adsabs.harvard.edu/abs/2010MNRAS.409..500O} {409, 500}

\bibitem[\protect\citeauthoryear{{Oppenheimer}, {Babul}, {Bah{\'e}}, {Butsky}
  \& {McCarthy}}{{Oppenheimer} et~al.}{2021}]{Oppenheimeretal21}
{Oppenheimer} B.~D.,  {Babul} A.,  {Bah{\'e}} Y.,  {Butsky} I.~S.,   {McCarthy}
  I.~G.,  2021, \mn@doi [Universe] {10.3390/universe7070209}, \href
  {https://ui.adsabs.harvard.edu/abs/2021Univ....7..209O} {7, 209}

\bibitem[\protect\citeauthoryear{{Pearson} et~al.,}{{Pearson}
  et~al.}{2019}]{Pearsonetal19}
{Pearson} W.~J.,  et~al., 2019, \mn@doi [\aap] {10.1051/0004-6361/201936337},
  \href {https://ui.adsabs.harvard.edu/abs/2019A&A...631A..51P} {631, A51}

\bibitem[\protect\citeauthoryear{{Peng}}{{Peng}}{2007}]{Peng07}
{Peng} C.~Y.,  2007, \mn@doi [\apj] {10.1086/522774}, \href
  {https://ui.adsabs.harvard.edu/abs/2007ApJ...671.1098P} {671, 1098}

\bibitem[\protect\citeauthoryear{{Peng} et~al.,}{{Peng}
  et~al.}{2010}]{Pengetal10}
{Peng} Y.-j.,  et~al., 2010, \mn@doi [\apj] {10.1088/0004-637X/721/1/193},
  \href {https://ui.adsabs.harvard.edu/abs/2010ApJ...721..193P} {721, 193}

\bibitem[\protect\citeauthoryear{{Pipino}, {Kaviraj}, {Bildfell}, {Babul},
  {Hoekstra}  \& {Silk}}{{Pipino} et~al.}{2009}]{Pipinoetal09}
{Pipino} A.,  {Kaviraj} S.,  {Bildfell} C.,  {Babul} A.,  {Hoekstra} H.,
  {Silk} J.,  2009, \mn@doi [MNRAS] {10.1111/j.1365-2966.2009.14534.x}, \href
  {http://adsabs.harvard.edu/abs/2009MNRAS.395..462P} {395, 462}

\bibitem[\protect\citeauthoryear{{Prasad}, {Sharma}  \& {Babul}}{{Prasad}
  et~al.}{2015}]{Prasadetal15}
{Prasad} D.,  {Sharma} P.,   {Babul} A.,  2015, \mn@doi [ApJ]
  {10.1088/0004-637X/811/2/108}, \href
  {http://adsabs.harvard.edu/abs/2015ApJ...811..108P} {811, 108}

\bibitem[\protect\citeauthoryear{{Prasad}, {Sharma}  \& {Babul}}{{Prasad}
  et~al.}{2017}]{Prasadetal17}
{Prasad} D.,  {Sharma} P.,   {Babul} A.,  2017, \mn@doi [MNRAS]
  {10.1093/mnras/stx1698}, \href
  {http://adsabs.harvard.edu/abs/2017MNRAS.471.1531P} {471, 1531}

\bibitem[\protect\citeauthoryear{{Prasad}, {Sharma}  \& {Babul}}{{Prasad}
  et~al.}{2018}]{Prasadetal18}
{Prasad} D.,  {Sharma} P.,   {Babul} A.,  2018, ApJ, \href
  {http://adsabs.harvard.edu/abs/2018arXiv180104282P} {863, 62}

\bibitem[\protect\citeauthoryear{{Pulido} et~al.,}{{Pulido}
  et~al.}{2018}]{Pulidoetal18}
{Pulido} F.~A.,  et~al., 2018, \mn@doi [ApJ] {10.3847/1538-4357/aaa54b}, \href
  {http://adsabs.harvard.edu/abs/2018ApJ...853..177P} {853, 177}

\bibitem[\protect\citeauthoryear{{Rafferty}, {McNamara}, {Nulsen}  \&
  {Wise}}{{Rafferty} et~al.}{2006}]{Raffertyetal06}
{Rafferty} D.~A.,  {McNamara} B.~R.,  {Nulsen} P.~E.~J.,   {Wise} M.~W.,  2006,
  \mn@doi [ApJ] {10.1086/507672}, \href
  {http://adsabs.harvard.edu/abs/2006ApJ...652..216R} {652, 216}

\bibitem[\protect\citeauthoryear{{Rafferty}, {McNamara}  \&
  {Nulsen}}{{Rafferty} et~al.}{2008}]{Raffertyetal08}
{Rafferty} D.~A.,  {McNamara} B.~R.,   {Nulsen} P.~E.~J.,  2008, \mn@doi [\apj]
  {10.1086/591240}, \href
  {https://ui.adsabs.harvard.edu/abs/2008ApJ...687..899R} {687, 899}

\bibitem[\protect\citeauthoryear{{Sabater} et~al.,}{{Sabater}
  et~al.}{2019}]{Sabateretal19}
{Sabater} J.,  et~al., 2019, \mn@doi [\aap] {10.1051/0004-6361/201833883},
  \href {https://ui.adsabs.harvard.edu/abs/2019A&A...622A..17S} {622, A17}

\bibitem[\protect\citeauthoryear{{Salim} et~al.,}{{Salim}
  et~al.}{2007}]{Salim07}
{Salim} S.,  et~al., 2007, \mn@doi [\apjs] {10.1086/519218}, \href
  {https://ui.adsabs.harvard.edu/abs/2007ApJS..173..267S} {173, 267}

\bibitem[\protect\citeauthoryear{{Salmon} et~al.,}{{Salmon}
  et~al.}{2015}]{Salmonetal15}
{Salmon} B.,  et~al., 2015, \mn@doi [\apj] {10.1088/0004-637X/799/2/183}, \href
  {https://ui.adsabs.harvard.edu/abs/2015ApJ...799..183S} {799, 183}

\bibitem[\protect\citeauthoryear{{Salpeter}}{{Salpeter}}{1955}]{Salpeter55}
{Salpeter} E.~E.,  1955, \mn@doi [\apj] {10.1086/145971}, \href
  {https://ui.adsabs.harvard.edu/abs/1955ApJ...121..161S} {121, 161}

\bibitem[\protect\citeauthoryear{{Saulder}, {van Kampen}, {Chilingarian},
  {Mieske}  \& {Zeilinger}}{{Saulder} et~al.}{2016}]{Saulderetal16}
{Saulder} C.,  {van Kampen} E.,  {Chilingarian} I.~V.,  {Mieske} S.,
  {Zeilinger} W.~W.,  2016, \mn@doi [\aap] {10.1051/0004-6361/201526711}, \href
  {https://ui.adsabs.harvard.edu/abs/2016A&A...596A..14S} {596, A14}

\bibitem[\protect\citeauthoryear{{Schellenberger} et~al.,}{{Schellenberger}
  et~al.}{2020}]{Schellenbergeretal20}
{Schellenberger} G.,  et~al., 2020, \mn@doi [ApJ] {10.3847/1538-4357/ab879c},
  \href {https://ui.adsabs.harvard.edu/abs/2020ApJ...894...72S} {894, 72}

\bibitem[\protect\citeauthoryear{{Schombert} \& {Smith}}{{Schombert} \&
  {Smith}}{2012}]{SchombertSmith12}
{Schombert} J.,  {Smith} A.~K.,  2012, \mn@doi [\pasa] {10.1071/AS11059}, \href
  {https://ui.adsabs.harvard.edu/abs/2012PASA...29..174S} {29, 174}

\bibitem[\protect\citeauthoryear{{Shulevski}, {Morganti}, {Oosterloo}  \&
  {Struve}}{{Shulevski} et~al.}{2012}]{Shulevskietal12}
{Shulevski} A.,  {Morganti} R.,  {Oosterloo} T.,   {Struve} C.,  2012, \mn@doi
  [A\&A] {10.1051/0004-6361/201219869}, \href
  {http://adsabs.harvard.edu/abs/2012A\%26A...545A..91S} {545, A91}

\bibitem[\protect\citeauthoryear{{Silk}}{{Silk}}{2013}]{Silk13}
{Silk} J.,  2013, \mn@doi [\apj] {10.1088/0004-637X/772/2/112}, \href
  {https://ui.adsabs.harvard.edu/abs/2013ApJ...772..112S} {772, 112}

\bibitem[\protect\citeauthoryear{{Silk} \& {Rees}}{{Silk} \&
  {Rees}}{1998}]{SilkRees98}
{Silk} J.,  {Rees} M.~J.,  1998, \aap, \href
  {https://ui.adsabs.harvard.edu/abs/1998A&A...331L...1S} {331, L1}

\bibitem[\protect\citeauthoryear{{Skibba} \& {Sheth}}{{Skibba} \&
  {Sheth}}{2009}]{SkibbaSheth09}
{Skibba} R.~A.,  {Sheth} R.~K.,  2009, \mn@doi [\mnras]
  {10.1111/j.1365-2966.2008.14007.x}, \href
  {https://ui.adsabs.harvard.edu/abs/2009MNRAS.392.1080S} {392, 1080}

\bibitem[\protect\citeauthoryear{{Somerville} \& {Dav{\'e}}}{{Somerville} \&
  {Dav{\'e}}}{2015}]{SomervilleDave15}
{Somerville} R.~S.,  {Dav{\'e}} R.,  2015, \mn@doi [\araa]
  {10.1146/annurev-astro-082812-140951}, \href
  {https://ui.adsabs.harvard.edu/abs/2015ARA&A..53...51S} {53, 51}

\bibitem[\protect\citeauthoryear{{Stern} et~al.,}{{Stern}
  et~al.}{2012}]{Sternetal12}
{Stern} D.,  et~al., 2012, \mn@doi [\apj] {10.1088/0004-637X/753/1/30}, \href
  {https://ui.adsabs.harvard.edu/abs/2012ApJ...753...30S} {753, 30}

\bibitem[\protect\citeauthoryear{{Tadhunter} et~al.,}{{Tadhunter}
  et~al.}{2011}]{Tadhunteretal11}
{Tadhunter} C.,  et~al., 2011, \mn@doi [\mnras]
  {10.1111/j.1365-2966.2010.17958.x}, \href
  {https://ui.adsabs.harvard.edu/abs/2011MNRAS.412..960T} {412, 960}

\bibitem[\protect\citeauthoryear{{Tang}, {Gu}, {Huang}  \& {Wang}}{{Tang}
  et~al.}{2009}]{Tangetal09b}
{Tang} Y.,  {Gu} Q.~S.,  {Huang} J.~S.,   {Wang} Y.~P.,  2009, \mn@doi [\mnras]
  {10.1111/j.1365-2966.2009.15038.x}, \href
  {https://ui.adsabs.harvard.edu/abs/2009MNRAS.397.1966T} {397, 1966}

\bibitem[\protect\citeauthoryear{{Tanimoto}, {Ueda}, {Kawamuro}  \&
  {Ricci}}{{Tanimoto} et~al.}{2016}]{Tanimotoetal16}
{Tanimoto} A.,  {Ueda} Y.,  {Kawamuro} T.,   {Ricci} C.,  2016, \mn@doi [\pasj]
  {10.1093/pasj/psw008}, \href
  {https://ui.adsabs.harvard.edu/abs/2016PASJ...68S..26T} {68, S26}

\bibitem[\protect\citeauthoryear{{Temi}, {Amblard}, {Gitti}, {Brighenti},
  {Gaspari}, {Mathews}  \& {David}}{{Temi} et~al.}{2018}]{Temietal17}
{Temi} P.,  {Amblard} A.,  {Gitti} M.,  {Brighenti} F.,  {Gaspari} M.,
  {Mathews} W.~G.,   {David} L.,  2018, \mn@doi [ApJ]
  {10.3847/1538-4357/aab9b0}, \href
  {http://adsabs.harvard.edu/abs/2018ApJ...858...17T} {858, 17}

\bibitem[\protect\citeauthoryear{{Tonry}, {Dressler}, {Blakeslee}, {Ajhar},
  {Fletcher}, {Luppino}, {Metzger}  \& {Moore}}{{Tonry}
  et~al.}{2001}]{Tonryetal01}
{Tonry} J.~L.,  {Dressler} A.,  {Blakeslee} J.~P.,  {Ajhar} E.~A.,  {Fletcher}
  A.~B.,  {Luppino} G.~A.,  {Metzger} M.~R.,   {Moore} C.~B.,  2001, ApJ, \href
  {http://adsabs.harvard.edu/cgi-bin/nph-bib_query?bibcode=2001ApJ...546..681T&db_key=AST}
  {546, 681}

\bibitem[\protect\citeauthoryear{{Tully}}{{Tully}}{1987}]{Tully87}
{Tully} R.~B.,  1987, ApJ, 321, 280

\bibitem[\protect\citeauthoryear{{Vaddi}, {O'Dea}, {Baum}, {Whitmore}, {Ahmed},
  {Pierce}  \& {Leary}}{{Vaddi} et~al.}{2016}]{Vaddietal16}
{Vaddi} S.,  {O'Dea} C.~P.,  {Baum} S.~A.,  {Whitmore} S.,  {Ahmed} R.,
  {Pierce} K.,   {Leary} S.,  2016, \mn@doi [\apj]
  {10.3847/0004-637X/818/2/182}, \href
  {https://ui.adsabs.harvard.edu/abs/2016ApJ...818..182V} {818, 182}

\bibitem[\protect\citeauthoryear{{Veale}, {Ma}, {Greene}, {Thomas},
  {Blakeslee}, {McConnell}, {Walsh}  \& {Ito}}{{Veale}
  et~al.}{2017}]{Vealeetal17}
{Veale} M.,  {Ma} C.-P.,  {Greene} J.~E.,  {Thomas} J.,  {Blakeslee} J.~P.,
  {McConnell} N.,  {Walsh} J.~L.,   {Ito} J.,  2017, \mn@doi [\mnras]
  {10.1093/mnras/stx1639}, \href
  {https://ui.adsabs.harvard.edu/abs/2017MNRAS.471.1428V} {471, 1428}

\bibitem[\protect\citeauthoryear{{Walker}, {Johnson}, {Gallagher}, {Hibbard},
  {Hornschemeier}, {Tzanavaris}, {Charlton}  \& {Jarrett}}{{Walker}
  et~al.}{2010}]{Walkeretal10}
{Walker} L.~M.,  {Johnson} K.~E.,  {Gallagher} S.~C.,  {Hibbard} J.~E.,
  {Hornschemeier} A.~E.,  {Tzanavaris} P.,  {Charlton} J.~C.,   {Jarrett}
  T.~H.,  2010, \mn@doi [\aj] {10.1088/0004-6256/140/5/1254}, \href
  {https://ui.adsabs.harvard.edu/abs/2010AJ....140.1254W} {140, 1254}

\bibitem[\protect\citeauthoryear{{Weigel}, {Schawinski}, {Treister},
  {Trakhtenbrot}  \& {Sanders}}{{Weigel} et~al.}{2018}]{Weigeletal18}
{Weigel} A.~K.,  {Schawinski} K.,  {Treister} E.,  {Trakhtenbrot} B.,
  {Sanders} D.~B.,  2018, \mn@doi [\mnras] {10.1093/mnras/sty383}, \href
  {https://ui.adsabs.harvard.edu/abs/2018MNRAS.476.2308W} {476, 2308}

\bibitem[\protect\citeauthoryear{{Werle} et~al.,}{{Werle}
  et~al.}{2020}]{Werleetal20}
{Werle} A.,  et~al., 2020, \mn@doi [\mnras] {10.1093/mnras/staa2217}, \href
  {https://ui.adsabs.harvard.edu/abs/2020MNRAS.497.3251W} {497, 3251}

\bibitem[\protect\citeauthoryear{{Wetzel}, {Tinker}  \& {Conroy}}{{Wetzel}
  et~al.}{2012}]{Wetzeletal12}
{Wetzel} A.~R.,  {Tinker} J.~L.,   {Conroy} C.,  2012, \mn@doi [\mnras]
  {10.1111/j.1365-2966.2012.21188.x}, \href
  {https://ui.adsabs.harvard.edu/abs/2012MNRAS.424..232W} {424, 232}

\bibitem[\protect\citeauthoryear{{Whitaker}, {van Dokkum}, {Brammer}  \&
  {Franx}}{{Whitaker} et~al.}{2012}]{Whitakeretal12}
{Whitaker} K.~E.,  {van Dokkum} P.~G.,  {Brammer} G.,   {Franx} M.,  2012,
  \mn@doi [\apjl] {10.1088/2041-8205/754/2/L29}, \href
  {https://ui.adsabs.harvard.edu/abs/2012ApJ...754L..29W} {754, L29}

\bibitem[\protect\citeauthoryear{{White} \& {Rees}}{{White} \&
  {Rees}}{1978}]{WhiteRees78}
{White} S.~D.~M.,  {Rees} M.~J.,  1978, \mn@doi [\mnras]
  {10.1093/mnras/183.3.341}, \href
  {https://ui.adsabs.harvard.edu/abs/1978MNRAS.183..341W} {183, 341}

\bibitem[\protect\citeauthoryear{{Wold}, {Lacy}  \& {Armus}}{{Wold}
  et~al.}{2007}]{Woldetal2007}
{Wold} M.,  {Lacy} M.,   {Armus} L.,  2007, \mn@doi [\aap]
  {10.1051/0004-6361:20065266}, \href
  {https://ui.adsabs.harvard.edu/abs/2007A&A...470..531W} {470, 531}

\bibitem[\protect\citeauthoryear{{Wright} et~al.,}{{Wright}
  et~al.}{2010}]{Wrightetal10}
{Wright} E.~L.,  et~al., 2010, \mn@doi [\aj] {10.1088/0004-6256/140/6/1868},
  \href {https://ui.adsabs.harvard.edu/abs/2010AJ....140.1868W} {140, 1868}

\bibitem[\protect\citeauthoryear{{Wuyts} et~al.,}{{Wuyts}
  et~al.}{2011}]{Wuytsetal11}
{Wuyts} S.,  et~al., 2011, \mn@doi [ApJ] {10.1088/0004-637X/742/2/96}, \href
  {http://adsabs.harvard.edu/abs/2011ApJ...742...96W} {742, 96}

\bibitem[\protect\citeauthoryear{{Wyder} et~al.,}{{Wyder}
  et~al.}{2005}]{Wyderetal05}
{Wyder} T.~K.,  et~al., 2005, \mn@doi [\apjl] {10.1086/424735}, \href
  {https://ui.adsabs.harvard.edu/abs/2005ApJ...619L..15W} {619, L15}

\bibitem[\protect\citeauthoryear{{Xu}, {Gu}, {Lu}, {Ge}, {Xiao}  \&
  {Contini}}{{Xu} et~al.}{2021}]{Keetal2021}
{Xu} K.,  {Gu} Q.,  {Lu} S.,  {Ge} X.,  {Xiao} M.,   {Contini} E.,  2021, arXiv
  e-prints, \href {https://ui.adsabs.harvard.edu/abs/2021arXiv211007935X} {p.
  arXiv:2110.07935}

\bibitem[\protect\citeauthoryear{{van der Wel}, {Rix}, {Holden}, {Bell}  \&
  {Robaina}}{{van der Wel} et~al.}{2009}]{vanderWeletal09}
{van der Wel} A.,  {Rix} H.-W.,  {Holden} B.~P.,  {Bell} E.~F.,   {Robaina}
  A.~R.,  2009, \mn@doi [\apjl] {10.1088/0004-637X/706/1/L120}, \href
  {https://ui.adsabs.harvard.edu/abs/2009ApJ...706L.120V} {706, L120}

\makeatother
\end{thebibliography}

\bsp	

\label{lastpage}

\end{document}